\documentclass[a4paper,11pt]{article}
\usepackage[left=20mm,right=20mm,top=20mm,bottom=20mm]{geometry}
\usepackage{amsmath}
\usepackage{amsfonts}
\usepackage{amssymb}
\usepackage{amsbsy}
\usepackage{color,graphics}
\usepackage{setspace}
\usepackage{epsfig}
\usepackage{cite}
\usepackage{graphicx}
\usepackage{latexsym,multicol}
\newtheorem{theorem}{Theorem}[section]

\newtheorem{definition}[theorem]{Definition}

\graphicspath{{./SEInhom-figs/}}
\begin{document}
\title{Likely cavitation and radial motion of stochastic elastic spheres}
\author{L. Angela Mihai\footnote{School of Mathematics, Cardiff University, Senghennydd Road, Cardiff, CF24 4AG, UK, Email: \texttt{MihaiLA@cardiff.ac.uk}}
	\qquad Thomas E. Woolley\footnote{School of Mathematics, Cardiff University, Senghennydd Road, Cardiff, CF24 4AG, UK, Email: \texttt{WoolleyT1@cardiff.ac.uk}}
	\qquad Alain Goriely\footnote{Mathematical Institute, University of Oxford, Woodstock Road, Oxford, OX2 6GG, UK, Email: \texttt{goriely@maths.ox.ac.uk}}
}	%\date{}
\maketitle

\begin{abstract}
The cavitation of solid elastic spheres is a classical problem of continuum mechanics. Here, we study this problem within the context of ``stochastic elasticity'' where the constitutive parameters are characterised by probability density functions. We consider homogeneous spheres of stochastic neo-Hookean material, composites with two concentric stochastic neo-Hookean phases, and inhomogeneous spheres of locally neo-Hookean material with a radially varying parameter. In all cases, we show that the material at the centre determines the critical load at which a spherical cavity forms there. However, while under dead-load traction, a supercritical bifurcation, with stable cavitation, is obtained in a static sphere of stochastic neo-Hookean material, for the composite and radially inhomogeneous spheres, a subcritical bifurcation, with snap cavitation, is also possible. For the dynamic spheres, oscillatory motions are produced under suitable dead-load traction, such that a cavity forms and expands to a maximum radius, then collapses again to zero periodically, but not under impulse traction. Under a surface impulse, a subcritical bifurcation is found in a static sphere of stochastic neo-Hookean material and also in an inhomogeneous sphere, whereas in composite spheres, supercritical bifurcations can occur as well. Given the non-deterministic material parameters, the results can be characterised in terms of probability distributions.\\
	
\end{abstract}

%\tableofcontents

\begin{quote}
	``It is a problem of mechanics, highly complicated and irrelevant to probability theory except insofar as it forces us to think a little more carefully about how probability theory must be formulated if it is to be applicable to real situations.'' - E.T. Jaynes (1996) \cite{Jaynes:2003}
\end{quote}

%%%%%%%%%%%%%%%%%%%%%%%%%%%%%%%%%%%%%%%%%%%%%%%%%%%%%%%%%%%%
%%%%%%%%%%%%%%%%%%%%   NEW SECTION  %%%%%%%%%%%%%%%%%%%%%%%%
%%%%%%%%%%%%%%%%%%%%%%%%%%%%%%%%%%%%%%%%%%%%%%%%%%%%%%%%%%%%
\section{Introduction}\label{sec:intro}

Cavitation in solids represents the void-formation within a solid under tensile loads, by analogy to the similar phenomenon observed in fluids. For rubber-like materials, this phenomenon was first reported by Gent and Lindley (1958) \cite{Gent:1959:GL}, whose experiments showed that rubber cylinders ruptured under relatively small tensile dead loads by opening an internal cavity. The nonlinear elastic analysis of Ball (1982) \cite{Ball:1982} provided the first theoretical explanation for the formation of a spherical cavity at the centre of a sphere of isotropic incompressible hyperelastic material in radially symmetric tension under prescribed surface displacements or dead loads. There, cavitation was treated as a bifurcation from the trivial state at a critical value of the surface traction or dis­placement, at which the trivial solution became unstable. This paved the way for numerous applied and theoretical studies devoted to this inherently nonlinear mechanical effect, which is not captured by the linear elasticity theory.

Many important results focusing on rubber-like materials are reviewed in \cite{Fond:2001,Gent:1991,Horgan:1995:HP}. Recent experimental studies regarding the onset, healing and growth of cavities in elastomers can be found in \cite{Hutchens:2016:HFC,Poulain:2017:PLLPRC,Poulain:2018:PLPRC,Raayai:2019:etal}. Finite element simulations are presented in \cite{Kang:2018:KWT}. While most of this work was carried out in the setting of static deformations, in general, the elastodynamics of nonlinear hyperelastic bodies has been less investigated. The governing equations for dynamic large strain deformations leading to finite amplitude oscillations of spherical and cylindrical shells of homogeneous isotropic incompressible hyperelastic material, treated as particular cases of \emph{quasi-equilibrated motions} defined by Truesdell (1962) \cite{Truesdell:1962}, were reviewed in \cite{TruesdellNoll:2004}. Under this type of dynamic deformation, the stress field is determined by the current configuration alone, and the body can be brought instantly to rest by applying a suitable pressure impulse on its boundary. Free and forced oscillations of spherical shells were first obtained in \cite{Heng:1963:HS,Knowles:1965:KJ,Wang:1965}, using a similar approach as introduced in \cite{Knowles:1960,Knowles:1962} for the axially symmetric radial oscillations of infinitely long, isotropic incompressible cylindrical tubes. Oscillatory motion caused by the dynamic cavitation of a neo-Hookean sphere was considered in \cite{ChouWang:1989b:CWH}. For a wide class of hyperelastic materials, the static and dynamic cavitation of homogeneous spheres were studied in \cite{Ball:1982}. For a hyperelastic sphere of Mooney-Rivlin material containing a cavity, the numerical solution to the nonlinear problem of large amplitude oscillations was presented in \cite{Balakrishnan:1978:BS}. Theoretical and experimental studies of cylindrical and spherical shells of rubber-like material under external pressure were proposed in \cite{Wang:1972:WE}. The finite amplitude radial oscillations of homogeneous isotropic incompressible hyperelastic spherical and cylindrical shells under a constant pressure difference between the inner and the outer surface were analysed theoretically in \cite{Calderer:1983}. Radial oscillations of non-homogeneous thick-walled cylindrical and spherical shells of neo-Hookean material, with a material parameter varying continuously along the radial direction, were treated in \cite{Ertepinar:1976:EA}. In \cite{Verron:1999:VKDR}, the dynamic inflation of hyperelastic spherical membranes of Mooney-Rivlin material subjected to a uniform step pressure was considered, and the absence of damping was discussed. It was concluded that, as the amplitude and period of oscillations are strongly influenced by the rate of internal pressure, if the pressure was suddenly imposed and the inflation process was short, then sustained oscillations due to the dominant elastic effects could be observed. However, in many systems where the pressure is slowly increasing, strong damping would generally preclude oscillations \cite{DePascalis:etal:2018}. The dynamic response of incompressible hyperelastic spherical shells subjected to periodic loading was examined in \cite{Ren:2009}. Radial oscillations of spherical shells of different hyperelastic materials were analysed in \cite{Beatty:2011}, where it was found that both the amplitude and period of oscillations decrease as the stiffness of the material increases. The influence of material constitutive law on the dynamic behaviour of spherical membranes was also considered in \cite{Martinez:2015:etal}. In \cite{Soares:etal:2019}, the nonlinear static and dynamic behaviour of a spherical membrane of neo-Hookean or Mooney-Rivlin material, subjected to a uniformly distributed radial pressure on the inner surface, was analysed, and the influence of the material constants was examined.

Despite these developments, there are important issues that remain unresolved. In particular, the quantification of uncertainties in material parameters and responses resulting from incomplete information remain largely unexplored \cite{Soize:2017}. However, for many materials, deterministic approaches, which are based on average data values, can greatly underestimate or overestimate their properties, and stochastic representations accounting also for data dispersion are needed to improve assessment and predictions \cite{Ghanem:2017:GHOR,Hughes:2010:HH,Kaminski:2018:KL,Oden:2018,Ostoja:2007,Sullivan:2015}.

For nonlinear elastic materials, stochastic hyperelastic models have recently been described by strain-energy densities where the parameters are characterised by probability density functions (see \cite{Staber:2015:SG,Staber:2016:SG,Staber:2017:SG,Staber:2018:SG,Staber:2019:SGSMI} and also \cite{Mihai:2018:MWG}). These are advanced phenomenological models, for which the parameters can be inferred from macroscopic load-deformation experiments, as shown, for example, in \cite{Fitt:2019:FWWM}. They rely on the finite elasticity theory as prior information, and on the notion of entropy (or uncertainty) \cite{Shannon:1948,Soni:2017:SG,Jaynes:1957a,Jaynes:1957b,Jaynes:2003} to enable the propagation of uncertainties from input data to output quantities of interest \cite{Soize:2013}, and can also be incorporated into Bayesian approaches \cite{Bayes:1763,McGrayne:2012} for model selection \cite{Fitt:2019:FWWM,Mihai:2018:MWG,Oden:2018,Robert:2007}.

To demonstrate the effect of probabilistic model parameters on predicted mechanical responses, for stochastic homogeneous incompressible hyperelastic bodies, we have analysed theoretically, so far, the static cavitation of a sphere under uniform tensile dead load \cite{Mihai:2019c:MDWG}, the inflation of pressurised spherical and cylindrical shells \cite{Mihai:2019a:MDWG}, the classic problem of the Rivlin cube \cite{Mihai:2019a:MWG}, the radial oscillatory motion of cylindrical and spherical shells under a uniform pressure impulse \cite{Mihai:2019b:MDWG}, and the rotation and perversion of stochastic incompressible anisotropic hyperelastic cylindrical tubes \cite{Mihai:2019b:MWG}. In \cite{Mihai:2019c:MDWG}, in addition to the well-known case of stable cavitation post-bifurcation at the critical dead load, we show, for the first time, the existence of unstable (snap) cavitation for some (deterministic or stochastic) isotropic incompressible materials satisfying Baker-Ericksen (BE) inequalities \cite{BakerEricksen:1954}. In general, these problems, for which the elastic solutions are given explicitly, can offer significant insight into how the uncertainties in input parameters are propagated to output quantities. To investigate the effect of probabilistic parameters in the case of more complex geometries and loading conditions, numerical approaches have been proposed in \cite{Staber:2018:SG,Staber:2019:SGSMI}. A similar stochastic methodology can be developed to study instabilities in other material systems (e.g., liquid crystal elastomers \cite{DeSimone:2009:dST,Fried:2004:FS,Warner:2007:WT}), and may lead to more accurate assessment and prediction in many application areas \cite{Cidonio:2019:CGDO,Holmes:2019}.

In this paper, we examine the static and dynamic cavitation and post-cavitation behaviour of an incompressible sphere of stochastic hyperelastic material, subject to either a uniform tensile surface dead load (which is constant in the reference configuration) or an impulse traction (which is kept constant in the current configuration), and each of them is applied uniformly in the radial direction. Here, cavitation represents the opening of a spherical cavity at the centre of the sphere as a bifurcated solution from the trivial solution, which becomes unstable at a critical value of the applied load. Moreover, the two different types of loads will lead to different static and dynamic solutions. First, in Section~\ref{sec:prerequisites}, we provide a summary of the stochastic elasticity prerequisites, where, in addition to the stochastic homogeneous material models, for which the elastic parameters are spatially-independent random variables at a macroscopic (non-molecular) level, we introduce a class of stochastic inhomogeneous models described by spatially-dependent strain-energy functions for which the parameters are non-Gaussian random fields. We then recall the notion of quasi-equilibrated motion, where the deformation field is circulation preserving and, at every time instant, the deformed state is possible also as a static equilibrium state under the same forces. Section~\ref{sec:sphere:NH} is devoted to the explicit analysis of cavitation and finite amplitude oscillations of a homogeneous sphere of stochastic neo-Hookean material. In Sections~\ref{sec:sphere:comp} and~\ref{sec:sphere:inhom}, we extend the analysis, respectively, to the cavitation and finite amplitude oscillations of a composite formed from two concentric homogeneous spheres of different stochastic neo-Hookean material, and to inhomogeneous spheres of locally neo-Hookean material with a radially varying parameter. In each case, the dynamic problem is considered first, then the static problem is treated by reducing the quasi-equilibrated motion to a static equilibrium state. In all cases, we show that the material at the centre determines the critical load at which a spherical cavity forms. However, there are important differences in the post-cavitation nonlinear elastic responses. While under dead-load traction, supercritical bifurcation, with stable cavitation, is obtained in a static sphere of stochastic neo-Hookean material, for the composite and radially inhomogeneous spheres, a subcritical bifurcation, with snap cavitation, is also possible. For the dynamic spheres, oscillatory motions are produced under suitable dead-load traction, such that a cavity forms and expands to a maximum radius, then collapses again to zero periodically, but not under impulse traction. Under a surface impulse, subcritical bifurcation is found in a static sphere of stochastic neo-Hookean material and also in an inhomogeneous sphere, whereas in composite spheres, supercritical bifurcation can occur as well. As the input material parameters are non-deterministic, the results can be characterised in terms of probability distributions. In Section~\ref{sec:conclusion}, we draw concluding remarks and highlight potential extensions of the stochastic approach. 

%%%%%%%%%%%%%%%%%%%%%%%%%%%%%%%%%%%%%%%%%%%%%%%%%%%%%%%%%%%%
%%%%%%%%%%%%%%%%%%%%   NEW SECTION  %%%%%%%%%%%%%%%%%%%%%%%%
%%%%%%%%%%%%%%%%%%%%%%%%%%%%%%%%%%%%%%%%%%%%%%%%%%%%%%%%%%%%
\section{Prerequisites}\label{sec:prerequisites}

In this section, we introduce a class of stochastic isotropic incompressible inhomogeneous hyperelastic models, where the constitutive parameter is spatially-dependent. This can be regarded as an extension of the stochastic homogeneous models in  \cite{Fitt:2019:FWWM,Mihai:2019a:MDWG,Mihai:2019b:MDWG,Mihai:2019c:MDWG,Mihai:2018:MWG,Mihai:2019a:MWG}, where both the mean value and variance of model parameters were constant in space, and also as a particular case of the inhomogeneous anisotropic ones discussed in \cite{Staber:2018:SG,Staber:2019:SGSMI}, which assumed heterogeneity in variance. 

As both static and dynamic spheres of either homogeneous or inhomogeneous stochastic material will be considered in the next sections, we also recall the notion of quasi-equilibrated motion defined in \cite{Truesdell:1962,TruesdellNoll:2004}. Quasi-equilibrated motions were first treated by us in \cite{Mihai:2019b:MDWG}, where the finite amplitude oscillations of cylindrical and spherical shells of stochastic isotropic incompressible homogeneous hyperelastic material under uniform pressure impulse were analysed. This type of motions can be reduced to static equilibrium states at any time instant under the given load.

%%%%%%%%%%%%%%%%%%%%%%%%%%%%%%%%%%%%%%%%%%%%%%%%%%%%%%%%%%%%
\subsection{Stochastic hyperelastic models}\label{sec:models}
We are concerned here with the large strain analysis of spherical bodies of stochastic incompressible isotropic hyperelastic material, and rely on the following physically realistic assumptions \cite{Fitt:2019:FWWM,Mihai:2019a:MDWG,Mihai:2019b:MDWG,Mihai:2019c:MDWG,Mihai:2018:MWG,Mihai:2019a:MWG}:
\begin{itemize}
	\item[(A1)] Material objectivity, stating that constitutive equations must be invariant under changes of frame of reference. This requires that the scalar strain-energy function, $W=W(\textbf{F})$, depending only on the deformation gradient $\textbf{F}$, with respect to the reference configuration, is unaffected by a superimposed rigid-body transformation (which involves a change of position) after deformation, i.e., $W(\textbf{R}^{T}\textbf{F})=W(\textbf{F})$, where $\textbf{R}\in SO(3)$ is a proper orthogonal tensor (rotation). Material objectivity is guaranteed by defining strain-energy functions in terms of the scalar invariants.
	
	\item[(A2)] Material isotropy, requiring that the strain-energy function is unaffected by a superimposed rigid-body transformation prior to deformation, i.e., $W(\textbf{F}\textbf{Q})=W(\textbf{F})$, where $\textbf{Q}\in SO(3)$. For isotropic materials, the  strain-energy  function is a symmetric function of the  principal stretches $\{\lambda_{i}\}_{i=1,2,3}$ of $\textbf{F}$, i.e., $W(\textbf{F})=\mathcal{W}(\lambda_{1},\lambda_{2},\lambda_{3})$.
	
	\item[(A3)] Baker-Ericksen (BE) inequalities, which state that \emph{the greater principal (Cauchy) stress occurs in the direction of the greater principal stretch} \cite{BakerEricksen:1954,Marzano:1983}, and take the equivalent form
	\begin{equation}\label{eq:W:BE}
	\left(\lambda_{i}\frac{\partial\mathcal{W}}{\partial\lambda_{i}}-\lambda_{j}\frac{\partial\mathcal{W}}{\partial\lambda_{j}}\right)\left(\lambda_{i}-\lambda_{j}\right)>0\quad \mbox{if}\quad \lambda_{i}\neq\lambda_{j},\quad i,j=1,2,3.
	\end{equation}
	When any two principal stretches are equal, the strict inequality ``$>$''  in \eqref{eq:W:BE} is replaced by ``$\geq$''.
	
	\item[(A4)] For any given finite deformation, at any point in the material, the shear modulus, $\mu$, and its inverse, $1/\mu$, are second-order random variables, i.e., they have finite mean value and finite variance \cite{Staber:2015:SG,Staber:2016:SG,Staber:2017:SG,Staber:2018:SG,Staber:2019:SGSMI}.
\end{itemize}
Assumptions (A1)-(A3) are general requirements in isotropic finite elasticity \cite{goriely17,Mihai:2017:MG,Ogden:1997,TruesdellNoll:2004}. In particular, (A3) implies that the shear modulus is always positive, i.e., $\mu>0$ \cite{Mihai:2017:MG}, while (A4) places random variables at the foundation on which to construct hyperelastic models \cite{Freiling:1986,Jaynes:2003,Mihai:2018:MWG,Mumford:2000,Soize:2017}. 

To make our approach analytically tractable, we restrict our attention to a class of inhomogeneous models defined by the strain-energy density
\begin{equation}\label{eq:W:stoch}
\mathcal{W}(\lambda_{1},\lambda_{2},\lambda_{3})=\frac{\mu}{2}\left(\lambda_{1}^2+\lambda_{2}^2+\lambda_{3}^2-3\right),
\end{equation}
where  the shear modulus, $\mu=\mu(R)>0$, is a random field variable depending on the radius $R$ of the undeformed sphere, and $\lambda_{1}$, $\lambda_{2}$, and $\lambda_{3}$ are the principal stretch ratios. When $\mu$ is independent of $R$, the material model \eqref{eq:W:stoch} reduces to the stochastic neo-Hookean model \cite{Staber:2015:SG,Mihai:2018:MWG}.

For the shear modulus, $\mu=\mu(R)$,  at any fixed $R$, we rely on the following available information
\begin{eqnarray}\label{eq:Emu1}\begin{cases}
E\left[\mu\right]=\underline{\mu}>0,&\\
E\left[\log\ \mu\right]=\nu,& \mbox{such that $|\nu|<+\infty$}.\label{eq:Emu2}\end{cases}
\end{eqnarray}
Under the constraints (\ref{eq:Emu1}), assumption (A4) is guaranteed \cite{Soize:2000,Soize:2001,Soize:2006,Soize:2017}. Then, by the maximum entropy principle, for any fixed $R$, $\mu$, follows a Gamma probability distribution with the shape and scale parameters,  $\rho_{1}=\rho_{1}(R)>0$ and $\rho_{2}=\rho_{2}(R)>0$, respectively, satisfying
\begin{equation}\label{eq:rho12}
\underline{\mu}=\rho_{1}\rho_{2},\qquad
\text{Var}[\mu]=\rho_{1}\rho_{2}^2,
\end{equation}
where $\underline{\mu}$ is the mean value, $\text{Var}[\mu]=\|\mu\|^2$ is the variance, and $\|\mu\|$ is the standard deviation of $\mu$. The corresponding probability density function takes the form \cite{Abramowitz:1964,Grimmett:2001:GS,Johnson:1994:JKB}
\begin{equation}\label{eq:mu:gamma}
g(\mu;\rho_{1},\rho_{2})=\frac{\mu^{\rho_{1}-1}e^{-\mu/\rho_{2}}}{\rho_{2}^{\rho_{1}}\Gamma(\rho_{1})},\qquad\mbox{for}\ \mu>0\ \mbox{and}\ \rho_{1}, \rho_{2}>0,
\end{equation}
where $\Gamma:\mathbb{R}^{*}_{+}\to\mathbb{R}$ is the complete Gamma function
\begin{equation}\label{eq:gamma}
\Gamma(z)=\int_{0}^{+\infty}t^{z-1}e^{-t}\text dt.
\end{equation}
The word `hyperparameters' is also used for the parameters $\rho_{1}$ and $\rho_{2}$ of the Gamma distribution, to distinguish them from the material parameter $\mu$ and other material constants \cite[p.~8]{Soize:2017}. Examples where $\mu=\mu(R)$ takes specific forms are discussed in Section~\ref{sec:sphere:inhom}.

%%%%%%%%%%%%%%%%%%%%%%%%%%%%%%%%%%%%%%%%%%%%%%%%%%%%%%%%%%%%
\subsection{Quasi-equilibrated motion}\label{sec:qemotion}

For the large strain dynamic deformation of an elastic solid, Cauchy's laws of motion (balance laws of linear and angular momentum) are governed by the following Eulerian field equations \cite[p.~40]{TruesdellNoll:2004},
\begin{eqnarray}
&&\rho\ddot{\textbf{x}}=\mathrm{div}\ \textbf{T}+\rho\textbf{b},\label{eq:1st}\\
&&\textbf{T}=\textbf{T}^{T},\label{eq:2nd}
\end{eqnarray}
where $\rho$ is the material density, which is assumed constant, $\textbf{x}=\chi(\textbf{X},t)$ is the motion of the elastic solid, with velocity $\dot{\textbf{x}}=\partial\chi(\textbf{X},t)/\partial t$ and acceleration $\ddot{\textbf{x}}=\partial^2\chi(\textbf{X},t)/\partial t^2$, $\textbf{b}=\textbf{b}(\textbf{x},t)$ is the body force, $\textbf{T}=\textbf{T}(\textbf{x},t)$ is the Cauchy stress tensor, and the superscript $T$ defines the transpose. 

To obtain possible dynamical solutions, one can solve Cauchy's equation for particular motions, or generalise known static solutions to dynamical forms using the universal notion of \textit{quasi-equilibrated motion}, which is defined as follows.

\begin{definition}\label{def:qem} \cite[p.~208]{TruesdellNoll:2004}
	A quasi-equilibrated motion, $\textbf{x}=\chi(\textbf{X},t)$, is the motion of an incompressible homogeneous elastic solid subject to a given body force, $\textbf{b}=\textbf{b}(\textbf{x},t)$, whereby, for each value of $t$, $\textbf{x}=\chi(\textbf{X},t)$ defines a static deformation that satisfies the equilibrium conditions under the body force $\textbf{b}=\textbf{b}(\textbf{x},t)$.
\end{definition}

\begin{theorem}\label{th:qem} \cite[p.~208]{TruesdellNoll:2004} (see also the proof in \cite{Mihai:2019b:MDWG})
	A quasi-equilibrated motion, $\textbf{x}=\chi(\textbf{X},t)$, of an incompressible homogeneous elastic solid subject to a given body force, $\textbf{b}=\textbf{b}(\textbf{x},t)$, is dynamically possible, subject to the same body force, if and only if the motion is circulation preserving with a single-valued acceleration potential $\xi$, i.e.,
	\begin{equation}\label{eq:cp}
	\ddot{\textbf{x}}=-\mathrm{grad}\ \xi.
	\end{equation}
	For the condition \eqref{eq:cp} to be satisfied, it is necessary that
	\begin{equation}\label{eq:ddotx:curl}
	\mathrm{curl}\ \ddot{\textbf{x}}=\textbf{0}.
	\end{equation}
	Then, the Cauchy stress tensor takes the form
	\begin{equation}\label{eq:T}
	\textbf{T}=-\rho\xi\textbf{I}+\textbf{T}^{(0)},
	\end{equation}
	where $\textbf{T}^{(0)}$ is the Cauchy stress for the equilibrium state at time $t$ and $\textbf{I}=\text{diag}(1,1,1)$ is the identity tensor. In this case, the stress field is determined by the present configuration alone. In particular, the shear stresses in the motion are the same as those of the equilibrium state at time $t$.
\end{theorem}

An immediate consequence of the above theorem is that a quasi-equilibrated motion is dynamically possible under a given body force in all elastic materials, if at every time instant the deformation is a possible equilibrium state under that body force in all those materials. Quasi-equilibrated motions of isotropic materials subject to surface tractions alone are obtained by taking the arbitrary constant in those deformations to be arbitrary functions of time. Under this type of motion, a body can be brought instantly to rest by applying a suitable pressure impulse on its boundary \cite[p.~209]{TruesdellNoll:2004}. Examples include the homogeneous motions that are possible in all homogeneous incompressible materials, and also those considered by us in the next sections. 

%%%%%%%%%%%%%%%%%%%%%%%%%%%%%%%%%%%%%%%%%%%%%%%%%%%%%%%%%%%%
\subsection{Radial motion of stochastic hyperelastic spheres}\label{sec:sphere:motion}

\begin{figure}[htbp]
	\begin{center}
		\includegraphics[width=0.75\textwidth]{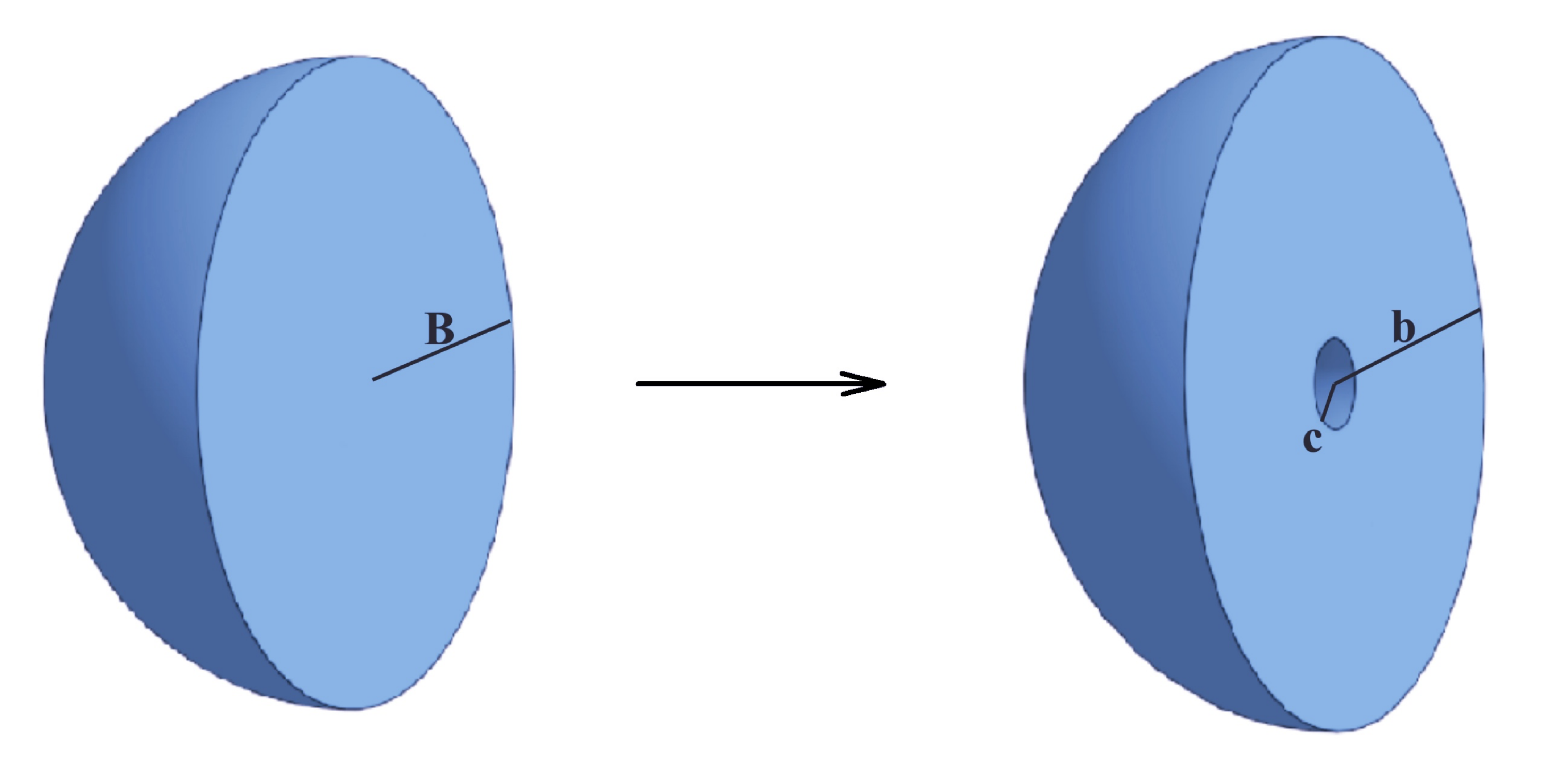}
		\caption{Schematic of cross-section of a sphere with undeformed radius $B$, showing the reference state (left), and the deformed state, with cavity radius $c$ and outer radius $b$ (right).}\label{fig:sphere-cave}
	\end{center}
\end{figure}

Throughout our analysis, we assume that a sphere of stochastic hyperelastic material defined by \eqref{eq:W:stoch} is subject to the following radially-symmetric dynamic deformation \cite{Mihai:2019b:MDWG},
\begin{equation}\label{eq:sphere:deform:qem}
r^3=R^3+c^3,\qquad \theta=\Theta,\qquad \phi=\Phi,
\end{equation}
where $(R,\Theta,\Phi)$ and $(r,\theta,\phi)$ are the spherical polar coordinates in the reference and current configuration, respectively, such that $0\leq R\leq B$, $B$ is the radius of the undeformed sphere, $c=c(t)\geq0$ is the cavity radius to be calculated, and $b=b(t)=\sqrt[3]{B^3+c(t)^3}$ is the radius of the deformed sphere at time $t$ (see Figure~\ref{fig:sphere-cave}).

%%%%%%%%%%%%%%
%%%%%%%%%%%%%

By the governing equations \eqref{eq:sphere:deform:qem},
\begin{equation}\label{eq:sphere:curl}
\textbf{0}=\mathrm{curl}\ \ddot{\textbf{x}}=
\left[
\begin{array}{c}
(\partial\ddot{\phi}/\partial\theta)/r-(\partial\ddot{\theta}/\partial\phi)/(r\sin\theta)\\
(\partial\ddot{r}/\partial\phi)/(r\sin\theta)-\partial\ddot{\phi}/\partial r\\
\partial\ddot{\theta}/\partial r-(\partial\ddot{r}/\partial\theta)/r
\end{array}
\right],
\end{equation}
i.e., the condition \eqref{eq:ddotx:curl} is valid for $\textbf{x}=(r,\theta,\phi)^{T}$. Therefore, \eqref{eq:sphere:deform:qem} describes a quasi-equilibrated motion, such that
\begin{equation}\label{eq:sphere:cp}
-\frac{\partial\xi}{\partial r}=\ddot{r}=\frac{2c\dot{c}^2+c^2\ddot{c}}{r^2}-\frac{2c^4\dot{c}^2}{r^5},
\end{equation}
where $\xi$ is the acceleration potential, satisfying \eqref{eq:cp}. Integrating \eqref{eq:sphere:cp} gives
\begin{equation}\label{eq:sphere:xi}
-\xi=-\frac{2c\dot{c}^2+a^2\ddot{c}}{r}+\frac{c^4\dot{c}^2}{2r^4}=-r\ddot{r}-\frac{3}{2}\dot{r}^2.
\end{equation}

For the deformation \eqref{eq:sphere:deform:qem}, the gradient tensor with respect to the polar coordinates $(R,\Theta,\Phi)$ is equal to
\begin{equation}\label{eq:sphere:F}
\textbf{F}=\mathrm{diag}\left(\frac{R^2}{r^2}, \frac{r}{R}, \frac{r}{R}\right),
\end{equation}
and the principal stretch ratios are
\begin{equation}\label{eq:sphere:lambda123}
\lambda_{1}=\frac{R^2}{r^2},\qquad \lambda_{2}=\lambda_{3}=\frac{r}{R}.
\end{equation}
The corresponding Cauchy-Green tensor is
\begin{equation}\label{eq:sphere:B}
\textbf{B}=\textbf{F}^2=\mathrm{diag}\left(\frac{R^4}{r^4}, \frac{r^2}{R^2}, \frac{r^2}{R^2}\right),
\end{equation}
with the principal invariants
\begin{equation}\label{eq:sphere:I123}
\begin{split}
I_{1}=&\mathrm{tr}\ (\textbf{B})=\frac{R^4}{r^4}+2\frac{r^2}{R^2},\\
I_{2}=&\frac{1}{2}\left[\left(\mathrm{tr}\,\textbf{B}\right)^{2}-\mathrm{tr}\left(\textbf{B}^{2}\right)\right]=\frac{r^4}{R^4}+2\frac{R^2}{r^2},\\
I_{3}=&\det\textbf{B}=1.
\end{split}
\end{equation}
The associated Cauchy stress tensor then takes the form \cite[pp.~87-91]{Green:1970:GA}
\begin{equation}\label{eq:sphere:T0}
\textbf{T}^{(0)}=-p^{(0)}\textbf{I}+\beta_{1}\textbf{B}+\beta_{-1}\textbf{B}^{-1},
\end{equation}
where the scalar $p^{(0)}$, which is commonly referred to as the arbitrary hydrostatic pressure \cite[pp.~286-287]{goriely17}, \cite[pp.~198-201]{Ogden:1997}, is the Lagrange multiplier for the internal constraint $I_{3}=1$ of incompressibility (i.e., all deformations are isochoric for incompressible materials) \cite[pp.~71-72]{TruesdellNoll:2004}, and the coefficients
\begin{equation}\label{eq:sphere:betas}
\beta_{1}={2}\frac{\partial W}{\partial I_{1}},\qquad \beta_{-1}=-2\frac{\partial W}{\partial I_{2}},
\end{equation}
are nonlinear material parameters, with $W=W(I_{1},I_{2},I_{3})=\mathcal{W}(\lambda_{1},\lambda_{2},\lambda_{2})$, and $I_{1}$, $I_{2}$, $I_{3}$ defined by \eqref{eq:sphere:I123}. For this stress tensor, the principal components at time $t$ are as follows,
\begin{equation}\label{eq:sphere:stresses}
\begin{split}
T^{(0)}_{rr}&=-p^{(0)}+\beta_{1}\frac{R^4}{r^4}+\beta_{-1}\frac{r^4}{R^4},\\
T^{(0)}_{\theta\theta}&=T^{(0)}_{rr}+\left(\beta_{1}-\beta_{-1}\frac{r^2}{R^2}\right)\left(\frac{r^2}{R^2}-\frac{R^4}{r^4}\right),\\
T^{(0)}_{\phi\phi}&=T^{(0)}_{\theta\theta}.
\end{split}
\end{equation}
As the stress components depend only on the radius $r$, the system of equilibrium equations reduces to
\begin{equation}\label{eq:sphere:equilibrium}
\frac{\partial T^{(0)}_{rr}}{\partial r}=2\frac{T^{(0)}_{\theta\theta}-T^{(0)}_{rr}}{r}.
\end{equation}
Hence, by \eqref{eq:sphere:stresses} and \eqref{eq:sphere:equilibrium}, the radial Cauchy stress for the equilibrium state at $t$ is equal to
\begin{equation}\label{eq:sphere:T0rr}
T^{(0)}_{rr}(r,t)=2\int\left(\beta_{1}-\beta_{-1}\frac{r^2}{R^2}\right)\left(\frac{r^2}{R^2}-\frac{R^4}{r^4}\right)\frac{dr}{r}+\psi(t),
\end{equation}
where $\psi=\psi(t)$ is an arbitrary function of time. Substitution of \eqref{eq:sphere:xi} and \eqref{eq:sphere:T0rr} into \eqref{eq:T} then gives the following principal Cauchy stresses at time $t$,
\begin{equation}\label{eq:sphere:Trr}
\begin{split}
T_{rr}(r,t)&=-\rho\left(\frac{c^2\ddot{c}+2c\dot{c}^2}{r}-\frac{c^4\dot{c}^2}{2r^4}\right)+2\int\left(\beta_{1}-\beta_{-1}\frac{r^2}{R^2}\right)\left(\frac{r^2}{R^2}-\frac{R^4}{r^4}\right)\frac{dr}{r}+\psi(t),\\
T_{\theta\theta}(r,t)&=T_{rr}(r,t)+\left(\beta_{1}-\beta_{-1}\frac{r^2}{R^2}\right)\left(\frac{r^2}{R^2}-\frac{R^4}{r^4}\right),\\
T_{\phi\phi}(r,t)&=T_{\theta\theta}(r,t).
\end{split}
\end{equation}
When $R^{2}/r^{2}\to 1$, all the stress components defined in \eqref{eq:sphere:Trr} are equal.

%%%%%%%%%%%%%%%%%%%%%%%%%%%%%%%%%%%%%%%%%%%%%%%%%%%%%%%%%%%%
%%%%%%%%%%%%%%%%%%%%   NEW SECTION  %%%%%%%%%%%%%%%%%%%%%%%%
%%%%%%%%%%%%%%%%%%%%%%%%%%%%%%%%%%%%%%%%%%%%%%%%%%%%%%%%%%%%
\section{Cavitation and radial motion of homogeneous spheres}\label{sec:sphere:NH}

We first analyse the static and dynamic radially-symmetric deformations of a stochastic homogeneous hyperelastic sphere, where the shear modulus $\mu$ of the stochastic model given by \eqref{eq:W:stoch} is a space-invariant random variable. Hence, the model is of stochastic neo-Hookean type \cite{Staber:2015:SG}, with the shear modulus $\mu$ characterised by the Gamma distribution defined by \eqref{eq:mu:gamma}. The radial motion of spheres of neo-Hookean material were treated deterministically in \cite{ChouWang:1989b:CWH}, while inhomogeneous cylindrical and spherical shells of neo-Hookean-like material with a radially varying material parameter were analysed in \cite{Ertepinar:1976:EA}. For homogeneous spheres of stochastic isotropic incompressible hyperelastic material, the static cavitation under uniform tensile dead load was investigated in \cite{Mihai:2019c:MDWG}, while radial oscillatory motions of homogeneous cylindrical and spherical shells of stochastic Mooney-Rivlin and neo-Hookean material, respectively, were treated in \cite{Mihai:2019b:MDWG}.

%%%%%%%%%%%%%%%%%%%%%%%%%%%%%%%%%%%%%%%%%%%%%%%%%%%%%%%%%%%%
\subsection{Oscillatory motion of a stochastic neo-Hookean sphere under dead-load traction}\label{sec:sphere:NH:osc}

For a sphere of stochastic neo-Hookean material subject to the quasi-equilibrated motion \eqref{eq:sphere:deform:qem}, we denote the inner and outer radial pressures acting on the curvilinear surfaces, $r=c(t)$ and $r=b(t)$ at time $t$, as $T_{1}(t)$ and $T_{2}(t)$, respectively \cite[pp.~217-219]{TruesdellNoll:2004}. 

Evaluating $T_{1}(t)=-T_{rr}(c,t)$ and $T_{2}(t)=-T_{rr}(b,t)$, using \eqref{eq:sphere:Trr}, with $r=c$ and $r=b$, respectively, and subtracting the results, gives
\begin{equation}\label{eq:sphere:T1T2:r}
\begin{split}
T_{1}(t)-T_{2}(t)&=\rho\left[\left(c^2\ddot{c}+2c\dot{c}^2\right)\left(\frac{1}{c}-\frac{1}{b}\right)-\frac{c^4\dot{c}^2}{2}\left(\frac{1}{c^4}-\frac{1}{b^4}\right)\right]+2\int_{c}^{b}\mu\left(\frac{r^2}{R^2}-\frac{R^4}{r^4}\right)\frac{dr}{r}\\
&=\rho\left[\left(c\ddot{c}+2\dot{c}^2\right)\left(1-\frac{c}{b}\right)-\frac{\dot{c}^2}{2}\left(1-\frac{c^4}{b^4}\right)\right]+2\int_{c}^{b}\mu\left(\frac{r^2}{R^2}-\frac{R^4}{r^4}\right)\frac{dr}{r}\\
&=\rho B^2\left[\left(\frac{c}{B}\frac{\ddot{c}}B+2\frac{\dot{c}^2}{B^2}\right)\left(1-\frac{c}{b}\right)-\frac{\dot{c}^2}{2B^2}\left(1-\frac{c^4}{b^4}\right)\right]+2\int_{c}^{b}\mu\left(\frac{r^2}{R^2}-\frac{R^4}{r^4}\right)\frac{dr}{r}.
\end{split}
\end{equation}
By setting the notation
\begin{equation}\label{eq:sphere:ux}
u=\frac{r^3}{R^3}=\frac{r^3}{r^3-c^3},\qquad
x=\frac{c}{B},\
\end{equation}
we can then rewrite
\[
\begin{split}
\left(\frac{c}{B}\frac{\ddot{c}}B+2\frac{\dot{c}^2}{B^2}\right)\left(1-\frac{c}{b}\right)&-\frac{\dot{c}^2}{2B^2}\left(1-\frac{c^4}{b^4}\right)\\
&=\left(\ddot{x}x+2\dot{x}^2\right)\left[1-\left(1+\frac{1}{x^3}\right)^{-1/3}\right]-\frac{\dot{x}^2}{2}\left[1-\left(1+\frac{1}{x^3}\right)^{-4/3}\right]\\
&=\left(\ddot{x}x+\frac{3}{2}\dot{x}^2\right)\left[1-\left(1+\frac{1}{x^3}\right)^{-1/3}\right]-\frac{\dot{x}^2}{2}\frac{1}{x^3}\left(1+\frac{1}{x^3}\right)^{-4/3}\\
&=\frac{1}{2x^2}\frac{d}{dx}\left\{\dot{x}^2x^3\left[1-\left(1+\frac{1}{x^3}\right)^{-1/3}\right]\right\}
\end{split}
\]
and
\[
\begin{split}
\int_{c}^{b}\mu\left(\frac{r^2}{R^2}-\frac{R^4}{r^4}\right)\frac{dr}{r}
&=\int_{c}^{b}\mu\left[{\left(\frac{r^3}{r^3-c^3}\right)^{2/3}}-\left(\frac{r^3-c^3}{r^3}\right)^{4/3}\right]\frac{dr}{r}\\
&=\frac{1}{3}\int_{x^3+1}^{\infty}\mu\frac{1+u}{u^{7/3}}du.
\end{split}
\]
Thus \eqref{eq:sphere:T1T2:r} is equivalent to
\begin{equation}\label{eq:sphere:T1T2:x}
2x^2\frac{T_{1}(t)-T_{2}(t)}{\rho B^2}
=\frac{d}{dx}\left\{\dot{x}^2x^3\left[1-\left(1+\frac{1}{x^3}\right)^{-1/3}\right]\right\}
+\frac{4x^2}{3\rho B^2}\int_{x^3+1}^{\infty}\mu\frac{1+u}{u^{7/3}}du.
\end{equation}
Next, assuming that the cavity surface is traction-free, $T_{1}(t)=0$ and  \eqref{eq:sphere:T1T2:x} is equivalent to
\begin{equation}\label{eq:sphere:T:x}
2x^2\frac{T_{rr}(b,t)}{\rho B^2}
=\frac{d}{dx}\left\{\dot{x}^2x^3\left[1-\left(1+\frac{1}{x^3}\right)^{-1/3}\right]\right\}
+\frac{4x^2}{3\rho B^2}\int_{x^3+1}^{\infty}\mu\frac{1+u}{u^{7/3}}du.
\end{equation}

We now denote
\begin{equation}\label{eq:sphere:Hint:NH}
H(x)=\frac{4}{3\rho B^2}\int_{0}^{x}\left(\zeta^2\int_{\zeta^3+1}^{\infty}\mu\frac{1+u}{u^{7/3}}du\right)d\zeta,
\end{equation}
and set the uniform dead-load traction (see also eq. (2.7) of \cite{ChouWang:1989b:CWH})
\begin{equation}\label{eq:sphere:dload}
P_{rr}(B)=\left(x^3+1\right)^{2/3} T_{rr}(b,t)=\left\{
\begin{array}{cc}
0 & \mbox{if}\ t\leq0,\\
p_{0}& \mbox{if}\ t>0,
\end{array}
\right.
\end{equation}
where $p_{0}$ is constant, and $x=x(t)$ is the dimensionless cavity radius, as denoted in \eqref{eq:sphere:ux}.

In this case, integrating \eqref{eq:sphere:T:x} once gives
\begin{equation}\label{eq:sphere:ode:C}
\dot{x}^2x^3\left[1-\left(1+\frac{1}{x^3}\right)^{-1/3}\right]+H(x)=\frac{2p_{0}}{\rho B^2}\left(x^3+1\right)^{1/3}+C,
\end{equation}
with $H(x)$ defined by \eqref{eq:sphere:Hint:NH}, and
\begin{equation}\label{eq:sphere:C}
C=\dot{x}_{0}^2x_{0}^3\left[1-\left(1+\frac{1}{x_{0}^3}\right)^{-1/3}\right]+H(x_{0})-\frac{2p_{0}}{\rho B^2}\left(x_{0}^3+1\right)^{1/3}.
\end{equation}
Then, after setting the initial conditions $x_{0}=x(0)=0$ and $\dot{x}_{0}=\dot{x}(0)=0$, equation \eqref{eq:sphere:ode:C} takes the form
\begin{equation}\label{eq:sphere:ode}
\dot{x}^2x^3\left[1-\left(1+\frac{1}{x^3}\right)^{-1/3}\right]+H(x)=\frac{2p_{0}}{\rho B^2}\left[\left(x^3+1\right)^{1/3}-1\right].
\end{equation}

From \eqref{eq:sphere:ode}, we obtain the velocity
\begin{equation}\label{eq:sphere:dotx:forced}
\dot{x}=\pm\sqrt{\frac{\frac{2p_{0}}{\rho B^2}\left[\left(x^3+1\right)^{1/3}-1\right]-H(x)}{x^3\left[1-\left(1+\frac{1}{x^3}\right)^{-1/3}\right]}}.
\end{equation}

%%%%%%%%%%%%%%%%
\begin{figure}[htbp]
	\begin{center}
		\includegraphics[width=0.6\textwidth]{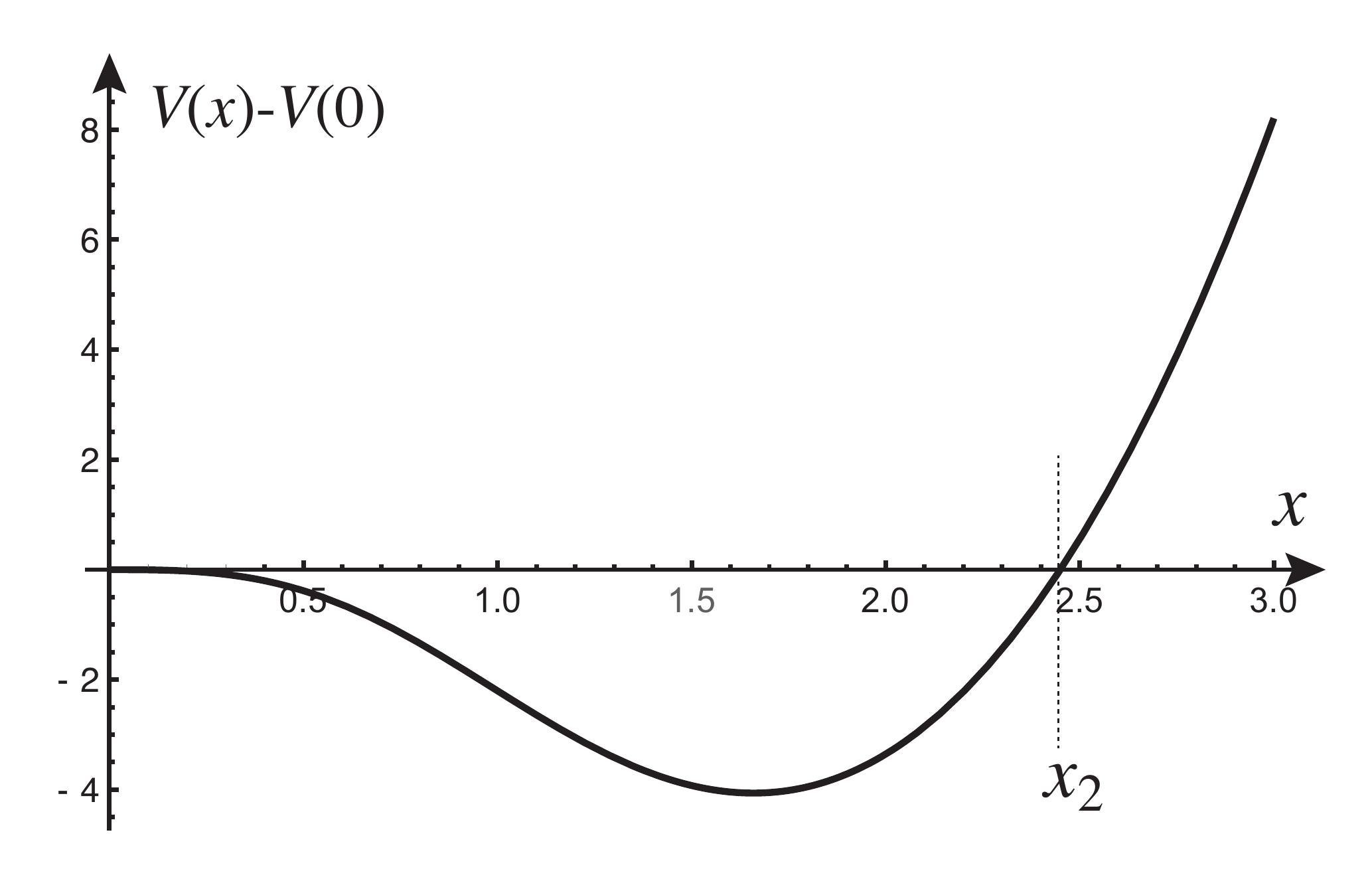}
		\caption{Example of potential $V(x)$ with $p_0=15$ and $\mu=4.05$, $\rho=B=1$. The periodic orbits lie between $x_1=0$ and $x_2\approx2.45$.}\label{fig:potential}
	\end{center}
\end{figure}
%%%%%%%%%%%%%%%%%

It is useful to note that this nonlinear elastic system is analogous to the motion of a point mass with energy
\begin{equation}
E=\frac{1}{2} m(x)\dot x^{2}+V(x),
\end{equation}
where the energy is $E=C$, the potential is given by $V(x)=H(x)-\frac{2p_{0}}{\rho B^2}\left(x^3+1\right)^{1/3}$ and the position-dependent mass is $m(x)=x^3\left[1-\left(1+\frac{1}{x^3}\right)^{-1/3}\right]$. Due to the constraints on the function $H$, the system has simple dynamics, and the only solutions of interest are either static or periodic solutions.

For the given sphere to undergo an oscillatory motion, the following equation,
\begin{equation}\label{eq:sphere:HC}
H(x)=\frac{2p_{0}}{\rho B^2}\left[\left(x^3+1\right)^{1/3}-1\right]
\end{equation}
must have exactly two finite distinct positive roots $x_{1},x_2$ such that $0\leq x_{1}<x_{2}<\infty$, as shown in Figure~\ref{fig:potential}. Then, the minimum and maximum radii of the cavity in the oscillation are $x_{1}B$ and $x_{2}B$, respectively, and the period of oscillation is equal to
\begin{equation}\label{eq:sphere:T}
T=2\left|\int_{x_{1}}^{x_{2}}\frac{dx}{\dot{x}}\right|=2\left|\int_{x_{1}}^{x_{2}}\sqrt{\frac{x^3\left[1-\left(1+\frac{1}{x^3}\right)^{-1/3}\right]}{\frac{2p_{0}}{\rho B^2}\left[\left(x^3+1\right)^{1/3}-1\right]-H(x)}}dx\right|.
\end{equation}
For the stochastic sphere, the amplitude and the period of the oscillation are random variables characterised by probability distributions.

Evaluating the integral in \eqref{eq:sphere:Hint:NH} gives
\begin{equation}\label{eq:sphere:H:NH}
H(x)=\frac{\mu}{\rho B^2}\left[2\left(x^3+1\right)^{2/3}-\frac{1}{\left(x^3+1\right)^{1/3}}-1\right],
\end{equation}
and assuming that the shear modulus, $\mu$, which is a random variable, is bounded from below, i.e.,
\begin{equation}\label{eq:sphere:shearmod:NH:eta}
\mu>\eta,
\end{equation}
for some constant $\eta>0$, it follows that $H(0)=0$ and $\lim_{x\to\infty}H(x)=\infty$.

When $p_{0}\neq0$, substitution of \eqref{eq:sphere:H:NH} in \eqref{eq:sphere:HC} gives
\begin{equation}\label{eq:sphere:p0:NH}
p_{0}\left[\left(x^3+1\right)^{1/3}-1\right]=\frac{\mu}{2}\left[2\left(x^3+1\right)^{2/3}-\frac{1}{\left(x^3+1\right)^{1/3}}-1\right].
\end{equation}
Equation \eqref{eq:sphere:p0:NH} has one solution at $x_{1}=0$, while the  second solution, $x_{2}$, is a root of 
\begin{equation}\label{eq:sphere:p0:NH:nonzero}
p_{0}=\frac{\mu}{2}\left[2\left(x^3+1\right)^{1/3}+\frac{1}{\left(x^3+1\right)^{1/3}}+2\right].
\end{equation}
As the right-hand side of equation \eqref{eq:sphere:p0:NH:nonzero} is an increasing function of $x$, this equation has a solution, $x_{2}>0$, if and only if (see also eq. (2.7) of \cite{ChouWang:1989b:CWH})
\begin{equation}\label{eq:sphere:p0:NH:bound}
p_{0}>\frac{5\mu}{2}=\lim_{x\to0_{+}}\frac{\mu}{2}\left[2\left(x^3+1\right)^{1/3}+\frac{1}{\left(x^3+1\right)^{1/3}}+2\right].
\end{equation}
Then,
\begin{equation}\label{eq:sphere:osc}
\frac{2p_{0}}{\rho B^2}\left[\left(x^3+1\right)^{1/3}-1\right]-H(x)
\left\{
\begin{array}{ll}
\geq0 & \mbox{if}\ x_{1}\leq x\leq x_{2},\\
<0 & \mbox{if}\ x>x_{2}.
\end{array}
\right.
\end{equation}
In this case, at the centre of the given sphere, a spherical cavity forms and expands until its radius reaches the value $c=x_{2}B$, where $x_{2}$ is the root of \eqref{eq:sphere:p0:NH:nonzero}, then contracts again to zero radius and repeats the cycle.

The critical dead load for the onset of cavitation is
\begin{equation}\label{eq:sphere:p0:NH:crit}
\lim_{x\to0_{+}}p_{0}=\frac{5\mu}{2}.
\end{equation}

%%%%%%%%%%%%%%
\begin{figure}[htbp]
	\begin{center}
		\includegraphics[width=0.6\textwidth]{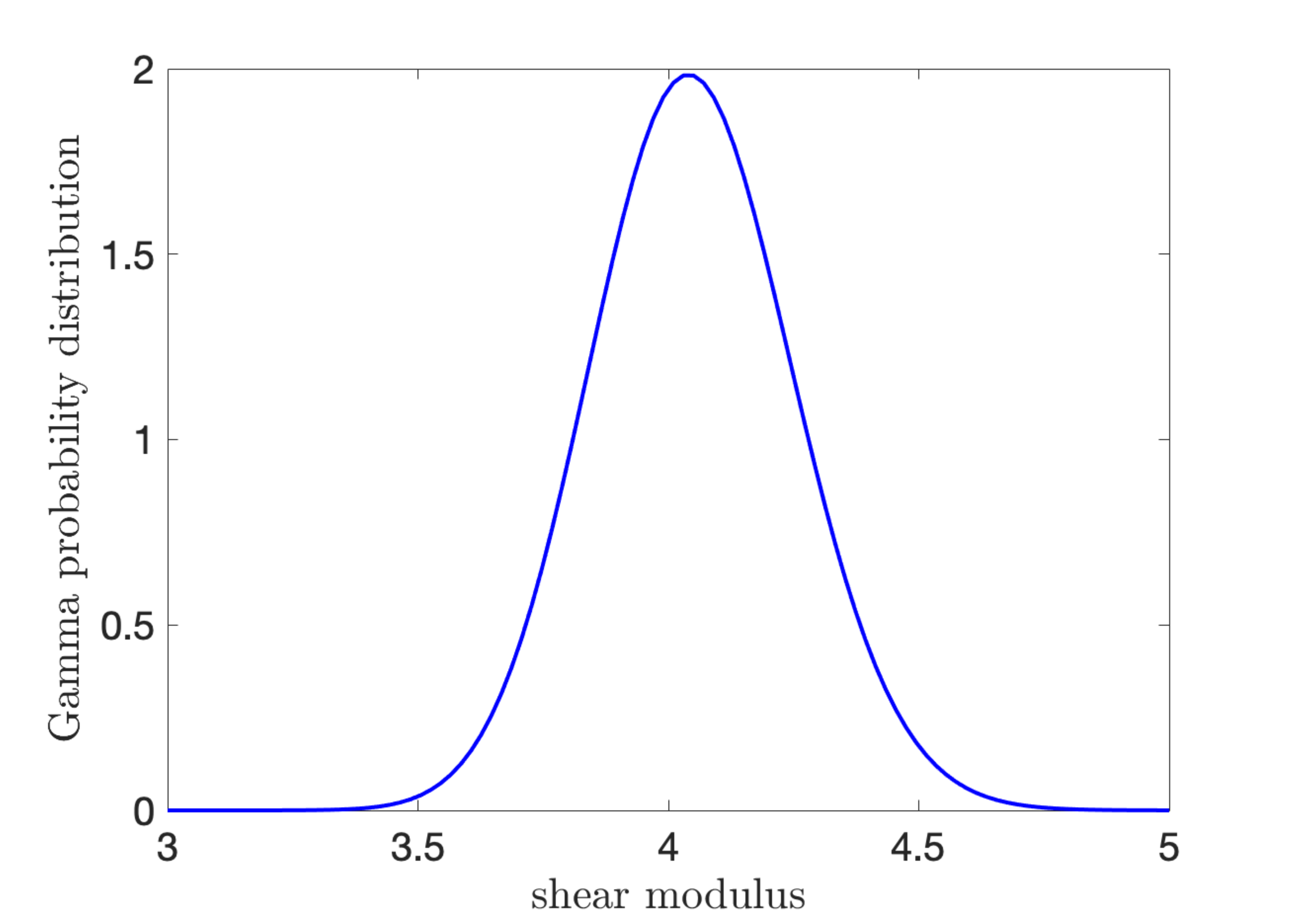}
		\caption{Example of Gamma distribution, defined by \eqref{eq:mu:gamma}, with hyperparameters $\rho_{1}=405$, $\rho_{2}=0.01$.}\label{fig:mu-gpdf}
	\end{center}
\end{figure}
%%%%%%%%%%%%

%%%%%%%%%%%%%%
\begin{figure}[htbp]
	\begin{center}
		\includegraphics[width=\textwidth]{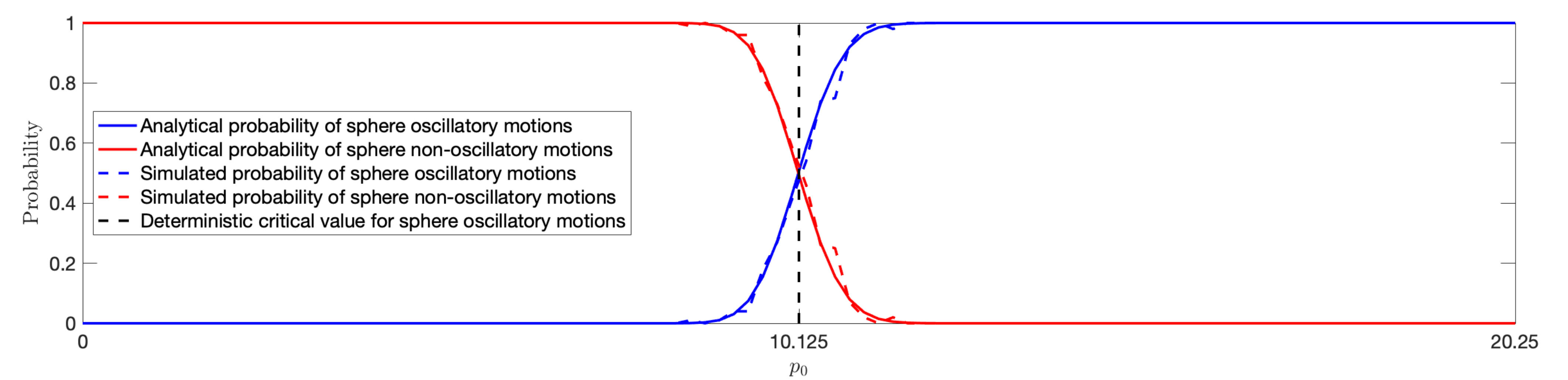}
		\caption{Probability distributions of whether oscillatory motions can occur or not for a sphere of stochastic neo-Hookean material under dead-load traction, when the shear modulus, $\mu$, follows a Gamma distribution with $\rho_{1}=405$, $\rho_{2}=0.01$. Continuous coloured lines represent analytically derived solutions, given by equations \eqref{eq:sphere:NH:P1}-\eqref{eq:sphere:NH:P2}, whereas the dashed versions represent stochastically generated data. The vertical line at the critical value, $p_{0}=10.125$, separates the expected regions based only on mean value, $\underline{\mu}=\rho_{1}\rho_{2}=4.05$. The probabilities were calculated from the average of 100 stochastic simulations.}\label{fig:intpdfs-NHsphere}
	\end{center}
\end{figure}
%%%%%%%%%%%%%%

%%%%%%%%%%%%%%
\begin{figure}[htbp]
\begin{center}
	\includegraphics[width=0.47\textwidth]{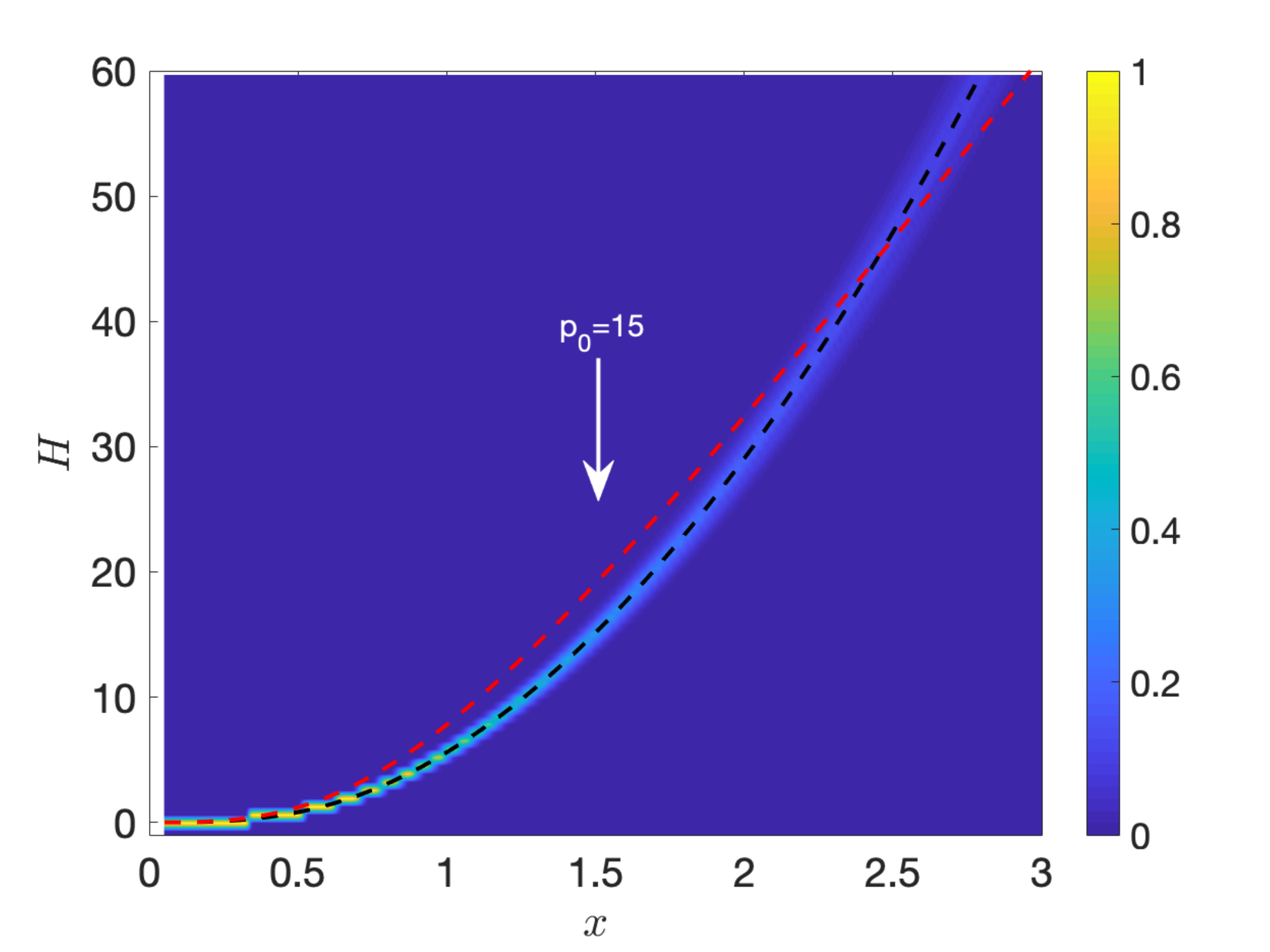}\qquad
	\includegraphics[width=0.47\textwidth]{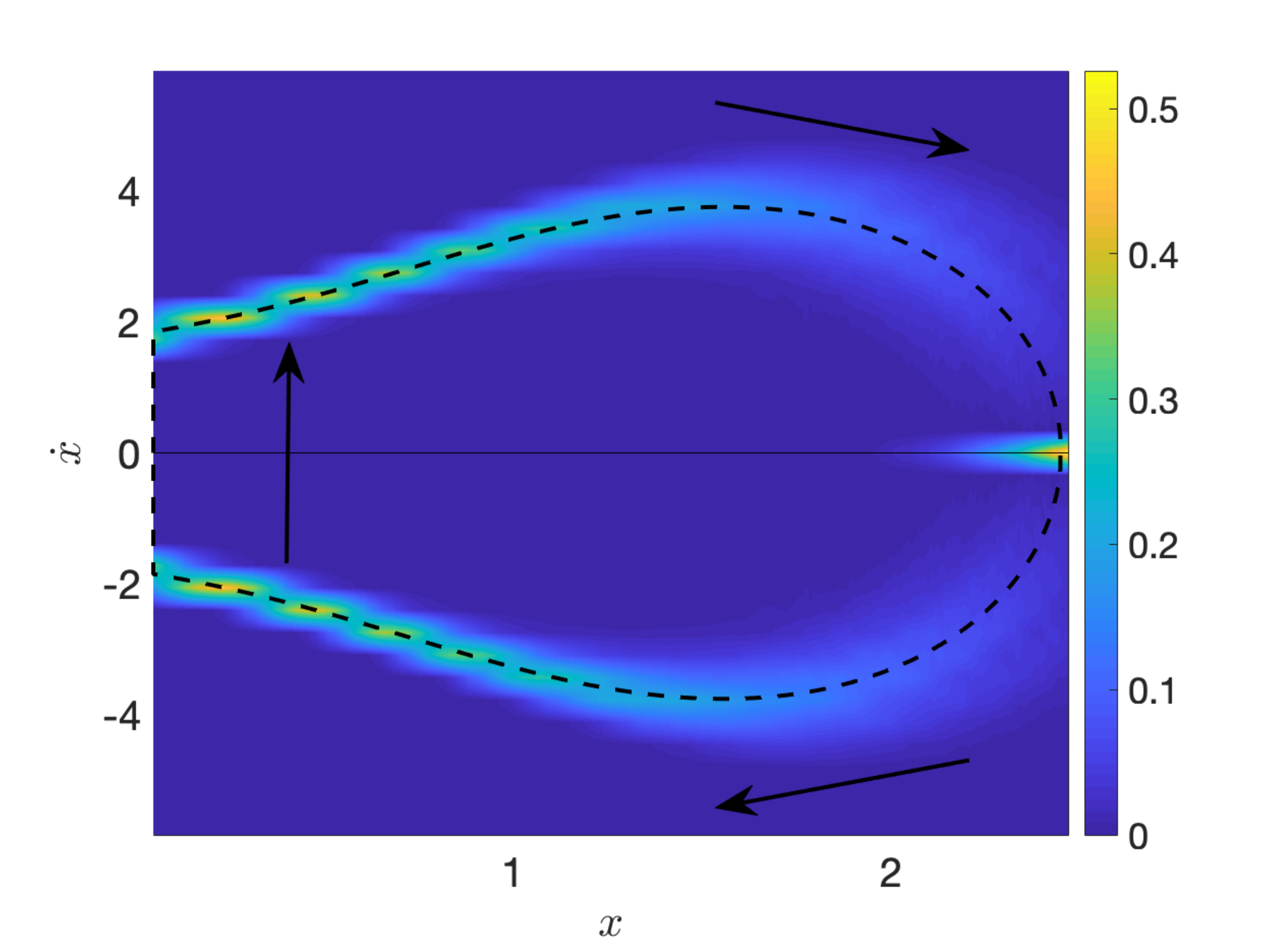}
	\caption{The function $H(x)$, defined by \eqref{eq:sphere:H:NH}, intersecting the (dashed red) curve $\frac{2p_{0}}{\rho B^2}\left[\left(x^3+1\right)^{1/3}-1\right]$, with $p_{0}=15$ (left), and the associated velocity, given by \eqref{eq:sphere:dotx:forced} (right), for a dynamic sphere of stochastic neo-Hookean material under dead-load traction, when $\rho=1$, $B=1$, and $\mu$ is drawn from the Gamma distribution with $\rho_{1}=405$, $\rho_{2}=0.01$. The dashed black lines correspond to the expected values based only on mean value, $\underline{\mu}=\rho_{1}\rho_{2}=4.05$. Each distribution was calculated from the average of $1000$ stochastic simulations.}\label{fig:stoch-NHsphere}
\end{center}
\end{figure}
%%%%%%%%%%%%

Regarding the random shear modulus, by \eqref{eq:sphere:shearmod:NH:eta} and \eqref{eq:sphere:p0:NH:bound}, for the motion to be oscillatory, the following condition must hold,
\begin{equation}\label{eq:sphere:shearmod:NH:bounds}
\eta<\mu<\frac{2p_{0}}{5}.
\end{equation}
Then, as $\mu$ follows a Gamma distribution, defined by \eqref{eq:mu:gamma}, the probability distribution of oscillatory motions occurring is
\begin{equation}\label{eq:sphere:NH:P1}
P_{1}(p_0)=\int_{0}^{\frac{2p_{0}}{5}}g(u;\rho_{1},\rho_{2})du,
\end{equation}
and that of non-oscillatory motions is
\begin{equation}\label{eq:sphere:NH:P2}
P_{2}(p_0)=1-P_{1}(p_0)=1-\int_{0}^{\frac{2p_{0}}{5}}g(u;\rho_{1},\rho_{2})du.
\end{equation}

For example, when $\mu$ satisfies the Gamma distribution with $\rho_{1}=405$ and $\rho_{2}=0.01$, shown in Figure~\ref{fig:mu-gpdf}, the probability distributions given by \eqref{eq:sphere:NH:P1}-\eqref{eq:sphere:NH:P2} are shown in Figure~\ref{fig:intpdfs-NHsphere} (with blue lines for $P_{1}$ and red lines for $P_{2}$). Specifically, the interval $(0,5\underline{\mu})$, where $\underline{\mu}=\rho_{1}\rho_{2}=4.05$ is the mean value of $\mu$, was divided into $100$ steps, then for each value of $p_{0}$, $100$ random values of $\mu$ were numerically generated from the specified Gamma distribution and compared with the inequalities defining the two intervals for values of $p_{0}$. For the deterministic elastic sphere, the critical value $p_{0}=5\underline{\mu}/2=10.125$ strictly divides the cases of oscillations occurring or not. For the stochastic problem, for the same critical value, there is, by definition, exactly 50\% chance that the motion is oscillatory, and 50\% chance that is not. To increase the probability of oscillatory motion ($P_{1}\approx 1$), one must apply a sufficiently small traction, $p_{0}$, below the expected critical point, whereas a non-oscillatory motion is certain to occur ($P_{2}\approx 1$) if $p_{0}$ is sufficiently large. However, the inherent variability in the probabilistic system means that there will also exist events where there is competition between the two cases.

We note that the Gamma distribution represented in Figure~\ref{fig:mu-gpdf} is approximately a normal distribution. This is because, when $\rho_{1}$ much larger than $\rho_{2}$, the Gamma probability distribution is approximated by a normal distribution (see, e.g., \cite{Fitt:2019:FWWM,Mihai:2019a:MDWG}). However, elastic moduli cannot be characterised by the normal distribution, since this distribution is defined on the entire real line whereas elastic moduli are typically positive. In practice, these moduli can meaningfully take on different values, corresponding to possible outcomes of experimental tests. Then, the principle of maximum entropy enables the construction of their probability distributions, given the available information. Approaches for the explicit derivation of probability distributions for the elastic parameters of stochastic homogeneous isotropic hyperelastic models calibrated to experimental data were developed in \cite{Mihai:2018:MWG,Staber:2017:SG}. 

In Figure~\ref{fig:stoch-NHsphere}, we illustrate the stochastic function $H(x)$, defined by \eqref{eq:sphere:H:NH}, intersecting the curve $\frac{2p_{0}}{\rho B^2}\left[\left(x^3+1\right)^{1/3}-1\right]$, with $p_{0}=15$, to find the two distinct solutions of equation \eqref{eq:sphere:HC}, and the associated velocity, given by \eqref{eq:sphere:dotx:forced},  assuming that $\rho=1$, $B=1$, and $\mu$ is drawn from the Gamma distribution with $\rho_{1}=405$ and $\rho_{2}=0.01$ (see Figure~\ref{fig:mu-gpdf}). Each figure displays a probability histogram at each value of $x$. The histogram comprises of 1000 stochastic simulations and the colour bar defines the probability of finding a given value of $H(x)$, or of the associated velocity, respectively, at a given value of $x$. The dashed black line corresponds to the expected values based only on the mean value $\underline{\mu}=\rho_{1}\rho_{2}=4.05$, of $\mu$.

Throughout this paper, different simulations were produced by fixing the parameters given in each figure caption, and repeatedly drawing random samples from the underlying distribution. Our computer simulations were run in Matlab 2018a where we made specific use of inbuilt functions for random number generation.

%%%%%%%%%%%%%%%%%%%%%%%%%%%%%%%%%%%%%%%%%%%%%%%%%%%%%%%%%%%%
\subsection{Static deformation of a stochastic neo-Hookean sphere under dead-load traction}\label{sec:sphere:NH:static}

In view of our subsequent analysis, we review here the cavitation of a static sphere of stochastic neo-Hookean material, with the shear modulus $\mu$ following the Gamma distribution defined by \eqref{eq:mu:gamma}. Incompressible spheres of different stochastic homogeneous hyperelastic material were treated in detail in \cite{Mihai:2019c:MDWG} (see also \cite{ChouWang:1989a:CWH} for the deterministic spheres). In this case, if the surface of the cavity is traction-free, then $T_{1}=0$ and \eqref{eq:sphere:T:x} reduces to
\begin{equation}\label{eq:sphere:T:NH}
T_{rr}(b)=\frac{2}{3}\int_{x^3+1}^{\infty}\mu\frac{1+u}{u^{7/3}}du.
\end{equation}
After evaluating the integral in \eqref{eq:sphere:T:NH}, the required uniform dead-load traction at the outer surface, $R=B$, in the reference configuration, takes the form
\begin{equation}\label{eq:sphere:P:NH}
P=\left(x^{3}+1\right)^{2/3}T_{rr}(b)=2\mu\left[\left(x^{3}+1\right)^{1/3}+\frac{1}{4\left(x^{3}+1\right)^{2/3}}\right],
\end{equation}
and increases as $x$ increases. The critical dead load for the onset of cavitation is then
\begin{equation}\label{eq:sphere:P0:NH}
P_{0}=\lim_{x\to0_{+}}P=\frac{5\mu}{2},
\end{equation}
and is equal to that given by \eqref{eq:sphere:p0:NH:crit} for the dynamic sphere \cite{Ball:1982,ChouWang:1989b:CWH}.

%%%%%%%%%%%%%%
\begin{figure}[htbp]
	\begin{center}
		\includegraphics[width=0.5\textwidth]{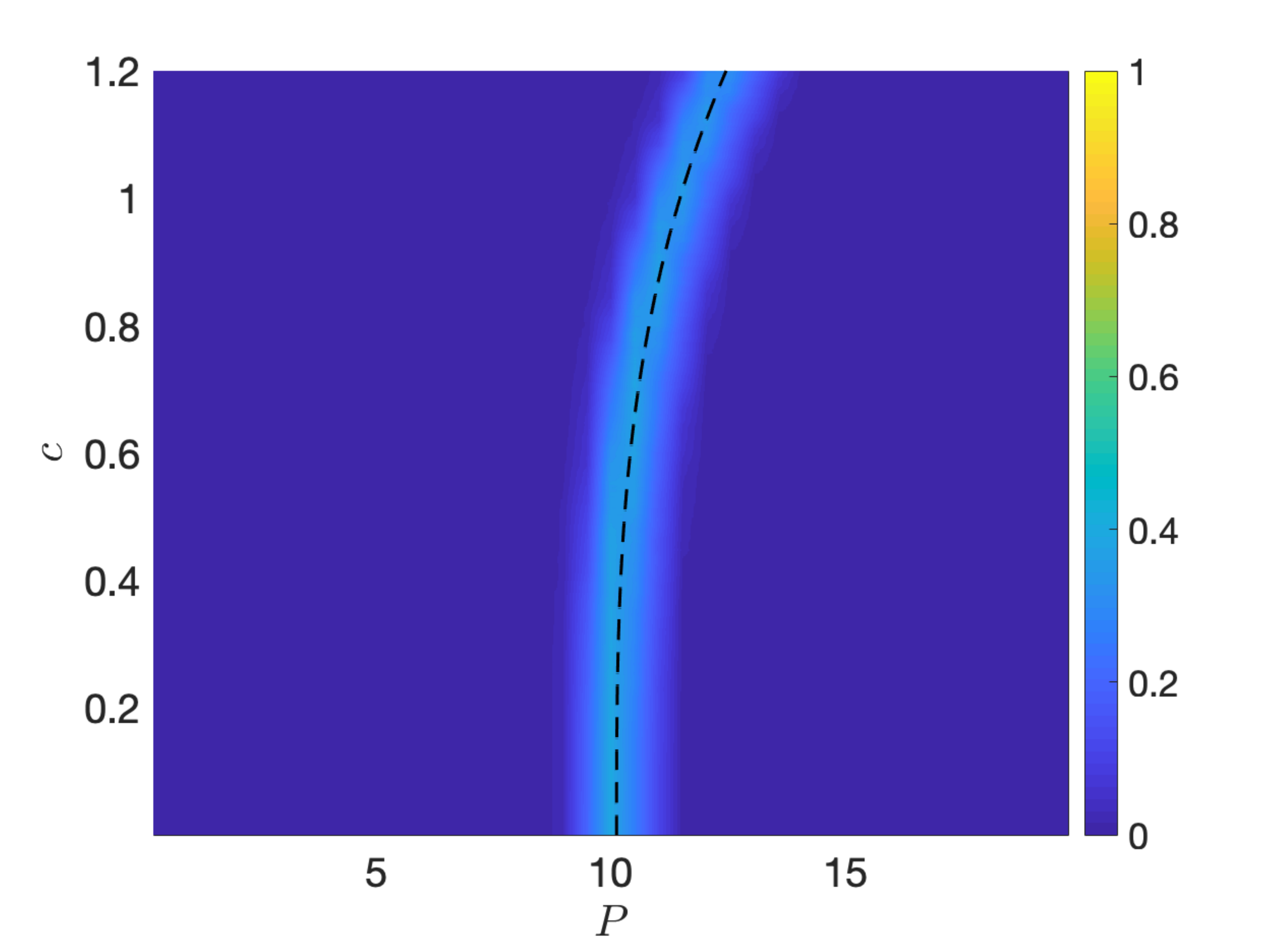}
		\caption{Probability distribution of the applied dead-load traction $P$ causing cavitation of radius $c$ in a static unit sphere (with $B=1$) of stochastic neo-Hookean material, when the shear modulus, $\mu$, follows a Gamma distribution with $\rho_{1}=405$, $\rho_{2}=4.05/\rho_{1}=0.01$. The dashed black line corresponds to the expected bifurcation based only on mean value, $\underline{\mu}=\rho_{1}\rho_{2}=4.05$.}\label{fig:Pstoch}
	\end{center}
\end{figure}
%%%%%%%%%%%%%

To analyse the stability of this cavitation, we study the behaviour of the cavity opening, with radius $c$ as a function of $P$, in a neighbourhood of $P_{0}$. After differentiating the function given by \eqref{eq:sphere:P:NH}, with respect to the dimensionless cavity radius $x=c/B$, we have
\begin{equation}\label{eq:sphere:dP:NH}
\frac{\text{d}P}{\text{d}x}=2\mu x^2\left[\frac{1}{\left(x^{3}+1\right)^{2/3}}-\frac{1}{2\left(x^{3}+1\right)^{5/3}}\right]>0,
\end{equation}
i.e., the cavitation is stable (with 100\% certainty), regardless of the material parameter $\mu$ \cite{Mihai:2019c:MDWG}. For example, the post-cavitation stochastic behaviour of the static unit sphere when the shear modulus, $\mu$, follows a Gamma distribution with $\rho_{1}=405$ and $\rho_{2}=4.05/\rho_{1}=0.01$ is shown in Figure~\ref{fig:Pstoch}.

%%%%%%%%%%%%%%%%%%%%%%%%%%%%%%%%%%%%%%%%%%%%%%%%%%%%%%%%%%%%
\subsection{Non-oscillatory motion of a stochastic neo-Hookean sphere under impulse traction}\label{sec:sphere:nosc}

For a sphere subject to the radially-symmetric dynamic deformation \eqref{eq:sphere:deform:qem},  an impulse (suddenly applied) traction, expressed  in terms of the Cauchy stresses, is prescribed as follows \cite{Mihai:2019b:MDWG},
\begin{equation}\label{eq:sphere:impulse}
2\frac{T_{rr}(b,t)}{\rho B^2}=\left\{
\begin{array}{cc}
0 & \mbox{if}\ t\leq0,\\
p_{0} & \mbox{if}\ t>0,
\end{array}
\right.
\end{equation}
where $p_{0}$ is constant in time. Introducing the dimensionless cavity radius $x(t)=c(t)/B$ and setting the initial conditions $x_{0}=x(0)=0$ and $\dot{x}_{0}=\dot{x}(0)=0$, we obtain the following differential equation
\begin{equation}\label{eq:sphere:ode:imp}
\dot{x}^2x^3\left[1-\left(1+\frac{1}{x^3}\right)^{-1/3}\right]+H(x)=\frac{p_{0}}{3}x^3,
\end{equation}
where $H(x)$ is given by \eqref{eq:sphere:H:NH}. From \eqref{eq:sphere:ode:imp}, we obtain the velocity
\begin{equation}\label{eq:sphere:dotx:forced:imp}
\dot{x}=\pm\sqrt{\frac{\frac{p_{0}}{3}x^3-H(x)}{x^3\left[1-\left(1+\frac{1}{x^3}\right)^{-1/3}\right]}},
\end{equation}
assuming that $\frac{p_{0}}{3}x^3-H(x)\geq0$. For an oscillatory motion to occur, the equation
\begin{equation}\label{eq:sphere:HC:imp}
H(x)=\frac{p_{0}}{3}x^3
\end{equation}
must have exactly two distinct solutions, $x=x_{1}$ and $x=x_{2}$, such that $0\leq x_{1}<x_{2}<\infty$.

%%%%%%%%%%%%%%
\begin{figure}[htbp]
	\begin{center}
		\includegraphics[width=0.5\textwidth]{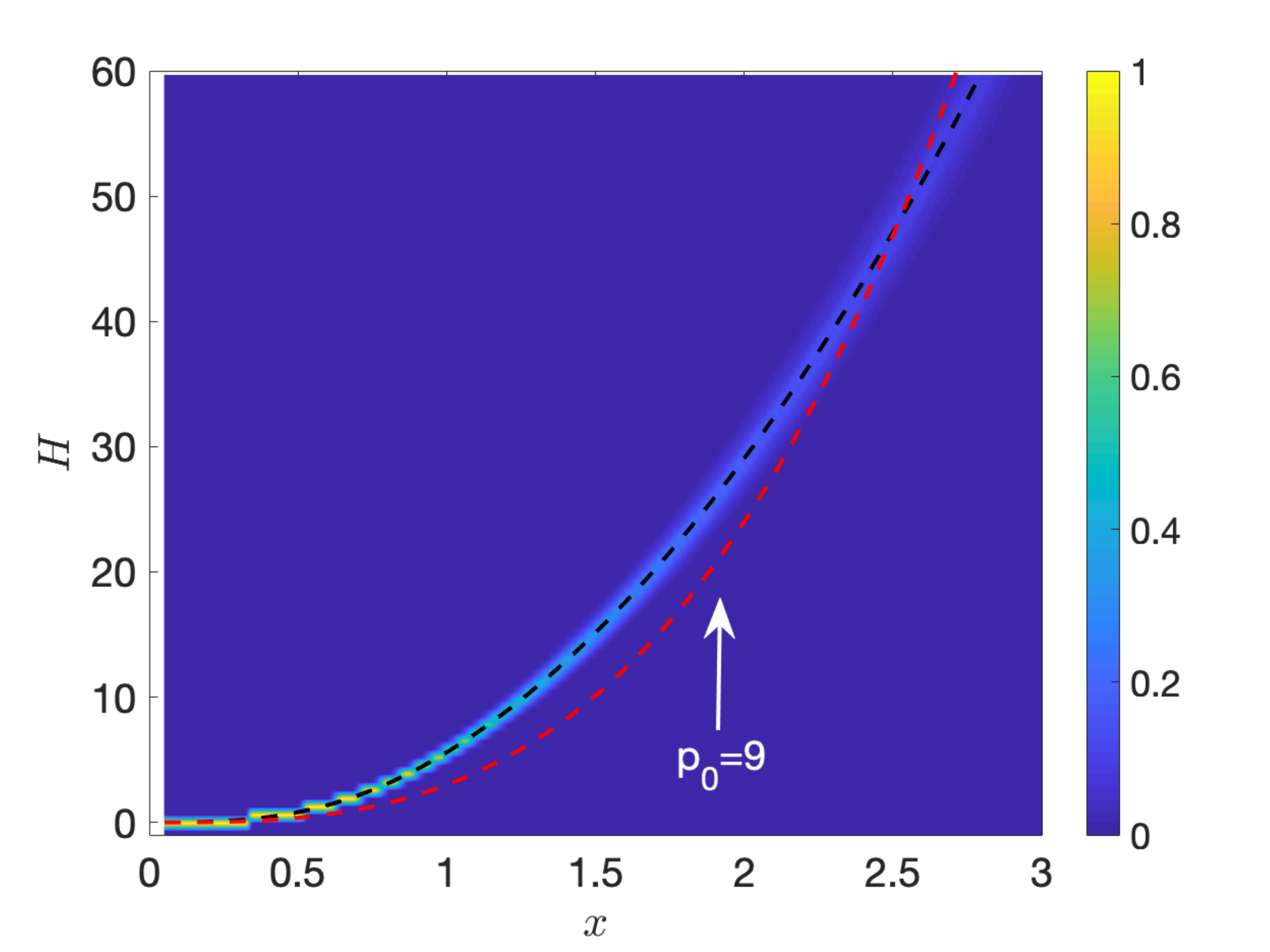}
		\caption{The function $H(x)$, defined by \eqref{eq:sphere:H:NH}, intersecting the (dashed red) curve $p_{0}x^3/3$, with $p_{0}=9$, for a dynamic sphere of stochastic neo-Hookean material under impulse traction, when $\rho=1$, $B=1$, and $\mu$ is drawn from the Gamma distribution with $\rho_{1}=405$ and $\rho_{2}=0.01$. The dashed black lines correspond to the expected values based only on mean value, $\underline{\mu}=\rho_{1}\rho_{2}=4.05$. Note that, between the points of intersection, the values of $H(x)$ are situated above the dashed red curve.}\label{fig:stoch-NHsphere-no}
	\end{center}
\end{figure}
%%%%%%%%%%%%

When $p_{0}\neq0$, substitution of \eqref{eq:sphere:H:NH} in \eqref{eq:sphere:HC:imp} gives
\begin{equation}\label{eq:sphere:p0:NH:imp}
\frac{p_{0}}{3}x^3=\frac{\mu}{\rho B^2}\left[2\left(x^3+1\right)^{2/3}-\frac{1}{\left(x^3+1\right)^{1/3}}-1\right],
\end{equation}
which has a solution at $x_{1}=0$. The  second solution, $x_{2}>0$, is a root of 
\begin{equation}\label{eq:sphere:p0:NH:nonzero:imp}
p_{0}=\frac{3\mu}{\rho B^2}\left[\frac{1}{\left(x^3+1\right)^{1/3}}+\frac{\left(x^3+1\right)^{1/3}+1}{\left(x^3+1\right)^{2/3}+\left(x^3+1\right)^{1/3}+1}\right].
\end{equation}
As the right-hand side of equation \eqref{eq:sphere:p0:NH:nonzero:imp} is a decreasing function of $x$, this equation has a solution, $x_{2}>0$, if and only if
\begin{equation}\label{eq:sphere:p0:NH:imp:left}
p_{0}>0=\lim_{x\to\infty}\frac{3\mu}{\rho B^2}\left[\frac{1}{\left(x^3+1\right)^{1/3}}+\frac{\left(x^3+1\right)^{1/3}+1}{\left(x^3+1\right)^{2/3}+\left(x^3+1\right)^{1/3}+1}\right]
\end{equation}
and
\begin{equation}\label{eq:sphere:p0:NH:imp:right}
p_{0}<\frac{5\mu}{\rho B^2}=\lim_{x\to0_{+}}\frac{3\mu}{\rho B^2}\left[\frac{1}{\left(x^3+1\right)^{1/3}}+\frac{\left(x^3+1\right)^{1/3}+1}{\left(x^3+1\right)^{2/3}+\left(x^3+1\right)^{1/3}+1}\right].
\end{equation}
However, since this implies
\begin{equation}\label{eq:sphere:nosc}
\frac{p_{0}}{3}x^3-H(x)\left\{
\begin{array}{ll}
<0 & \mbox{if}\ x_{1}< x< x_{2},\\
>0 & \mbox{if}\ x>x_{2},
\end{array}
\right.
\end{equation}
the sphere cannot oscillate when $p_{0}$ is constant in time, as assumed in \eqref{eq:sphere:impulse} (see Figure~\ref{fig:stoch-NHsphere-no}).

By \eqref{eq:sphere:impulse} and \eqref{eq:sphere:p0:NH:nonzero:imp}, the tensile traction, $T=T_{rr}(b,t)$, takes the form
\begin{equation}\label{eq:sphere:T:NH:imp}
T=\frac{3\mu}{2}\left[\frac{1}{\left(x^3+1\right)^{1/3}}+\frac{\left(x^3+1\right)^{1/3}+1}{\left(x^3+1\right)^{2/3}+\left(x^3+1\right)^{1/3}+1}\right],
\end{equation}
and decreases as $x$ increases. Thus, the critical tension for the onset of cavitation is
\begin{equation}\label{eq:sphere:T0:NH:imp}
T_{0}=\lim_{x\to0_{+}}T=\frac{5\mu}{2},
\end{equation}
as found also under dead loading.

%%%%%%%%%%%%%%%%%%%%%%%%%%%%%%%%%%%%%%%%%%%%%%%%%%%%%%%%%%%%
\subsection{Static deformation of a stochastic neo-Hookean sphere under impulse traction}\label{sec:sphere:NH:static:imp}

For the static sphere subject to a uniform constant surface load in the current configuration, given in terms of the Cauchy stresses, we have
\begin{equation}\label{eq:sphere:Tst:NH}
T=2\mu\left[\frac{1}{\left(x^{3}+1\right)^{1/3}}+\frac{1}{4\left(x^{3}+1\right)^{4/3}}\right],
\end{equation}
which decreases as $x$ increases. Then, the critical tension for cavitation initiation is also given by \eqref{eq:sphere:T0:NH:imp}.

%%%%%%%%%%%%%%
\begin{figure}[htbp]
	\begin{center}
		\includegraphics[width=0.5\textwidth]{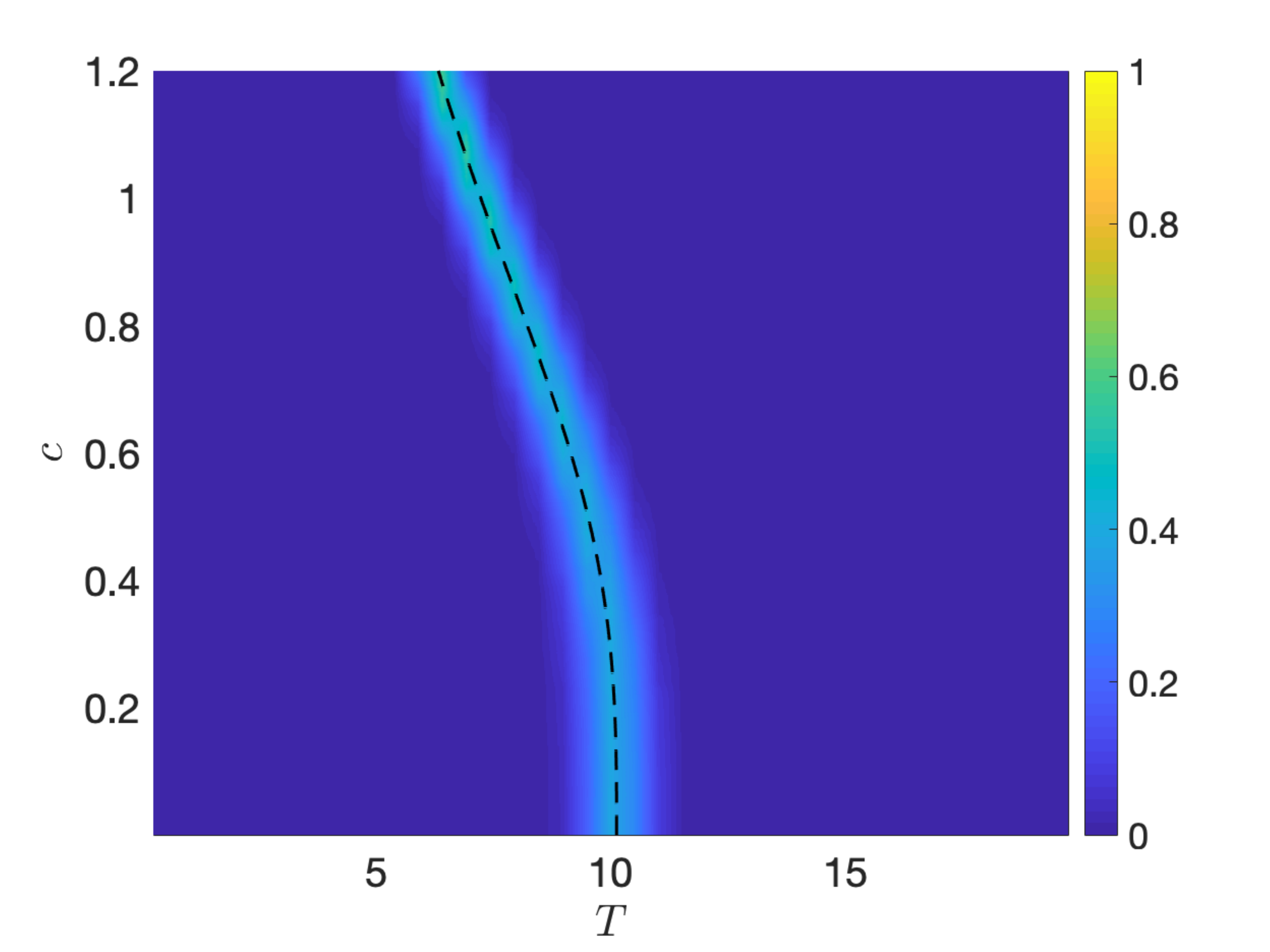}
		\caption{Probability distribution of the applied impulse traction $T$ causing cavitation of radius $c$ in a static unit sphere (with $B=1)$ of stochastic neo-Hookean material, when the shear modulus, $\mu$, follows a Gamma distribution with $\rho_{1}=405$ and $\rho_{2}=0.01$. The dashed black line corresponds to the expected bifurcation based only on the mean value, $\underline{\mu}=\rho_{1}\rho_{2}=4.05$.}\label{fig:Tstoch}
	\end{center}
\end{figure}
%%%%%%%%%%%%%

The post-cavitation behaviour can be inferred from the sign of the derivative of $T$ in a neighbourhood of $T_{0}$.  By differentiating \eqref{eq:sphere:Tst:NH} with respect to $x$, we obtain
\begin{equation}\label{Aeq:sphere:dTst:NH:imp}
\frac{\text{d}T}{\text{d}x}=-2\mu x^2\left[\frac{1}{\left(x^{3}+1\right)^{4/3}}+\frac{1}{\left(x^{3}+1\right)^{7/3}}\right]<0,
\end{equation}
i.e., the cavitation is unstable, regardless of the material parameter $\mu$ \cite{Ball:1982}. The post-cavitation stochastic behaviour of the static homogeneous sphere is shown in Figure~\ref{fig:Tstoch}, for the unit sphere (with $B=1$) of stochastic neo-Hookean material, where the shear modulus, $\mu$, is drawn from a Gamma distribution with shape and scale parameters $\rho_{1}=405$ and $\rho_{2}=4.05/\rho_{1}=0.01$, respectively (see also Figure~2 of \cite{Ball:1982}).

%%%%%%%%%%%%%%%%%%%%%%%%%%%%%%%%%%%%%%%%%%%%%%%%%%%%%%%%%%%%
%%%%%%%%%%%%%%%%%%%%   NEW SECTION  %%%%%%%%%%%%%%%%%%%%%%%%
%%%%%%%%%%%%%%%%%%%%%%%%%%%%%%%%%%%%%%%%%%%%%%%%%%%%%%%%%%%%
\section{Cavitation and radial motion of concentric homogeneous spheres}\label{sec:sphere:comp}

\begin{figure}[htbp]
	\begin{center}
		\includegraphics[width=0.75\textwidth]{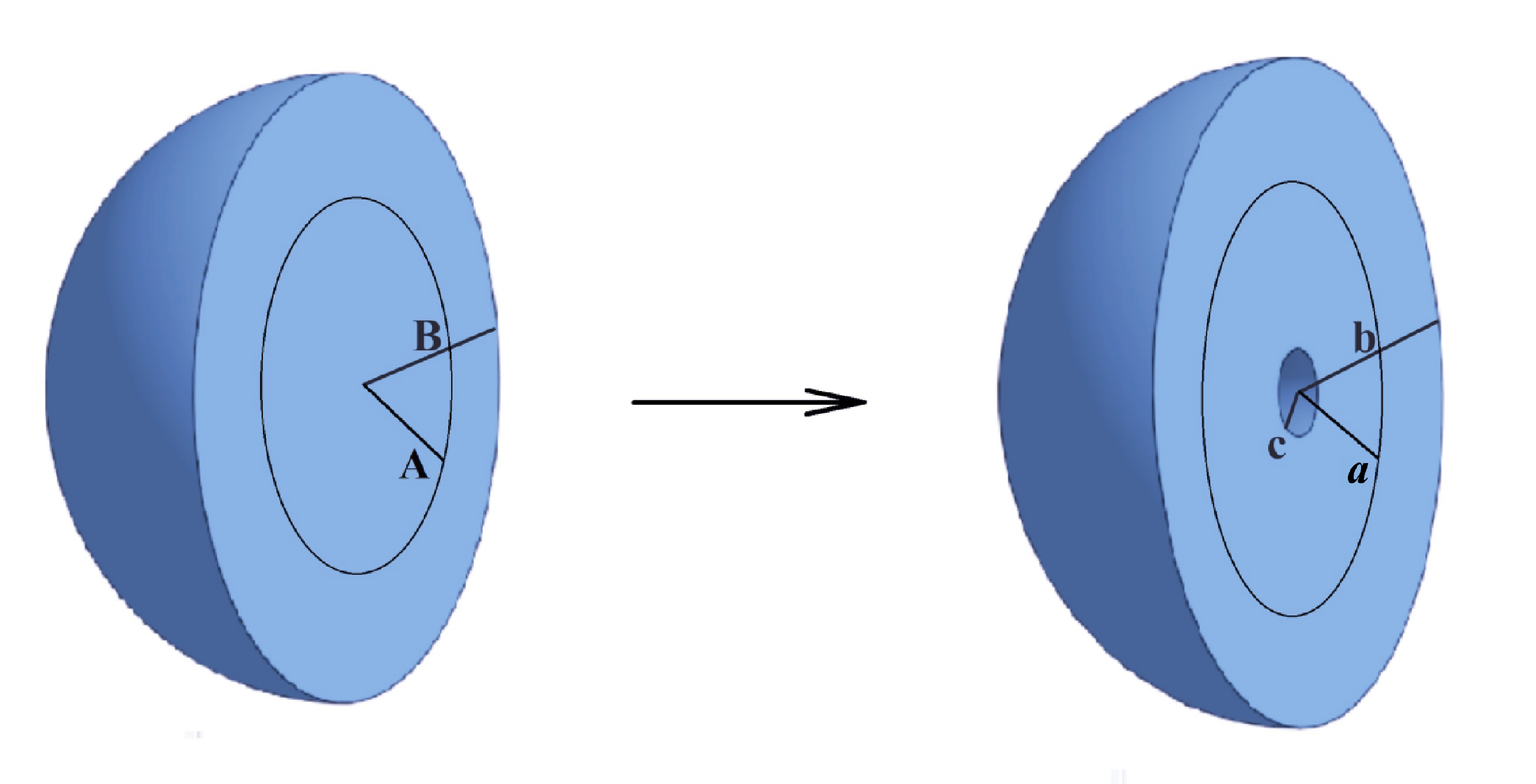}
		\caption{Schematic of cross-section of a composite sphere made of two concentric homogeneous spheres, with undeformed radii $A$ and $B$, respectively, showing the reference state (left), and the deformed state, with cavity radius $c$ and radii $a$ and $b$, respectively (right).}\label{fig:comp-sphere-cave}
	\end{center}
\end{figure}

Next, we extend the approach developed in the previous section to investigate the behaviour under the quasi-equilibrated radial motion \eqref{eq:sphere:deform:qem} of a composite formed from two concentric homogeneous spheres (see Figure~\ref{fig:comp-sphere-cave}). We restrict our attention to composite spheres with two stochastic neo-Hookean phases, similar to those containing two concentric spheres of different neo-Hookean material treated deterministically in \cite{Horgan:1989:HP} and \cite{Sivaloganathan:1991}. In this case, we define the following strain-energy function,
\begin{equation}\label{eq:sphere:two}
\mathcal{W}(\lambda_{1},\lambda_{2},\lambda_{3})=\left\{
\begin{array}{cc}
\frac{\mu^{(1)}}{2}\left(\lambda_{1}^{2}+\lambda_{2}^{2}+\lambda_{3}^{2}-3\right), & 0<R<A,\\
\frac{\mu^{(2)}}{2}\left(\lambda_{1}^{2}+\lambda_{2}^{2}+\lambda_{3}^{2}-3\right), & A<R<B,
\end{array}
\right.
\end{equation}
where $0<R<A$ and $A<R<B$ denote the radii of the inner and outer sphere in the reference configuration, and the corresponding shear moduli $\mu^{(1)}$ and $\mu^{(2)}$ are spatially-independent random variables characterised by the Gamma distributions $g(u;\rho^{(1)}_{1},\rho^{(1)}_{2})$ and $g(u;\rho^{(2)}_{1},\rho^{(2)}_{2})$, defined by \eqref{eq:mu:gamma}, respectively.

%%%%%%%%%%%%%%
%%%%%%%%%%%%%

%%%%%%%%%%%%%%%%%%%%%%%%%%%%%%%%%%%%%%%%%%%%%%%%%%%%%%%%%%%%
\subsection{Oscillatory motion of a sphere with two stochastic neo-Hookean phases under dead-load traction}\label{sec:sphere:comp:osc}

For the composite sphere, we denote the radial pressures acting on the curvilinear surfaces $r=c(t)$ and $r=b(t)$ at time $t$ as $T_{1}(t)$ and $T_{2}(t)$, respectively, and impose the continuity condition for the stress components across their interface, $r=a(t)$. By analogy to \eqref{eq:sphere:T1T2:r}, we obtain
\begin{equation}\label{eq:sphere:T1T2:r:comp}
\begin{split}
T_{1}(t)-T_{2}(t)
&=\rho B^2\left[\left(\frac{c}{B}\frac{\ddot{c}}B+2\frac{\dot{c}^2}{B^2}\right)\left(1-\frac{c}{b}\right)-\frac{\dot{c}^2}{2B^2}\left(1-\frac{c^4}{b^4}\right)\right]\\
&+2\int_{c}^{a}\mu^{(1)}\left(\frac{r^2}{R^2}-\frac{R^4}{r^4}\right)\frac{dr}{r}+2\int_{a}^{b}\mu^{(2)}\left(\frac{r^2}{R^2}-\frac{R^4}{r^4}\right)\frac{dr}{r}.
\end{split}
\end{equation}
Using the notation \eqref{eq:sphere:ux}, we can write \eqref{eq:sphere:T1T2:r:comp} equivalently as follows,
\begin{equation}\label{eq:sphere:T1T2:x:comp}
\begin{split}
2x^2\frac{T_{1}(t)-T_{2}(t)}{\rho B^2}
&=\frac{d}{dx}\left\{\dot{x}^2x^3\left[1-\left(1+\frac{1}{x^3}\right)^{-1/3}\right]\right\}\\
&+\frac{4x^2}{3\rho B^2}\int_{x^3+1}^{x^3B^3/A^3+1}\mu^{(2)}\frac{1+u}{u^{7/3}}du+\frac{4x^2}{3\rho B^2}\int_{x^3B^3/A^3+1}^{\infty}\mu^{(1)}\frac{1+u}{u^{7/3}}du.
\end{split}
\end{equation}

In this case, we denote
\begin{equation}\label{eq:sphere:Hint:comp}
\begin{split}
H(x)&=\frac{4}{3\rho B^2}\int_{0}^{x}\left(\zeta^2\int_{\zeta^3+1}^{\zeta^3B^3/A^3+1}\mu^{(2)}\frac{1+u}{u^{7/3}}du\right)d\zeta\\
&+\frac{4}{3\rho B^2}\int_{0}^{x}\left(\zeta^2\int_{\zeta^3B^3/A^3+1}^{\infty}\mu^{(1)}\frac{1+u}{u^{7/3}}du\right)d\zeta,
\end{split}
\end{equation}
and, after evaluating the integrals, we obtain the equivalent form
\begin{equation}\label{eq:sphere:H:comp}
\begin{split}
H(x)&=\frac{\mu^{(2)}}{\rho B^2}\left[2\left(x^3+1\right)^{2/3}-\frac{1}{\left(x^3+1\right)^{1/3}}-1\right]\\
&+\frac{\mu^{(1)}-\mu^{(2)}}{\rho B^2}\frac{A^3}{B^3}\left[2\left(x^3\frac{B^3}{A^3}+1\right)^{2/3}-\frac{1}{\left(x^3\frac{B^3}{A^3}+1\right)^{1/3}}-1\right].
\end{split}
\end{equation}

Then, substitution of \eqref{eq:sphere:H:comp} in \eqref{eq:sphere:HC} gives
\begin{equation}\label{eq:sphere:p0:comp}
\begin{split}
p_{0}\left[\left(x^3+1\right)^{1/3}-1\right]&=\frac{\mu^{(2)}}{2}\left[2\left(x^3+1\right)^{2/3}-\frac{1}{\left(x^3+1\right)^{1/3}}-1\right]\\
&+\frac{\mu^{(1)}-\mu^{(2)}}{2}\frac{A^3}{B^3}\left[2\left(x^3\frac{B^3}{A^3}+1\right)^{2/3}-\frac{1}{\left(x^3\frac{B^3}{A^3}+1\right)^{1/3}}-1\right].
\end{split}
\end{equation}
This equation has one solution at $x_{1}=0$, while the  second solution, $x_{2}>0$, is a root of 
\begin{equation}\label{eq:sphere:p0:comp:nonzero}
\begin{split}
p_{0}&=\frac{\mu^{(2)}}{2}\left[2\left(x^3+1\right)^{1/3}+\frac{1}{\left(x^3+1\right)^{1/3}}+2\right]\\\
&-\frac{\mu^{(2)}}{2}\frac{\left(x^3+1\right)^{2/3}+\left(x^3+1\right)^{1/3}+1}{\left(x^3\frac{B^3}{A^3}+1\right)^{2/3}+\left(x^3\frac{B^3}{A^3}+1\right)^{1/3}+1}\left[2\left(x^3\frac{B^3}{A^3}+1\right)^{1/3}+\frac{1}{\left(x^3\frac{B^3}{A^3}+1\right)^{1/3}}+2\right]\\
&+\frac{\mu^{(1)}}{2}\frac{\left(x^3+1\right)^{2/3}+\left(x^3+1\right)^{1/3}+1}{\left(x^3\frac{B^3}{A^3}+1\right)^{2/3}+\left(x^3\frac{B^3}{A^3}+1\right)^{1/3}+1}\left[2\left(x^3\frac{B^3}{A^3}+1\right)^{1/3}+\frac{1}{\left(x^3\frac{B^3}{A^3}+1\right)^{1/3}}+2\right].
\end{split}
\end{equation}
Next, we expand $p_{0}$, given by \eqref{eq:sphere:p0:comp:nonzero}, to the first order in $x^3$ to obtain
\begin{equation}\label{eq:sphere:p0:comp:nonzero:app}
p_{0}\approx\frac{5\mu^{(1)}}{2}+\frac{x^3}{3}\left[2\left(\frac{B^3}{A^3}-1\right)\left(\mu^{(2)}-\mu^{(1)}\right)+\frac{\mu^{(1)}}{2}\right].
\end{equation}
The critical tension for cavity initiation is then
\begin{equation}\label{eq:sphere:p0:comp:crit}
\lim_{x\to0_{+}}p_{0}=\frac{5\mu^{(1)}}{2},
\end{equation}
and a comparison with \eqref{eq:sphere:p0:NH:crit} shows that it is the same as for the homogeneous sphere made of the same material as the inner sphere.

%%%%%%%%%%%%%%
\begin{figure}[htbp]
	\begin{center}
		\includegraphics[width=0.5\textwidth]{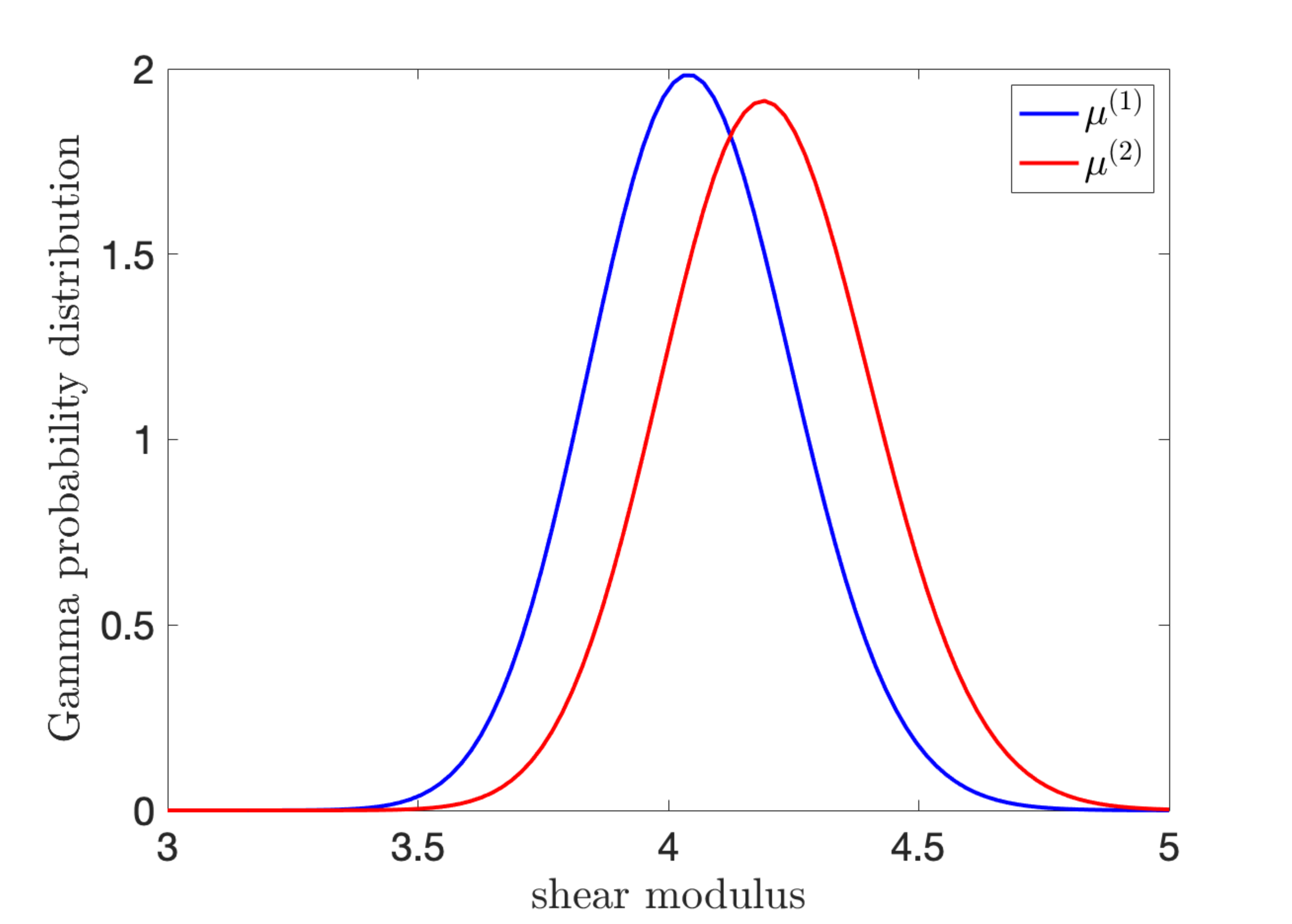}
		\caption{Examples of Gamma distributions with hyperparameters $\rho^{(1)}_{1}=405$, $\rho^{(1)}_{2}=4.05/\rho^{(1)}_{1}=0.01$ for $\mu^{(1)}>0$, and $\rho^{(1)}_{1}=405$, $\rho^{(2)}_{2}=4.2/\rho^{(2)}_{1}$ for $\mu^{(2)}>0$.}\label{fig:mu12-gpdf}
	\end{center}
\end{figure}
%%%%%%%%%%%%%%

%%%%%%%%%%%%%%
\begin{figure}[htbp]
	\begin{center}
		\includegraphics[width=0.47\textwidth]{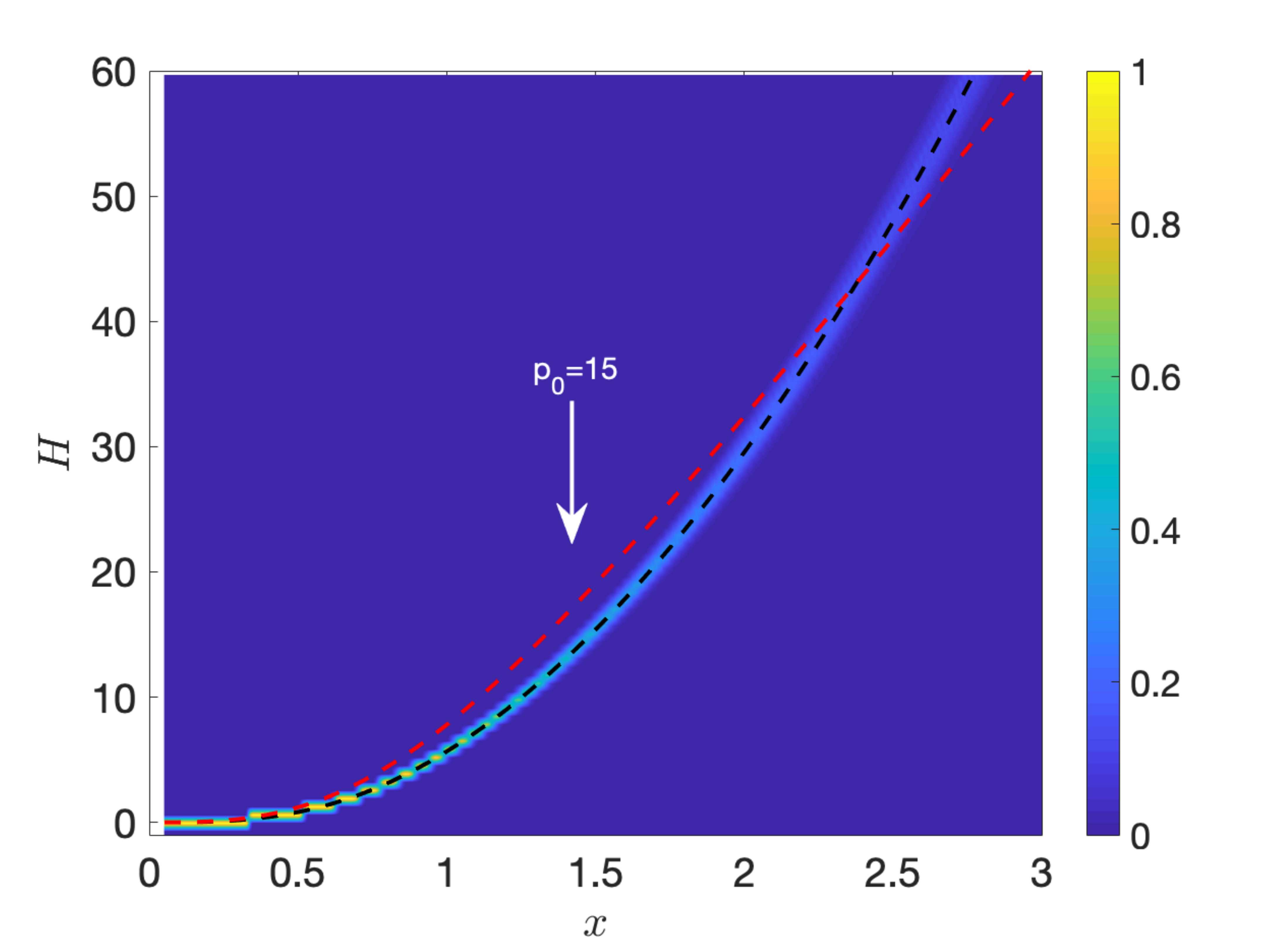}\qquad
		\includegraphics[width=0.47\textwidth]{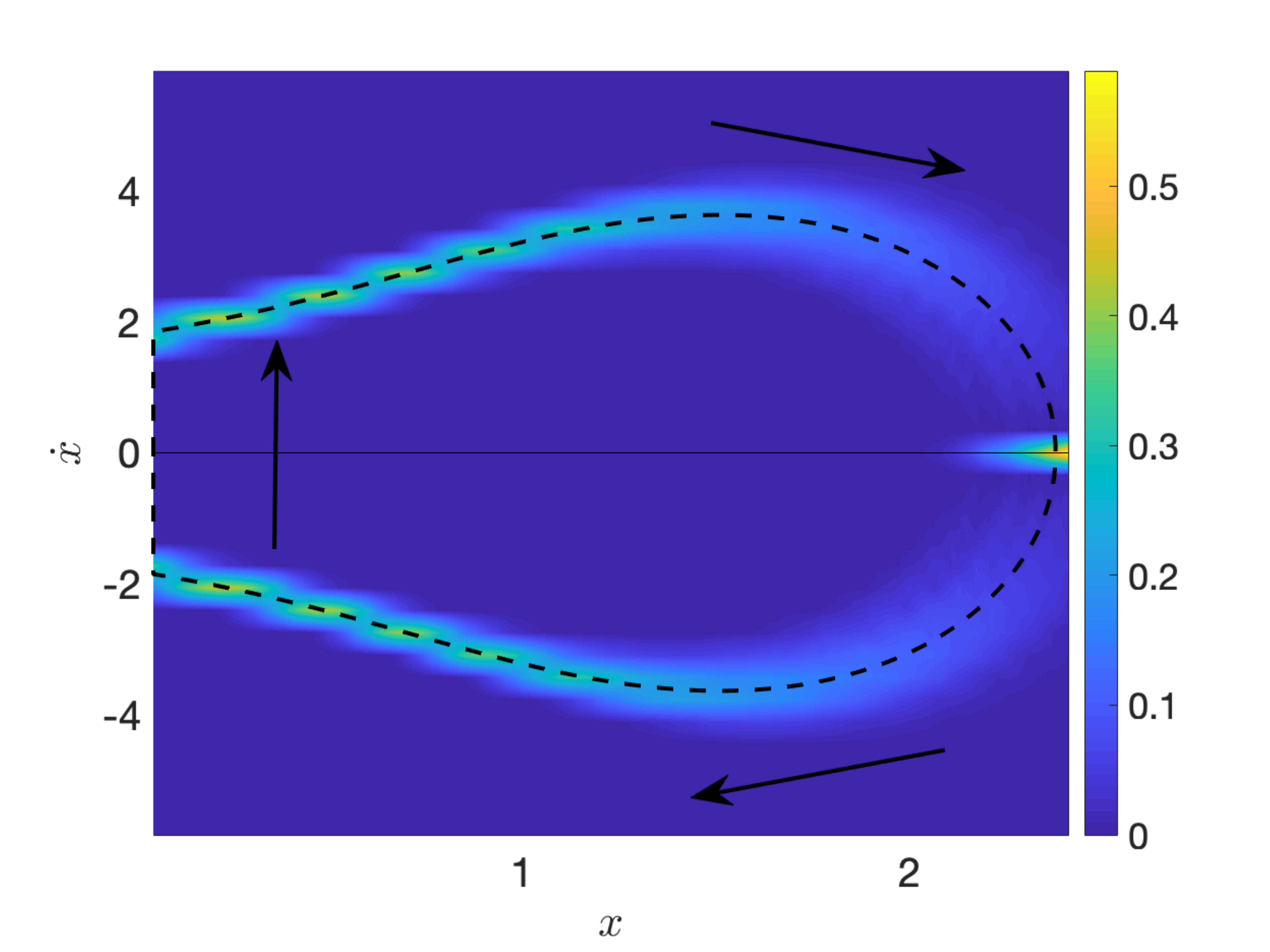}
		\caption{The function $H(x)$, defined by \eqref{eq:sphere:H:comp}, intersecting the (dashed red) curve $\frac{2p_{0}}{\rho B^2}\left[\left(x^3+1\right)^{1/3}-1\right]$, with $p_{0}=15$ (left), and the associated velocity, given by \eqref{eq:sphere:dotx:forced} (right), for a dynamic composite sphere with two concentric stochastic neo-Hookean phases, with radii $A=1/2$ and $B=1$, respectively, under dead-load traction, when the shear modulus of the inner phase, $\mu^{(1)}$, follows a Gamma distribution with $\rho^{(1)}_{1}=405$, $\rho^{(1)}_{2}=4.05/\rho^{(1)}_{1}=0.01$, while the shear modulus of the outer phase, $\mu^{(2)}$, is drawn from a Gamma distribution with $\rho^{(2)}_{1}=405$, $\rho^{(2)}_{2}=4.2/\rho^{(2)}_{1}$. The dashed black lines correspond to the expected values based only on mean values, $\underline{\mu}^{(1)}=4.05$ and $\underline{\mu}^{(2)}=4.2$. Each distribution was calculated from the average of $1000$ stochastic simulations.}\label{fig:stoch-compsphere}
	\end{center}
\end{figure}
%%%%%%%%%%%%

Considering the random shear parameters of the two concentric spheres, we distinguish the following two cases:
\begin{itemize}
\item[(i)] If
\begin{equation}\label{eq:sphere:comp:supercrit}
\frac{\mu^{(2)}}{\mu^{(1)}}>1-\frac{1}{4}\left(\frac{B^3}{A^3}-1\right)^{-1},
\end{equation}
then the right-hand side of \eqref{eq:sphere:p0:comp:nonzero:app} is an increasing function of $x$, and
\begin{equation}\label{eq:sphere:p0:comp:bound}
p_{0}>\lim_{x\to 0_{+}}\frac{5\mu^{(1)}}{2}+\frac{x^3}{3}\left[2\left(\frac{B^3}{A^3}-1\right)\left(\mu^{(2)}-\mu^{(1)}\right)+\frac{\mu^{(1)}}{2}\right]=\frac{5\mu^{(1)}}{2}.
\end{equation}
By \eqref{eq:sphere:p0:comp:bound}, for the motion of the composite sphere to be oscillatory, the shear modulus of the inner sphere must satisfy
\begin{equation}\label{eq:sphere:shearmod:comp:bounds}
\eta<\mu^{(1)}<\frac{2p_{0}}{5},
\end{equation}
where the upper bound follows from \eqref{eq:sphere:p0:comp:bound}, while the lower bound is assumed to be strictly positive, i.e., $\eta>0$. Thus, the probability distribution of oscillatory motions occurring is given by \eqref{eq:sphere:NH:P1} and that of non-oscillatory motions by \eqref{eq:sphere:NH:P2}, with $g(u;\rho^{(1)}_{1},\rho^{(1)}_{2})$ instead of $g(u;\rho_{1},\rho_{2})$ (see Figure~\ref{fig:intpdfs-NHsphere}). 

As the material parameters, $\mu^{(1)}$ and $\mu^{(2)}$, are positive, \eqref{eq:sphere:comp:supercrit} is possible for any values $0<A<B$. In particular,  if
\begin{equation}
\label{eq:sphere:comp:supercrit:BvsA}
\frac{B}{A}<\left(\frac{5}{4}\right)^{1/3},
\end{equation}
then \eqref{eq:sphere:comp:supercrit} always holds, regardless of the material constants.

\item[(ii)] If
\begin{equation}\label{eq:sphere:comp:subcrit}
\frac{\mu^{(2)}}{\mu^{(1)}}<1-\frac{1}{4}\left(\frac{B^3}{A^3}-1\right)^{-1},
\end{equation}
then the right-hand side in \eqref{eq:sphere:p0:comp:nonzero:app} decreases as $x$ increases, hence
\begin{equation}\label{eq:sphere:p0:comp:bounds}
0<p_{0}<\frac{5\mu^{(1)}}{2}.
\end{equation}
As the material constants, $\mu^{(1)}$ and $\mu^{(2)}$, are positive, \eqref{eq:sphere:comp:subcrit} is possible if and only if
\begin{equation}
\label{eq:sphere:comp:subcrit:BvsA}
\frac{B}{A}>\left(\frac{5}{4}\right)^{1/3}.
\end{equation}
However, in this case,
\begin{equation}\label{eq:sphere:comp:osc-no}
\frac{2p_{0}}{\rho B^2}\left[\left(x^3+1\right)^{1/3}-1\right]-H(x)\left\{
\begin{array}{ll}
<0 & \mbox{if}\ x_{1}< x< x_{2},\\
>0 & \mbox{if}\ x>x_{2},
\end{array}
\right.
\end{equation}
and the sphere cannot oscillate.
\end{itemize}

In Figure~\ref{fig:stoch-compsphere}, we represent the stochastic function $H(x)$, defined by \eqref{eq:sphere:H:comp}, intersecting the curve $\frac{2p_{0}}{\rho B^2}\left[\left(x^3+1\right)^{1/3}-1\right]$ , with $p_{0}=15$, to find the two distinct solutions to \eqref{eq:sphere:HC}, and the associated velocity, given by \eqref{eq:sphere:dotx:forced}, for a composite sphere with two concentric stochastic neo-Hookean phases, with radii $A=1/2$ and $B=1$, respectively, when the shear modulus of the inner phase, $\mu^{(1)}$, follows a Gamma distribution with $\rho^{(1)}_{1}=405$, $\rho^{(1)}_{2}=4.05/\rho^{(1)}_{1}=0.01$, while the shear modulus of the outer phase, $\mu^{(2)}$, is drawn from a Gamma distribution with $\rho^{(2)}_{1}=405$, $\rho^{(2)}_{2}=4.2/\rho^{(2)}_{1}$ (see Figure~\ref{fig:mu12-gpdf}).

%%%%%%%%%%%%%%%%%%%%%%%%%%%%%%%%%%%%%%%%%%%%%%%%%%%%%%%%%%%%
\subsection{Static deformation of a sphere with two stochastic neo-Hookean phases under dead-load traction}\label{sec:sphere:comp:static}

For the static composite sphere, if the surface of the cavity is traction-free, then $T_{1}=0$ and \eqref{eq:sphere:T1T2:x:comp} reduces to
\begin{equation}\label{eq:sphere:T:comp}
T_{rr}(b)=\frac{2}{3}\int_{x^3+1}^{x^3B^3/A^3+1}\mu^{(2)}\frac{1+u}{u^{7/3}}du+\frac{2}{3}\int_{x^3B^3/A^3+1}^{\infty}\mu^{(1)}\frac{1+u}{u^{7/3}}du.
\end{equation}
After evaluating the integrals in \eqref{eq:sphere:T:comp}, the required dead-load traction at the outer surface, $R=B$, in the reference configuration, is equal to
\begin{equation}\label{eq:sphere:P:comp}
\begin{split}
P&=\left(x^3+1\right)^{2/3}T_{rr}(b)\\
&=2\mu^{(2)}\left[\left(x^3+1\right)^{1/3}+\frac{1}{4\left(x^3+1\right)^{2/3}}\right]\\
&+2\left(\mu^{(1)}-\mu^{(2)}\right)\left(x^3+1\right)^{2/3}\left[\frac{1}{\left(x^3\frac{B^3}{A^3}+1\right)^{1/3}}+\frac{1}{4\left(x^3\frac{B^3}{A^3}+1\right)^{4/3}}\right].
\end{split}
\end{equation}
The critical dead load for the cavity formation is then
\begin{equation}\label{eq:sphere:P0:comp}
P_{0}=\lim_{x\to0_{+}}P=\frac{5\mu^{(1)}}{2},
\end{equation}
and is the same as for the homogeneous sphere made entirely of the material of the inner sphere \cite{Horgan:1989:HP}, as found also in the dynamic case.

To study the stability of this cavitation, we examine the behaviour of the cavity opening, with radius $c$ as a function of $P$, in a neighbourhood of $P_{0}$. Expanding the expression of $P$, given by \eqref{eq:sphere:P:comp}, to the first order in $x^3$, we obtain \cite{Horgan:1989:HP}
\begin{equation}\label{eq:sphere:P:comp:app}
P\approx\frac{5\mu^{(1)}}{2}+\frac{x^3}{3}\left[4\left(\frac{B^3}{A^3}-1\right)\left(\mu^{(2)}-\mu^{(1)}\right)+\mu^{(1)}\right].
\end{equation}
After differentiating the above function with respect to the dimensionless cavity radius $x=c/B$, we have
\begin{equation}\label{eq:sphere:dP:comp:app}
\frac{\text{d}P}{\text{d}x}\approx x^2\left[4\left(\frac{B^3}{A^3}-1\right)\left(\mu^{(2)}-\mu^{(1)}\right)+\mu^{(1)}\right].
\end{equation}

Then, $\text{d}P/\text{d}x\to 0$ as $x\to 0_{+}$, and the bifurcation at the critical dead load, $P_{0}$, is supercritical (respectively, subcritical) if $\text{d}P/\text{d}x>0$ (respectively, $\text{d}P/\text{d}x<0$) for arbitrarily small $x$. From \eqref{eq:sphere:dP:comp:app}, the following two cases are distinguished (see also the discussion in \cite{Horgan:1989:HP}):
\begin{itemize}
\item[(i)] If \eqref{eq:sphere:comp:supercrit} is valid, then the solution presents itself as a supercritical bifurcation, and the cavitation is stable, in the sense that a new bifurcated solution exists locally for values of $P>P_{0}$, and the cavity radius monotonically increases with the applied load post-bifurcation. In particular, when $\mu^{(1)}=\mu^{(2)}$, the problem reduces to the case of a homogeneous sphere made of neo-Hookean material, for which stable cavitation is known to occur \cite{Mihai:2019c:MDWG}.

\item[(ii)] If \eqref{eq:sphere:comp:subcrit} holds, then the bifurcation is subcritical, with the cavitation being unstable, in the sense that the required dead load starts to decrease post-bifurcation, causing a snap cavitation, i.e., a sudden jump in the cavity opening.
\end{itemize}

Thus, on the one hand, when \eqref{eq:sphere:comp:supercrit:BvsA} is satisfied, \eqref{eq:sphere:comp:supercrit} is valid regardless of the values of $\mu^{(1)}$ and $\mu^{(2)}$, and the cavitation is guaranteed to be stable (with 100\% certainty). On the other hand, if \eqref{eq:sphere:comp:subcrit:BvsA} holds, then, as either of the two inequalities \eqref{eq:sphere:comp:supercrit} and \eqref{eq:sphere:comp:subcrit} is possible, the probability distribution of stable cavitation in the composite sphere is equal to
\begin{equation}\label{eq:sphere:comp:P1}
P_{1}(\mu^{(2)})=\int_{0}^{\mu^{(2)}/\left[1-\frac{1}{4}\left(\frac{B^3}{A^3}-1\right)^{-1}\right]}g(u;\rho^{(1)}_{1},\rho^{(1)}_{2})du,
\end{equation}
while that of unstable (snap) cavitation is
\begin{equation}\label{eq:sphere:comp:P2}
P_{2}(\mu^{(2)})=1-P_{1}(\mu^{(2)})=1-\int_{0}^{\mu^{(2)}/\left[1-\frac{1}{4}\left(\frac{B^3}{A^3}-1\right)^{-1}\right]}g(u;\rho^{(1)}_{1},\rho^{(1)}_{2})du.
\end{equation}

%%%%%%%%%%%%%%
\begin{figure}[htbp]
	\begin{center}
		\includegraphics[width=\textwidth]{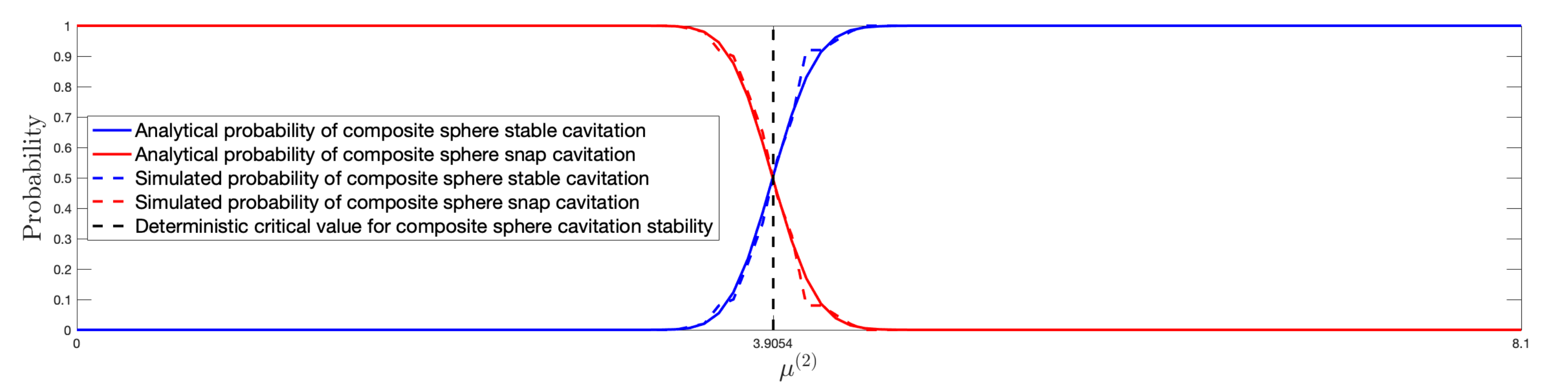}
		\caption{Probability distributions of stable or unstable cavitation in a static sphere made of two concentric stochastic neo-Hookean phases, with radii $A=1/2$ and $B=1$, respectively, under dead-load traction, when the shear modulus of the inner sphere, $\mu^{(1)}$, follows a Gamma distribution with $\rho^{(1)}_{1}=405$, $\rho^{(1)}_{2}=0.01$. Continuous coloured lines represent analytically derived solutions, given by equations \eqref{eq:sphere:comp:P1}-\eqref{eq:sphere:comp:P2}, and the dashed versions represent stochastically generated data. The vertical line at the critical value, $27\underline{\mu}^{(1)}/28=3.9054$, separates the expected regions based only on mean value, $\underline{\mu}=\rho^{(1)}_{1}\rho^{(1)}_{2}=4.05$. The probabilities were calculated from the average of 100 stochastic simulations.}\label{fig:comp-sphere-pdfs}
	\end{center}
\end{figure}
%%%%%%%%%%%%%%%

%%%%%%%%%%%%%%
\begin{figure}[htbp]
	\begin{center}
		\includegraphics[width=0.47\textwidth]{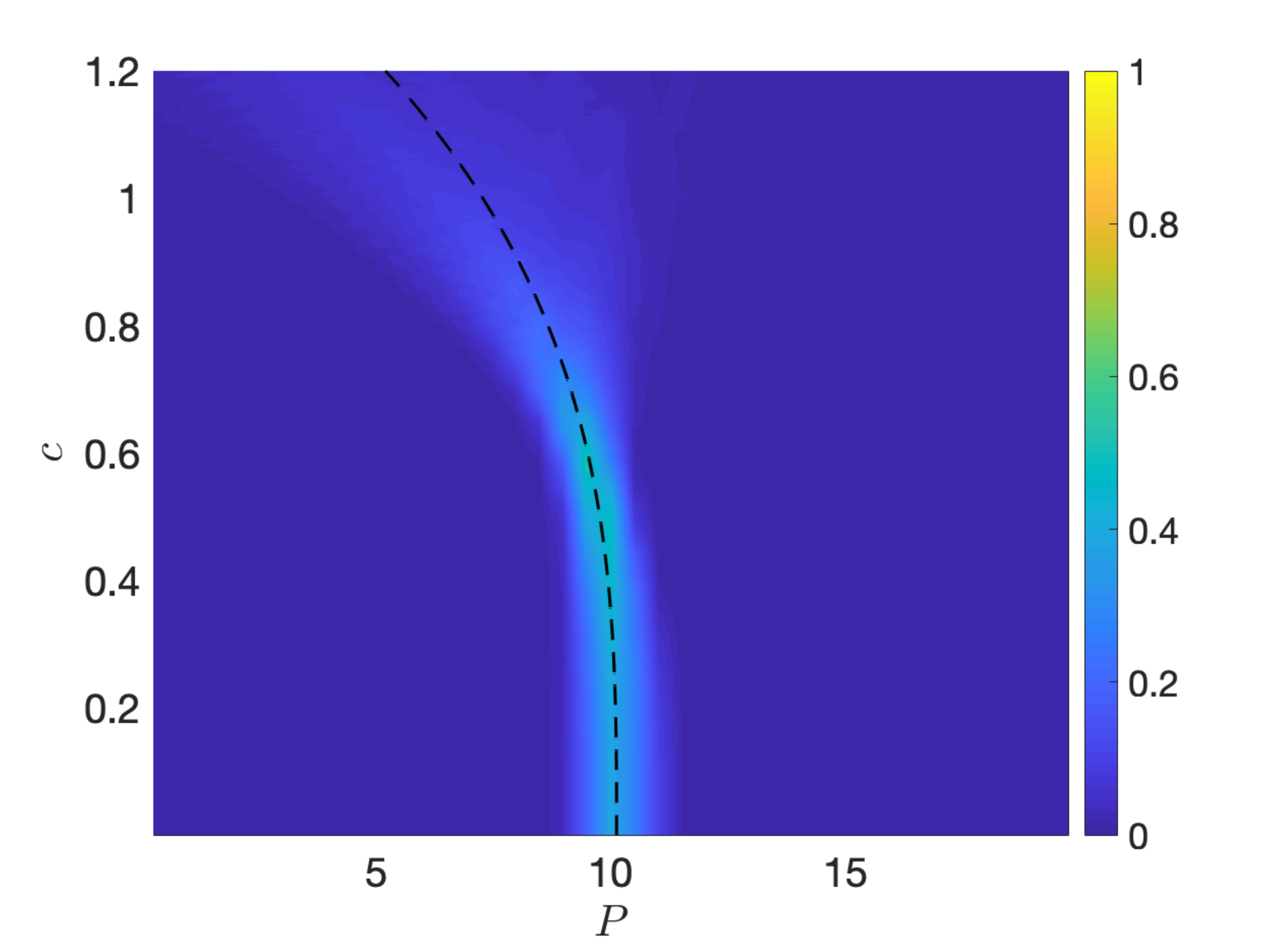}\qquad
		\includegraphics[width=0.47\textwidth]{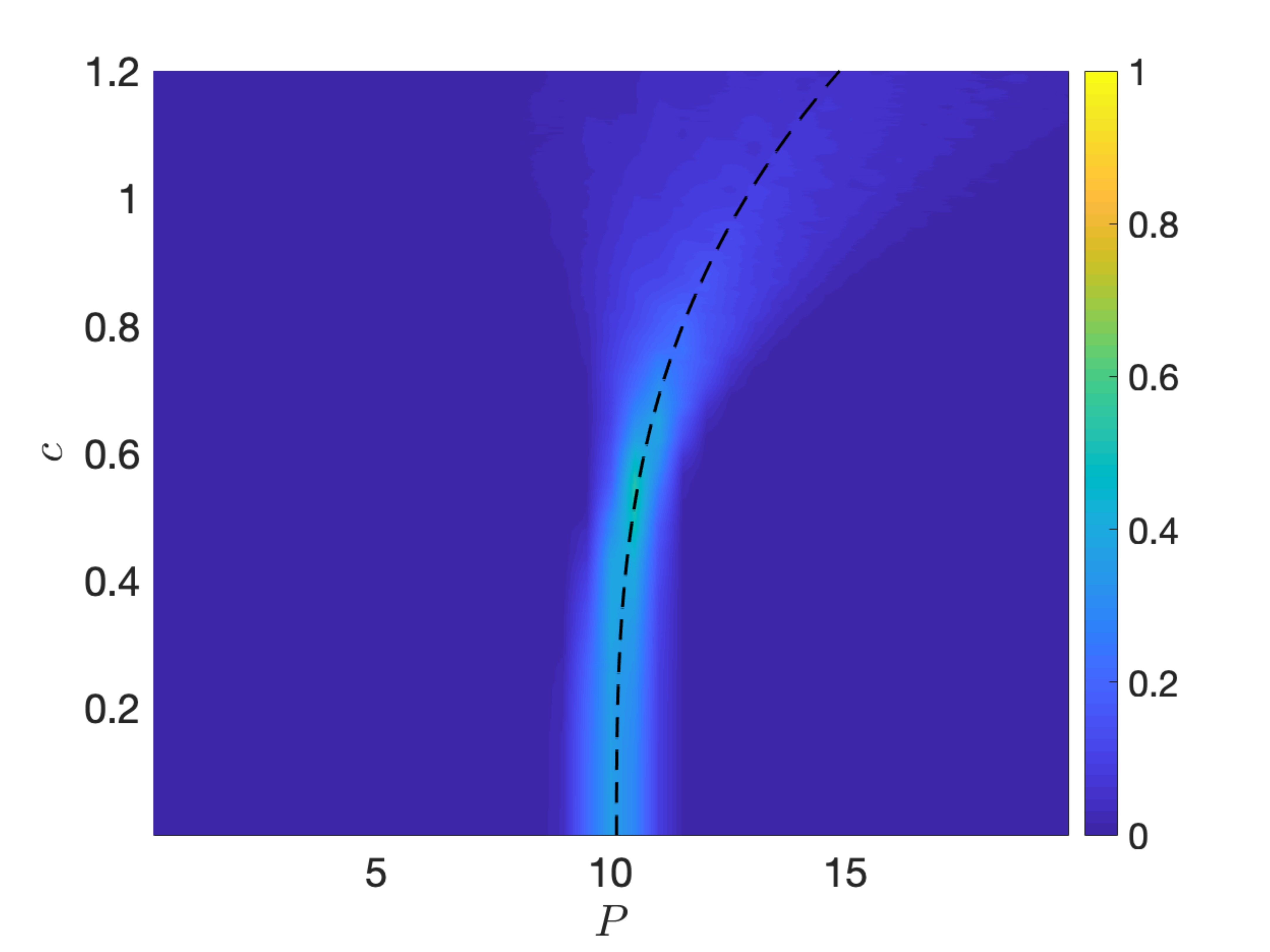}
		\caption{Probability distribution of the applied dead-load traction $P$ causing cavitation of radius $c$ in a static sphere of two concentric stochastic neo-Hookean phases, with radii $A=1/2$ and $B=1$, respectively, when the shear modulus of the inner phase, $\mu^{(1)}$, follows a Gamma distribution with $\rho^{(1)}_{1}=405$, $\rho^{(1)}_{2}=4.05/\rho^{(1)}_{1}=0.01$, while the shear modulus of the outer phase, $\mu^{(2)}$, is drawn from Gamma distribution with $\rho^{(2)}_{1}=405$, $\rho^{(2)}_{2}=3.6/\rho^{(2)}_{1}$ (left) or $\rho^{(2)}_{1}=405$, $\rho^{(2)}_{2}=4.2/\rho^{(2)}_{1}$ (right). The dashed black lines correspond to the expected bifurcation based only on mean parameter values.}\label{fig:Pcompstoch}
	\end{center}
\end{figure}
%%%%%%%%%%%%%

For example, setting $A=1/2$ and $B=1$, if $\rho^{(1)}_{1}=405$, $\rho^{(1)}_{2}=4.05/\rho^{(1)}_{1}= 0.01$ for the inner sphere, and $\rho^{(2)}_{1}=405$, $\rho^{(2)}_{2}=4.2/\rho^{(2)}_{1}$ for the outer sphere (see Figure~\ref{fig:mu12-gpdf}), then the probability distributions given by equations \eqref{eq:sphere:comp:P1}-\eqref{eq:sphere:comp:P2} are illustrated numerically in Figure~\ref{fig:comp-sphere-pdfs} (with blue lines for $P_{1}$ and red lines for $P_{2}$). For the numerical realisations in these plots, the interval $(0,8.1)=\left(0,2\underline{\mu}^{(1)}\right)$ was discretised into $100$ representative points, then for each value of $\mu^{(2)}$, $100$ random values of $\mu^{(1)}$ were numerically generated from the specified Gamma distribution and compared with the inequalities defining the two intervals for values of $\mu^{(2)}$. For the deterministic elastic case, which is based on the mean value of the shear modulus, $\underline{\mu}^{(1)}=\rho^{(1)}_{1}\rho^{(1)}_{2}=4.05$, the critical value of $27\underline{\mu}^{(1)}/28=3.9054$ strictly separates the cases where cavitation instability can, and cannot, occur. However, in the stochastic case, the two states compete. For example, at the same critical value, there is, by definition, exactly 50\% chance that the cavitation is stable, and 50\% chance that is not.

The post-cavitation stochastic behaviours shown in Figure~\ref{fig:Pcompstoch} correspond to two different static composite sphere where the shear modulus of the inner phase, $\mu^{(1)}$, follows a Gamma distribution with $\rho^{(1)}_{1}=405$, $\rho^{(1)}_{2}=4.05/\rho^{(1)}_{1}=0.01$, while the shear modulus of the outer phase, $\mu^{(2)}$, is drawn from a Gamma distribution with $\rho^{(2)}_{1}=405$, $\rho^{(2)}_{2}=3.6/\rho^{(2)}_{1}$  (in Figure~\ref{fig:Pcompstoch}-left) or $\rho^{(2)}_{1}=405$, $\rho^{(2)}_{2}=4.2/\rho^{(2)}_{1}$ (in Figure~\ref{fig:Pcompstoch}-right). In each case, unstable or stable cavitation is expected, respectively, but there is also about 5\% chance that the opposite behaviour is presented.

%%%%%%%%%%%%%%%%%%%%%%%%%%%%%%%%%%%%%%%%%%%%%%%%%%%%%%%%%%%%
\subsection{Non-oscillatory motion of a sphere with two stochastic neo-Hookean phases under impulse traction}\label{sec:sphere:comp:nosc}

In this case, setting the initial conditions $x_{0}=x(0)=0$ and $\dot{x}_{0}=\dot{x}(0)=0$, we obtain equation \eqref{eq:sphere:ode:imp}, where $H(x)$ is given by \eqref{eq:sphere:H:comp}. Then, substitution of \eqref{eq:sphere:H:comp} in \eqref{eq:sphere:HC:imp} gives
\begin{equation}\label{eq:sphere:p0:comp:imp}
\begin{split}
\frac{p_{0}}{3}x^3&=\frac{\mu^{(2)}}{\rho B^2}\left[2\left(x^3+1\right)^{2/3}-\frac{1}{\left(x^3+1\right)^{1/3}}-1\right]\\
&+\frac{\mu^{(1)}-\mu^{(2)}}{\rho B^2}\frac{A^3}{B^3}\left[2\left(x^3\frac{B^3}{A^3}+1\right)^{2/3}-\frac{1}{\left(x^3\frac{B^3}{A^3}+1\right)^{1/3}}-1\right].
\end{split}
\end{equation}
Equivalently, by \eqref{eq:sphere:impulse}, in terms of the tensile traction $T=T_{rr}(b,t)$, 
\begin{equation}\label{eq:sphere:T:comp:imp}
\begin{split}
\frac{2T}{3}x^3&=\mu^{(2)}\left[2\left(x^3+1\right)^{2/3}-\frac{1}{\left(x^3+1\right)^{1/3}}-1\right]\\
&+\left(\mu^{(1)}-\mu^{(2)}\right)\frac{A^3}{B^3}\left[2\left(x^3\frac{B^3}{A^3}+1\right)^{2/3}-\frac{1}{\left(x^3\frac{B^3}{A^3}+1\right)^{1/3}}-1\right].
\end{split}
\end{equation}
This equation has one solution at $x_{1}=0$, while the  second solution, $x_{2}>0$, is a root of
\begin{equation}\label{eq:sphere:T:comp:nonzero}
\begin{split}
T&=\frac{3\mu^{(2)}}{2}\left[\frac{\left(x^3+1\right)^{1/3}+1}{\left(x^3+1\right)^{2/3}+\left(x^3+1\right)^{1/3}+1}+\frac{1}{\left(x^3+1\right)^{1/3}}\right]\\
&+\frac{3\left(\mu^{(1)}-\mu^{(2)}\right)}{2}\left[\frac{\left(x^3\frac{B^3}{A^3}+1\right)^{1/3}+1}{\left(x^3\frac{B^3}{A^3}+1\right)^{2/3}+\left(x^3\frac{B^3}{A^3}+1\right)^{1/3}+1}+\frac{1}{\left(x^3\frac{B^3}{A^3}+1\right)^{1/3}}\right].
\end{split}
\end{equation}
After expanding $T$, given by \eqref{eq:sphere:T:comp:nonzero}, to the first order in $x^3$, we have
\begin{equation}\label{eq:sphere:T:comp:app}
T\approx\frac{5\mu^{(1)}}{2}+x^3\left[\frac{B^3}{A^3}\left(\mu^{(2)}-\mu^{(1)}\right)-\mu^{(2)}\right].
\end{equation}
The associated critical tension for cavity initiation is
\begin{equation}\label{eq:sphere:T0:comp}
T_{0}=\lim_{x\to0_{+}}T=\frac{5\mu^{(1)}}{2},
\end{equation}
and a comparison with that given by \eqref{eq:sphere:T0:NH:imp} shows that it is the same as for the homogeneous sphere made of the same material as the inner sphere.

%%%%%%%%%%%%%%
\begin{figure}[htbp]
	\begin{center}
		\includegraphics[width=0.45\textwidth]{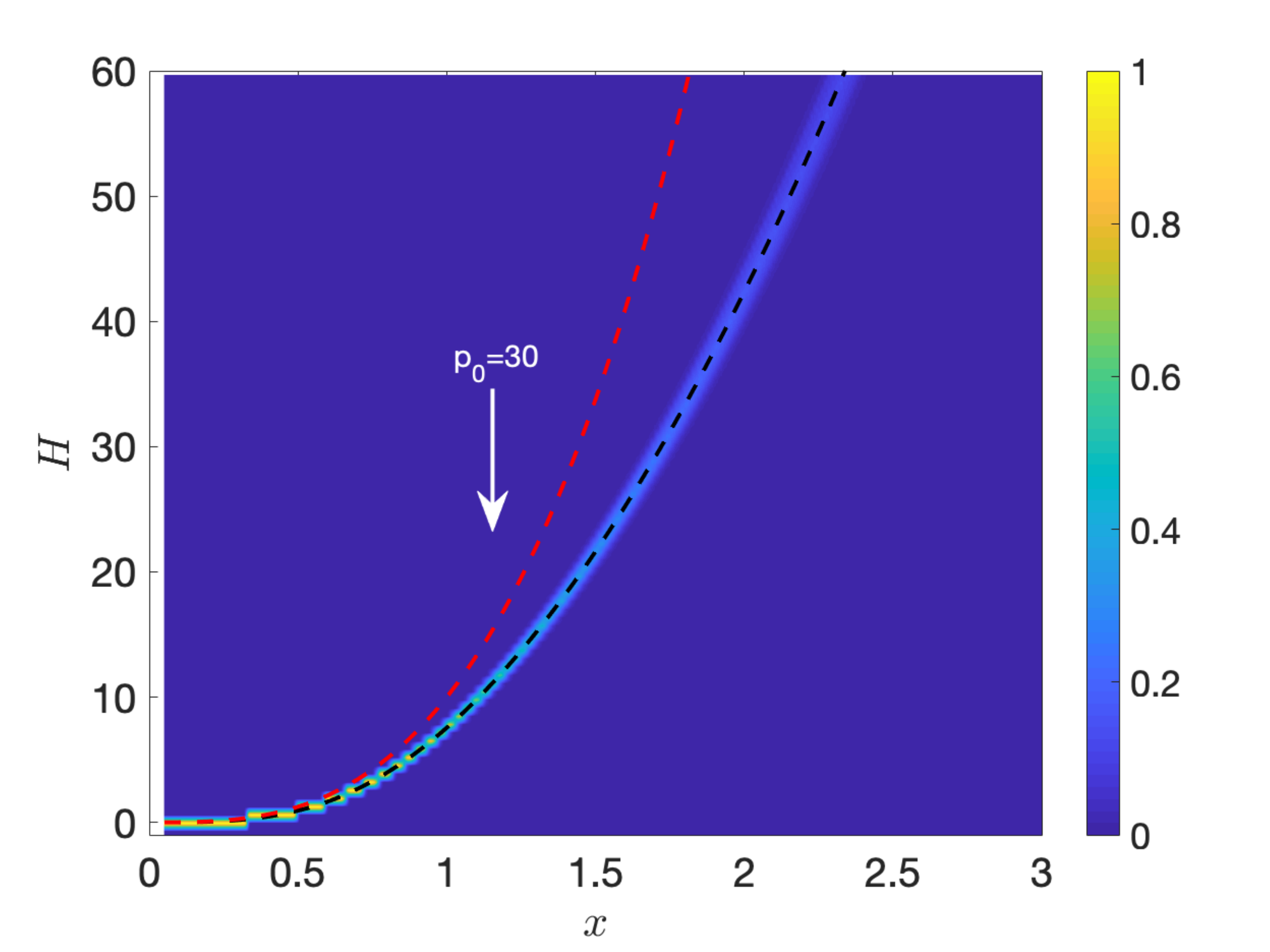}\qquad
		\includegraphics[width=0.45\textwidth]{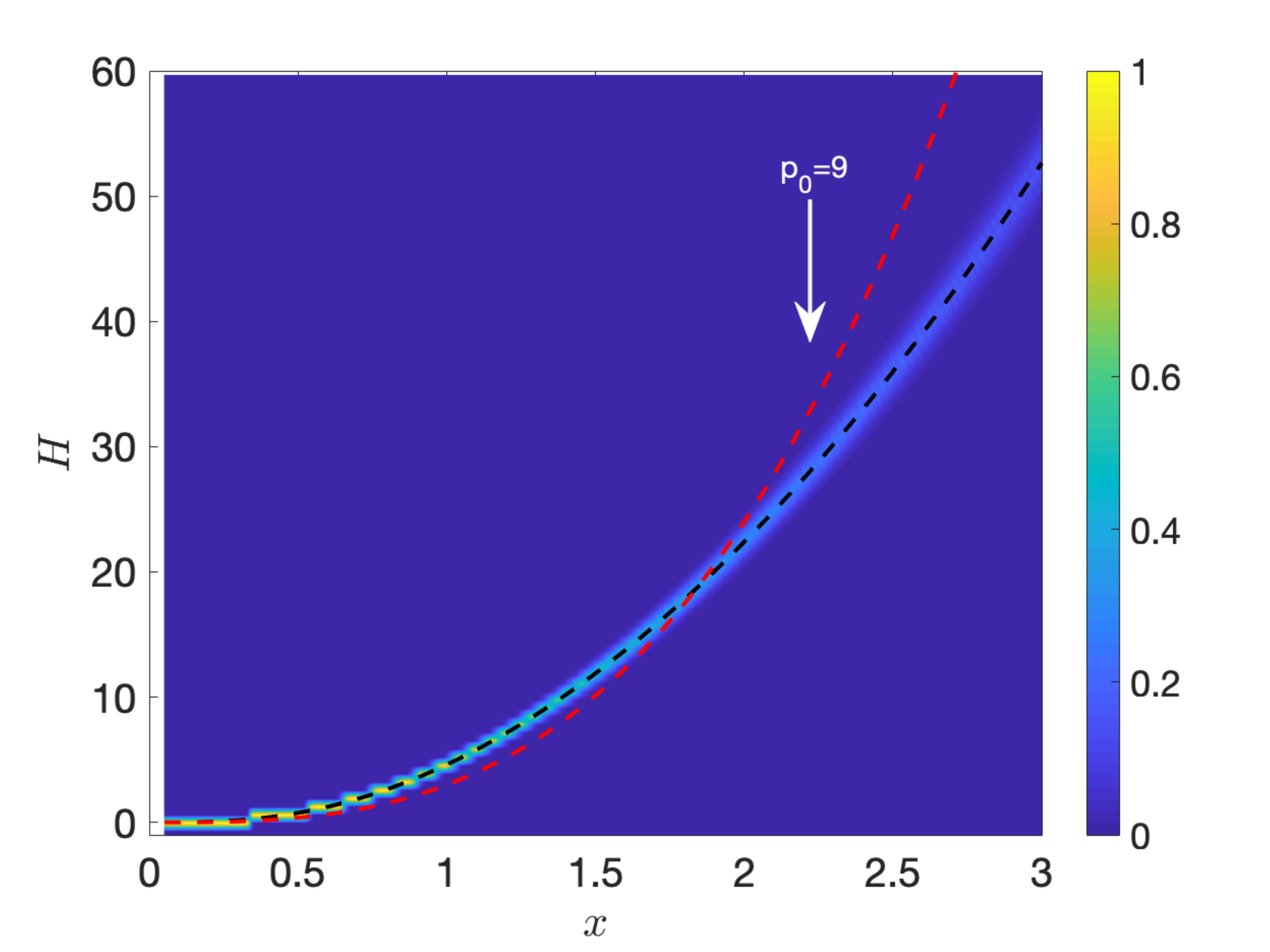}
		\caption{The function $H(x)$, defined by \eqref{eq:sphere:H:comp}, and the (dashed red) curve $\frac{p_{0}x^3}{3}$, with different values of $p_{0}$, for a dynamic sphere made of two concentric stochastic neo-Hookean phases, with radii $A=1/2$ and $B=1$, respectively, under impulse traction, when $\rho=1$ and the shear modulus of the inner sphere, $\mu^{(1)}$, follows a Gamma distribution with $\rho^{(1)}_{1}=405$, $\rho^{(1)}_{2}=0.01$, while for the outer sphere, the shear modulus, $\mu^{(2)}$, follows a Gamma distribution with $\rho^{(2)}_{1}=405$, $\rho^{(1)}_{2}=0.02$ (left) or $\rho^{(2)}_{1}=405$, $\rho^{(1)}_{2}=0.005$ (right). The dashed black lines correspond to the expected values based only on mean value parameters. }\label{fig:stoch-compsphere-no}
	\end{center}
\end{figure}
%%%%%%%%%%%%

We now distinguish the following two cases:
\begin{itemize}
	\item[(i)] If
	\begin{equation}\label{eq:sphere:comp:supercrit:imp}
	\frac{\mu^{(2)}}{\mu^{(1)}}>\left(1-\frac{A^3}{B^3}\right)^{-1},
	\end{equation}
	then the right-hand side of \eqref{eq:sphere:T:comp:app} is an increasing function of $x$, hence
	\begin{equation}\label{eq:sphere:T:comp:bound}
	T>\lim_{x\to 0_{+}}\frac{5\mu^{(1)}}{2}+x^3\left[\frac{B^3}{A^3}\left(\mu^{(2)}-\mu^{(1)}\right)-\mu^{(2)}\right]=\frac{5\mu^{(1)}}{2}.
	\end{equation}
	By \eqref{eq:sphere:T:comp:bound}, for the motion of the composite sphere to be oscillatory, the shear modulus of the inner sphere must satisfy
	\begin{equation}\label{eq:sphere:shearmod:comp:bounds:imp}
	\eta<\mu^{(1)}<\frac{2T}{5},
	\end{equation}
	where the lower bound is assumed to be strictly positive, i.e., $\eta>0$. Equivalently, by \eqref{eq:sphere:impulse}, in terms of the constant $p_{0}$, 
	\begin{equation}\label{eq:sphere:shearmod:comp:bounds:p0}
	\eta<\mu^{(1)}<\frac{p_{0}}{5}\rho B^2,
	\end{equation}
	However, in this case,
	\begin{equation}\label{eq:sphere:comp:osc-no1}
	\frac{p_{0}}{3}x^3-H(x)>0,
	\end{equation}
	for all $x>0$, and the sphere does not oscillate (see Figure~\ref{fig:stoch-compsphere-no}-left).
	
	\item[(ii)] If
	\begin{equation}\label{eq:sphere:comp:subcrit:imp}
	\frac{\mu^{(2)}}{\mu^{(1)}}<\left(1-\frac{A^3}{B^3}\right)^{-1},
	\end{equation}
	then the right-hand side of \eqref{eq:sphere:T:comp:app} decreases as $x$ increases, hence
	\begin{equation}\label{eq:sphere:T:comp:bounds}
	0<T<\frac{5\mu^{(1)}}{2},
	\end{equation}
	or equivalently, by \eqref{eq:sphere:impulse},
	\begin{equation}\label{eq:sphere:T:comp:bounds:p0}
	0<p_{0}<\frac{5\mu^{(1)}}{\rho B^2}.
	\end{equation}
	However, in this case,
	\begin{equation}\label{eq:sphere:comp:osc-no2}
	\frac{p_{0}}{3}x^3-H(x)\left\{
	\begin{array}{ll}
	<0 & \mbox{if}\ x_{1}< x< x_{2},\\
	>0 & \mbox{if}\ x>x_{2},
	\end{array}
	\right.
	\end{equation}
	and the sphere cannot oscillate (see Figure~\ref{fig:stoch-compsphere-no}-right).
\end{itemize}

Note that, since the material parameters, $\mu^{(1)}$ and $\mu^{(2)}$, are positive, both \eqref{eq:sphere:comp:supercrit:imp}  and  \eqref{eq:sphere:comp:subcrit:imp} are possible for any values $0<A<B$. 

%%%%%%%%%%%%%%%%%%%%%%%%%%%%%%%%%%%%%%%%%%%%%%%%%%%%%%%%%%%%
\subsection{Static deformation of a sphere with two stochastic neo-Hookean phases under impulse traction}\label{sec:sphere:comp:static:imp}

For the static composite sphere subject to a uniform constant surface load in the current configuration, given in terms of the Cauchy stresses, the tensile traction takes the form
\begin{equation}\label{eq:sphere:Tst:comp}
\begin{split}
T&=2\mu^{(2)}\left[\frac{1}{\left(x^{3}+1\right)^{1/3}}+\frac{1}{4\left(x^{3}+1\right)^{4/3}}\right]\\
&+2\left(\mu^{(1)}-\mu^{(2)}\right)\left[\frac{1}{\left(x^3\frac{B^3}{A^3}+1\right)^{1/3}}+\frac{1}{4\left(x^3\frac{B^3}{A^3}+1\right)^{4/3}}\right].
\end{split}
\end{equation}
The critical traction for the onset of cavitation is thus also given by \eqref{eq:sphere:T0:comp}.

%%%%%%%%%%%%%%
\begin{figure}[htbp]
	\begin{center}
		\includegraphics[width=\textwidth]{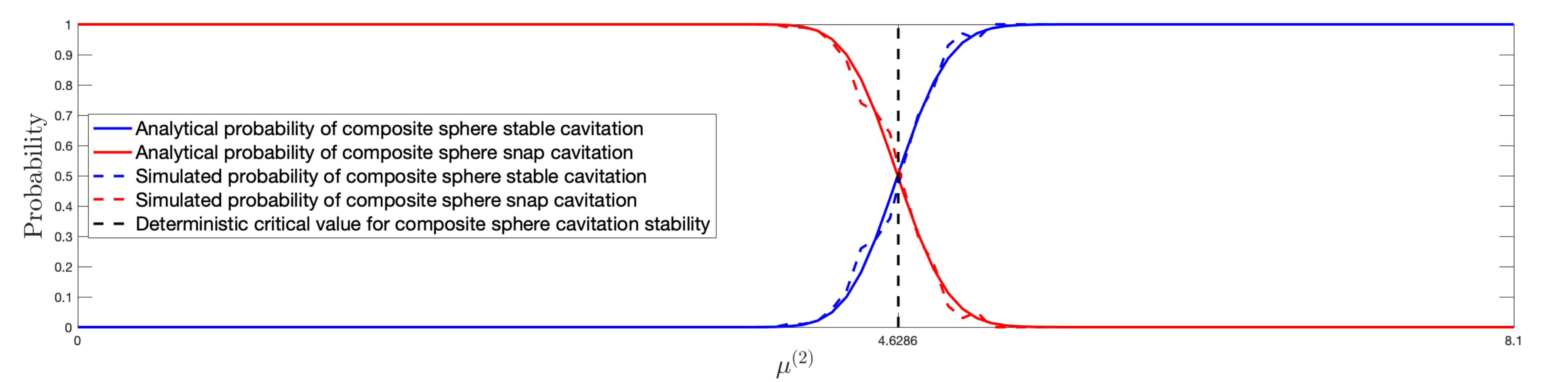}
		\caption{Probability distributions of stable or unstable cavitation in a static sphere made of two concentric stochastic neo-Hookean phases, with radii $A=1/2$ and $B=1$, respectively, under impulse traction, when the shear modulus of the inner sphere, $\mu^{(1)}$, follows a Gamma distribution with $\rho^{(1)}_{1}=405$, $\rho^{(1)}_{2}=0.01$. Continuous coloured lines represent analytically derived solutions, given by equations \eqref{eq:sphere:comp:P1:imp}-\eqref{eq:sphere:comp:P2:imp}, and the dashed versions represent stochastically generated data. The vertical line at the critical value, $8\underline{\mu}^{(1)}/7=4.6286$, separates the expected regions based only on mean value, $\underline{\mu}=\rho^{(1)}_{1}\rho^{(1)}_{2}=4.05$. The probabilities were calculated from the average of 100 stochastic simulations.}\label{fig:comp-sphere-pdfs-no}
	\end{center}
\end{figure}
%%%%%%%%%%%%%%%

%%%%%%%%%%%%%%
\begin{figure}[htbp]
	\begin{center}
		\includegraphics[width=0.45\textwidth]{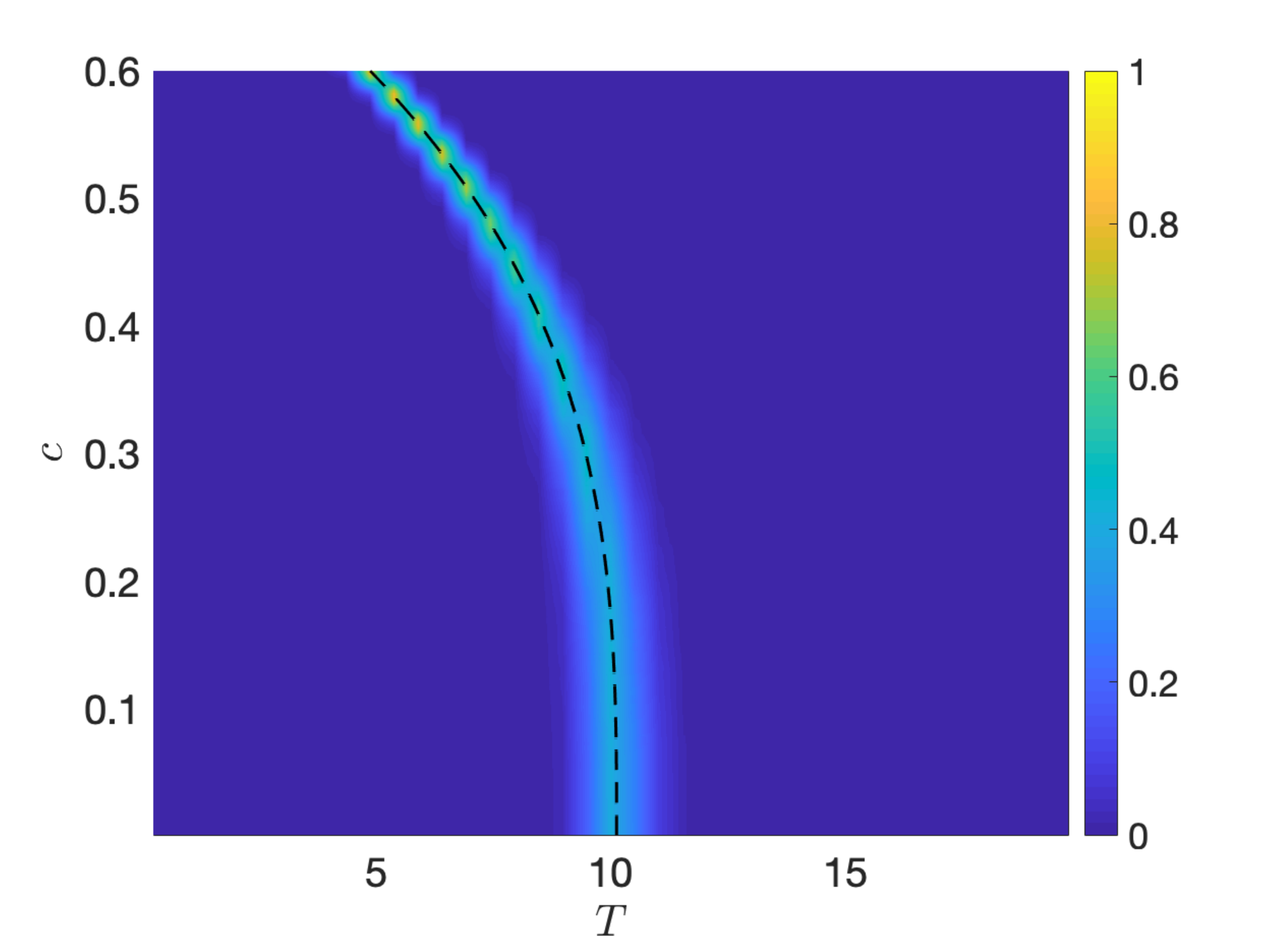}\qquad
		\includegraphics[width=0.45\textwidth]{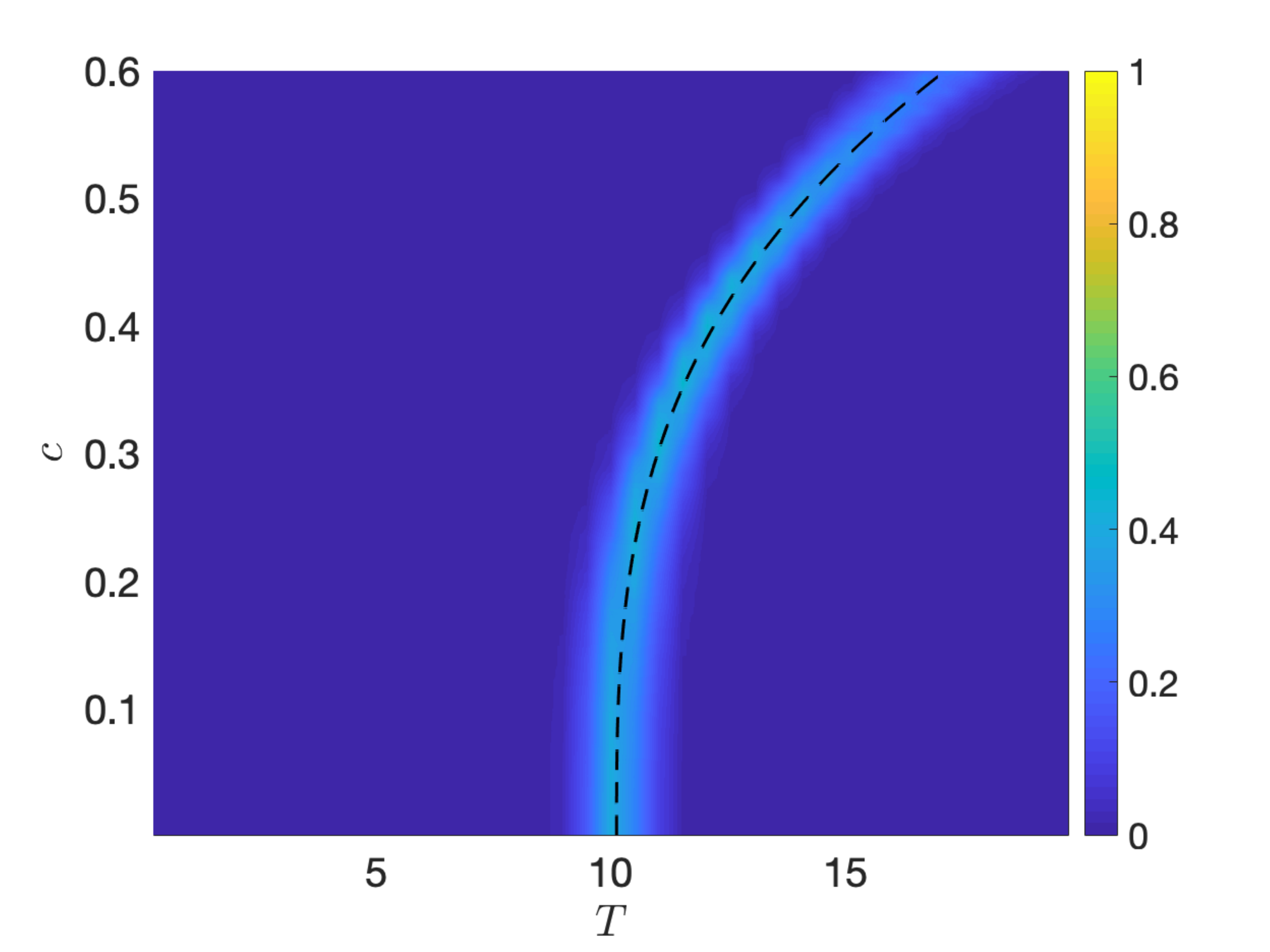}
		\caption{Probability distribution of the applied impulse traction $T$ causing cavitation of radius $c$ in a static sphere of two concentric stochastic neo-Hookean phases, with radii $A=1/2$ and $B=1$, respectively, when the shear modulus of the inner phase, $\mu^{(1)}$, follows a Gamma distribution with $\rho^{(1)}_{1}=405$, $\rho^{(1)}_{2}=4.05/\rho^{(1)}_{1}=0.01$, while the shear modulus of the outer phase, $\mu^{(2)}$, is drawn Gamma distribution with $\rho^{(2)}_{1}=405$, $\rho^{(2)}_{2}=0.005$ (left) or $\rho^{(2)}_{1}=405$, $\rho^{(2)}_{2}=0.02$ (right). The dashed black lines correspond to the expected bifurcation based only on mean parameter values.}\label{fig:Tcompstoch}
	\end{center}
\end{figure}
%%%%%%%%%%%%%

To investigate the stability of this cavitation, we examine the behaviour of the cavity opening, with scaled radius $x$ as a function of $T$, in a neighbourhood of $T_{0}$. Expanding the expression of $T$, given by \eqref{eq:sphere:Tst:comp}, to the first order in $x^3$, we have
\begin{equation}\label{eq:sphere:Tst:comp:app}
T\approx\frac{5\mu^{(1)}}{2}+\frac{4x^3}{3}\left[\frac{B^3}{A^3}\left(\mu^{(2)}-\mu^{(1)}\right)-\mu^{(2)}\right],
\end{equation}
and
\begin{equation}\label{eq:sphere:dTst:comp:app}
\frac{\text{d}T}{\text{d}x}\approx 4x^2\left[\frac{B^3}{A^3}\left(\mu^{(2)}-\mu^{(1)}\right)-\mu^{(2)}\right].
\end{equation}
Then, $\text{d}T/\text{d}x\to 0$ as $x\to 0_{+}$, and the bifurcation at the critical impulse load, $T_{0}$, is supercritical (respectively, subcritical) if $\text{d}T/\text{d}x>0$ (respectively, $\text{d}T/\text{d}x<0$) for arbitrarily small $x$. From \eqref{eq:sphere:dTst:comp:app}, the following two cases arise:
\begin{itemize}
	\item[(i)] If \eqref{eq:sphere:comp:supercrit:imp} holds, then the solution after bifurcation is supercritical, and the cavitation is stable, in the sense that a new bifurcated solution exists locally for values of $T>T_{0}$, and the cavity radius monotonically increases with the applied load post-bifurcation. 
	
	\item[(ii)] If \eqref{eq:sphere:comp:subcrit:imp} holds, then the bifurcated solution is subcritical, with the cavitation being unstable, in the sense that the required dead load starts to decrease post-bifurcation, causing a snap cavitation, i.e., a sudden jump in the cavity opening. In particular, when $\mu^{(1)}=\mu^{(2)}$, the problem reduces to the case of a homogeneous sphere of neo-Hookean material, for which unstable cavitation always occurs (see Section~\ref{sec:sphere:NH}).
\end{itemize}
As either of the two inequalities \eqref{eq:sphere:comp:supercrit:imp} and \eqref{eq:sphere:comp:subcrit:imp} is possible, the probability distribution of stable cavitation in the composite sphere is equal to (see Figure~\ref{fig:comp-sphere-pdfs-no})
\begin{equation}\label{eq:sphere:comp:P1:imp}
P_{1}(\mu^{(2)})=\int_{0}^{\mu^{(2)}\left(1-\frac{A^3}{B^3}\right)}g(u;\rho^{(1)}_{1},\rho^{(1)}_{2})\ du,
\end{equation}
while that of unstable (snap) cavitation is
\begin{equation}\label{eq:sphere:comp:P2:imp}
P_{2}(\mu^{(2)})=1-P_{1}(\mu^{(2)})=1-\int_{0}^{\mu^{(2)}\left(1-\frac{A^3}{B^3}\right)}g(u;\rho^{(1)}_{1},\rho^{(1)}_{2})\ du.
\end{equation}

The post-cavitation stochastic behaviours shown in Figure~\ref{fig:Tcompstoch} correspond to two different static composite sphere where the shear modulus of the inner phase, $\mu^{(1)}$, follows a Gamma distribution with $\rho^{(1)}_{1}=405$, $\rho^{(1)}_{2}=4.05/\rho^{(1)}_{1}=0.01$, while the shear modulus of the outer phase, $\mu^{(2)}$, is drawn from a Gamma distribution with $\rho^{(2)}_{1}=405$, $\rho^{(2)}_{2}=0.005$  (in Figure~\ref{fig:Tcompstoch}-left) or $\rho^{(2)}_{1}=405$, $\rho^{(2)}_{2}=0.02$ (in Figure~\ref{fig:Tcompstoch}-right). Note that, in each case, unstable or stable cavitation is expected, respectively, but there is other values of the parameters lead to the opposite behaviour.

%%%%%%%%%%%%%%%%%%%%%%%%%%%%%%%%%%%%%%%%%%%%%%%%%%%%%%%%%%%%
%%%%%%%%%%%%%%%%%%%%   NEW SECTION  %%%%%%%%%%%%%%%%%%%%%%%%
%%%%%%%%%%%%%%%%%%%%%%%%%%%%%%%%%%%%%%%%%%%%%%%%%%%%%%%%%%%%
\section{Cavitation and radial motion of inhomogeneous spheres}\label{sec:sphere:inhom}

We further examine the cavitation of radially inhomogeneous, incompressible spheres of stochastic hyperelastic material. Cavitation of radially inhomogeneous hyperelastic spheres of compressible or incompressible material in static equilibrium was investigated in \cite{Sivaloganathan:1991}. Here, we adopt a neo-Hookean-like model, where the constitutive parameter varies continuously along the radial direction. Our inhomogeneous model is similar to those proposed in \cite{Ertepinar:1976:EA}, where the dynamic inflation of spherical shells was treated explicitly.

In particular, we define the class of stochastic inhomogeneous hyperelastic models \eqref{eq:W:stoch}, with the shear modulus taking the form (see also eq. (29) of \cite{Ertepinar:1976:EA})
\begin{equation}\label{eq:mur}
\mu(R)=C_{1}+C_{2}\frac{R^3}{B^3},
\end{equation}
where $\mu(R)>0$, for all $0\leq R\leq B$. For simplicity, we further assume that, for any fixed $R$, the mean value $\underline{\mu}$ of $\mu=\mu(R)$ is independent of $R$, (namely the expected value of $C_2$ is 0), whereas its variance, $\text{Var}[\mu]$ changes with $R$. Then, $C_{1}=\mu(0)>0$ is a single-valued (deterministic) constant and $C_{2}$ is a random value, defined by a given probability distribution. 

Alternative modelling formulations where both the mean value and variance of the shear modulus may vary with $R$ are discussed at the end of this section.

%%%%%%%%%%%%%%
\begin{figure}[htbp]
	\begin{center}
		\includegraphics[width=0.5\textwidth]{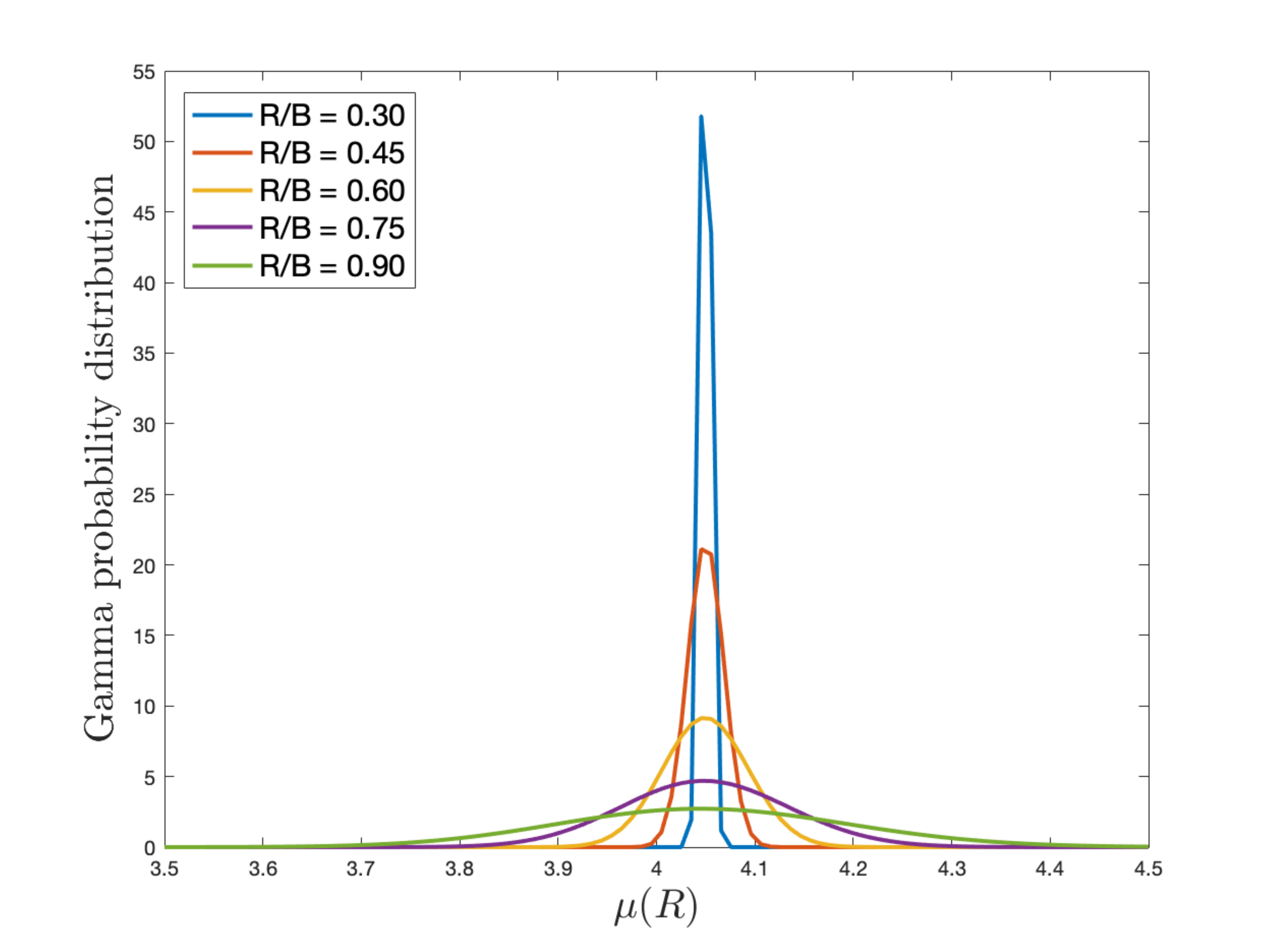}
		\caption{Examples of Gamma distribution, with hyperparameters $\rho_1=405\cdot B^6/R^6$ and $\rho_2=0.01\cdot R^6/B^6$, for the shear modulus $\mu(R)$, given by \eqref{eq:mur}. In this case, $C_{1}=\underline{\mu}=\rho_{1}\rho_{2}=4.05$ and $C_{2}=\mu(B)-C_{1}$.}\label{fig:muR-gpdf}
	\end{center}
\end{figure}
%%%%%%%%%%%%%

When the mean value of the shear modulus $\mu(R)$, described by \eqref{eq:mur}, does not depend on $R$, it follows that, for any fixed $R$,
\begin{equation}\label{eq:mur:C12}
\underline{\mu}=C_{1},\qquad \text{Var}[\mu]=\text{Var}[C_{2}]\frac{R^6}{B^6},
\end{equation}
where $\text{Var}[C_{2}]$ is the variance of $C_{2}$. Note that, at $R=0$, the value $\mu(0)=C_{1}$ represents the material property at the centre of the sphere, which is a unique  point, whereas at any fixed $R>0$, the standard deviation $\|\mu\|=\sqrt{\text{Var}[\mu]}$ is proportional to the volume $4\pi R^3/3$ of the sphere with radius $R$.

By \eqref{eq:rho12} and \eqref{eq:mur:C12}, the hyperparameters of the corresponding Gamma distribution, defined by \eqref{eq:mu:gamma}, take the form
\begin{equation}\label{eq:rho12:C12}
\rho_{1}=\frac{C_{1}}{\rho_{2}},\qquad \rho_{2}=\frac{\text{Var}[\mu]}{C_{1}}=\frac{\text{Var}[C_{2}]}{C_{1}}\frac{R^6}{B^6}.
\end{equation}
For example, one can choose two constant values, $C_{0}>0$ and $C_{1}>0$, and set the hyperparameters for the Gamma distribution at any given $R$ as follows,
\begin{equation}\label{eq:rho12:C01}
\rho_{1}=\frac{C_{1}}{C_{0}}\frac{B^6}{R^6},\qquad \rho_{2}=C_{0}\frac{R^6}{B^6}.
\end{equation}
By \eqref{eq:mur}, $C_{2}=\left(\mu(R)-C_{1}\right)B^3/R^3$ is the shifted Gamma-distributed random variable with mean value $\underline{C}_{2}=0$ and variance $\text{Var}[C_{2}]=\rho_{1}\rho_{2}^2B^6/R^6=C_{0}C_{1}$. When $R$ decreases towards zero, by \eqref{eq:rho12:C01}, $\rho_{1}$ increases, while $\rho_{2}$ decreases, hence, the Gamma distribution converges to a normal distribution \cite{Fitt:2019:FWWM,Mihai:2019a:MDWG}, which degenerates at the centre of the sphere, where $R=0$, as the shear modulus there takes the deterministic value $C_{1}$. In Figure~\ref{fig:muR-gpdf}, we show these distributions when $\rho_1=405\cdot B^6/R^6$ and $\rho_2=0.01\cdot R^6/B^6$. For this case, by \eqref{eq:mur}, $C_{1}=\underline{\mu}=\rho_{1}\rho_{2}=4.05$ and $C_{2}=\mu(B)-C_{1}$.

%%%%%%%%%%%%%%%%%%%%%%%%%%%%%%%%%%%%%%%%%%%%%%%%%%%%%%%%%%%%
\subsection{Oscillatory motion of a stochastic radially inhomogeneous sphere under dead-load traction}\label{sec:sphere:inhom:osc}

The shear modulus of the form given by \eqref{eq:mur} can be expressed equivalently as follows,
\begin{equation}\label{eq:sphere:muu}
\mu=C_{1}+C_{2}\frac{x^3}{u-1},
\end{equation}
where $u=r^3/R^3$ and $x=c/B$, as denoted in \eqref{eq:sphere:ux}. 

Also, writing the invariants given by \eqref{eq:sphere:I123} in the equivalent forms
\begin{equation}\label{eq:sphere:I123:inhom}
I_{1}=u^{-4/3}+2u^{2/3},\ I_{2}=u^{4/3}+2u^{-2/3},\ 
I_{3}=1,
\end{equation}
and substituting these in \eqref{eq:sphere:betas} gives
\begin{equation}\label{eq:sphereW12:inhom}
\begin{split}
&\beta_{1}=2\frac{\partial W}{\partial I_{1}}=\mu(u)+\frac{d\mu}{du}\frac{\text{d}u}{\text{d}I_{1}}\left(I_{1}-3\right),\\
&\beta_{-1}=-2\frac{\partial W}{\partial I_{2}}=-\frac{\text{d}\mu}{\text{d}u}\frac{\text{d}u}{\text{d}I_{2}}\left(I_{1}-3\right).
\end{split}
\end{equation}
Hence,
\begin{equation}\label{eq:sphere:betas:inhom}
\begin{split}
\beta_{1}&=C_{1}+\frac{3C_{2}}{4}\frac{x^3}{u-1}\left[\frac{4}{3}-\frac{2u^3-3u^{7/3}+u}{\left(u-1\right)^2\left(u+1\right)}\right],\\
\beta_{-1}&=\frac{3C_{2}}{4}\frac{x^3}{u-1}\frac{2u^{7/3}-3u^{5/3}+u^{1/3}}{\left(u-1\right)^2\left(u+1\right)},
\end{split}
\end{equation}
and
\begin{equation}\label{eq:mu:inhom}
\beta_{1}-\beta_{-1}\frac{r^2}{R^2}=C_{1}+\frac{3C_{2}}{2}\frac{x^3}{u-1}\left[\frac{2}{3}-\frac{2u^3-3u^{7/3}+u}{\left(u-1\right)^2\left(u+1\right)}\right].
\end{equation}

After calculations analogous to those in Section~\ref{sec:sphere:NH}, but with $\beta_{1}-\beta_{-1}\frac{r^2}{R^2}$ instead of $\mu$ in \eqref{eq:sphere:T1T2:r}, \eqref{eq:sphere:T1T2:x}, and \eqref{eq:sphere:T:x}, we denote
\begin{equation}\label{eq:sphere:H:inhom:betas}
H(x)=\frac{4}{3\rho B^2}\int_{0}^{x}\left[\zeta^2\int_{\zeta^3+1}^{\infty}\left(\beta_{1}-\beta_{-1}\frac{r^2}{R^2}\right)\frac{1+u}{u^{7/3}}du\right]d\zeta.
\end{equation}
Equivalently, by \eqref{eq:mu:inhom},
\begin{equation}\label{eq:sphere:H:inhom}
\begin{split}
H(x)&=\frac{4C_{1}}{3\rho B^2}\int_{0}^{x}\left(\zeta^2\int_{\zeta^3+1}^{\infty}\frac{1+u}{u^{7/3}}du\right)d\zeta\\
&+\frac{2C_{2}}{\rho B^2}\int_{0}^{x}\left\{\zeta^5\int_{\zeta^3+1}^{\infty}\frac{1+u}{u^{7/3}\left(u-1\right)}\left[\frac{2}{3}-\frac{2u^3-3u^{7/3}+u}{\left(u-1\right)^2\left(u+1\right)}\right]du\right\}d\zeta\\
&=\frac{C_{1}}{\rho B^2}\left[2\left(x^3+1\right)^{2/3}-\frac{1}{\left(x^3+1\right)^{1/3}}-1\right]\\
&-\frac{C_{2}}{\rho B^2}\left[\sqrt{3}\arctan\frac{2\left(x^3+1\right)^{1/3}+1}{\sqrt{3}}-\frac{\pi}{\sqrt{3}}\right]\\
&+\frac{3C_{2}}{2\rho B^2}\left\{\ln\left[\left(x^3+1\right)^{2/3}+\left(x^3+1\right)^{1/3}+1\right]-\ln 3\right\}\\
&-\frac{C_{2}}{\rho B^2}\left[\left(x^3+1\right)^{2/3}+\frac{1}{\left(x^3+1\right)^{1/3}}-2\right].
\end{split}
\end{equation}

Then, by setting the dead load \eqref{eq:sphere:dload} and substituting \eqref{eq:sphere:H:inhom} in \eqref{eq:sphere:HC}, we obtain
\begin{equation}\label{eq:sphere:p0:inhom}
\begin{split}
p_{0}\left[\left(x^3+1\right)^{1/3}-1\right]&=\frac{C_{1}}{2}\left[2\left(x^3+1\right)^{2/3}-\frac{1}{\left(x^3+1\right)^{1/3}}-1\right]\\
&-\frac{C_{2}}{2}\left[\sqrt{3}\arctan\frac{2\left(x^3+1\right)^{1/3}+1}{\sqrt{3}}-\frac{\pi}{\sqrt{3}}\right]\\
&+\frac{3C_{2}}{4}\left\{\ln\left[\left(x^3+1\right)^{2/3}+\left(x^3+1\right)^{1/3}+1\right]-\ln 3\right\}\\
&-\frac{C_{2}}{2}\left[\left(x^3+1\right)^{2/3}+\frac{1}{\left(x^3+1\right)^{1/3}}-2\right].
\end{split}
\end{equation}
Equation \eqref{eq:sphere:p0:inhom} has a solution at $x_{1}=0$, while the  second solution, $x_{2}>0$, is a root of 
\begin{equation}\label{eq:sphere:p0:inhom:nonzero}
\begin{split}
p_{0}&=\frac{C_{1}}{2}\left[2\left(x^3+1\right)^{1/3}+\frac{1}{\left(x^3+1\right)^{1/3}}+2\right]\\
&-\frac{C_{2}}{2}\left[\left(x^3+1\right)^{1/3}-\frac{1}{\left(x^3+1\right)^{1/3}}\right].
\end{split}
\end{equation}
To obtain the right-hand side of the above equation, we have assumed that, in \eqref{eq:sphere:p0:inhom}, $x$ is sufficiently small, such that, after expanding the respective functions to the second order in $(x^3+1)^{1/3}$,
\begin{equation}
\left[\sqrt{3}\arctan\frac{2\left(x^3+1\right)^{1/3}+1}{\sqrt{3}}-\frac{\pi}{\sqrt{3}}\right]\left[\left(x^3+1\right)^{1/3}-1\right]^{-1}\approx\frac{3-\left(x^3+1\right)^{1/3}}{4}
\end{equation}
and
\begin{equation}
\frac{\ln\left[\left(x^3+1\right)^{2/3}+\left(x^3+1\right)^{1/3}+1\right]-\ln 3}{\left(x^3+1\right)^{1/3}-1}\approx\frac{7-\left(x^3+1\right)^{1/3}}{6}.
\end{equation}
Next, expanding the right-hand side of the equation \eqref{eq:sphere:p0:inhom:nonzero} to the first order in $x^3$ gives
\begin{equation}\label{eq:sphere:p0:inhom:nonzero:app}
p_{0}\approx\frac{5C_{1}}{2}+\frac{x^3}{6}\left(C_{1}-2C_{2}\right).
\end{equation}
The critical dead load for the onset of cavitation is then
\begin{equation}\label{eq:sphere:p0:inhom:crit}
\lim_{x\to0_{+}}p_{0}=\frac{5C_{1}}{2},
\end{equation}
and is the same as for the homogeneous sphere made entirely from the material found at the centre of the inhomogeneous sphere.

%%%%%%%%%%%%%%
\begin{figure}[htbp]
	\begin{center}
		\includegraphics[width=0.47\textwidth]{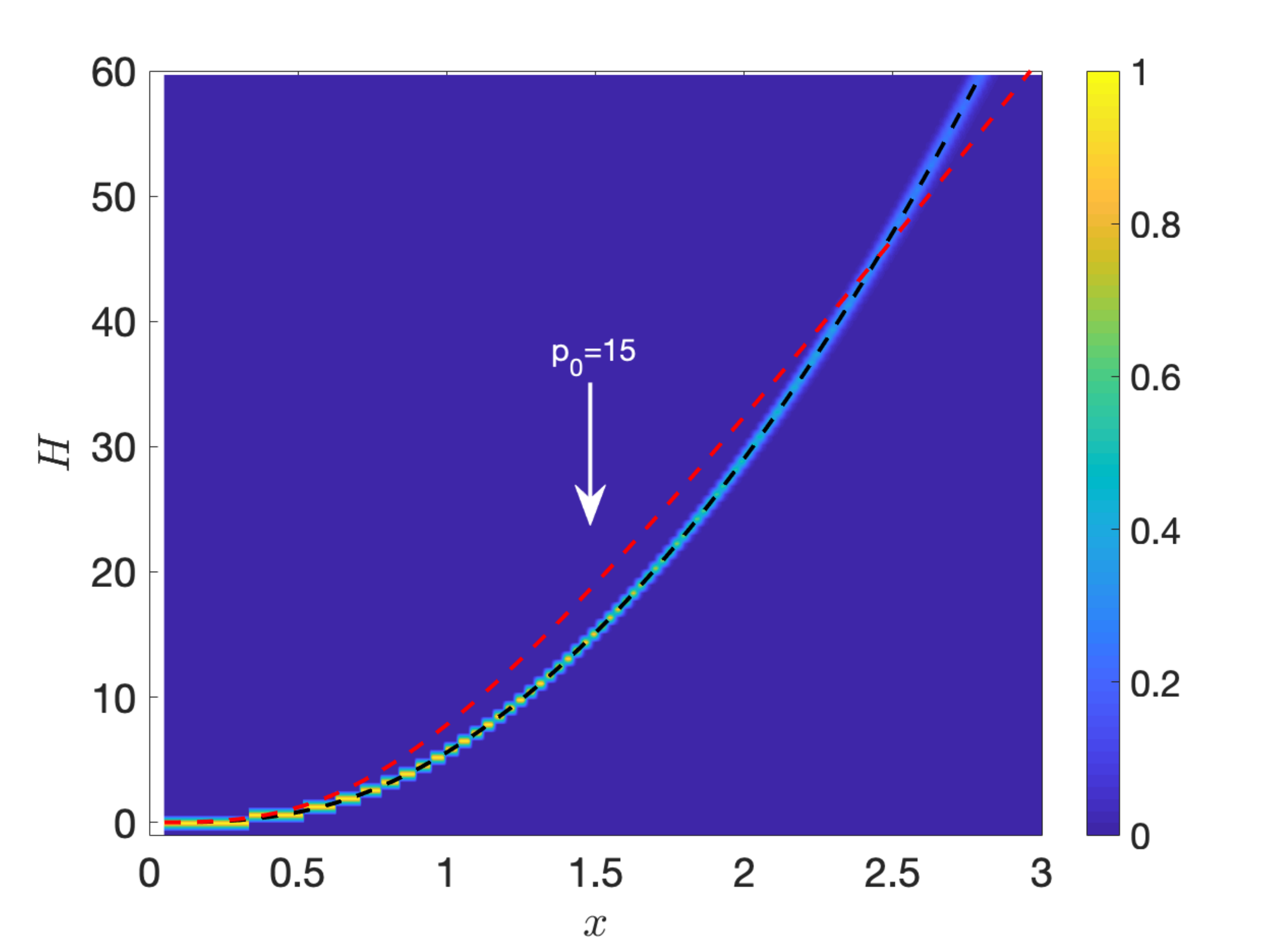}\qquad
		\includegraphics[width=0.47\textwidth]{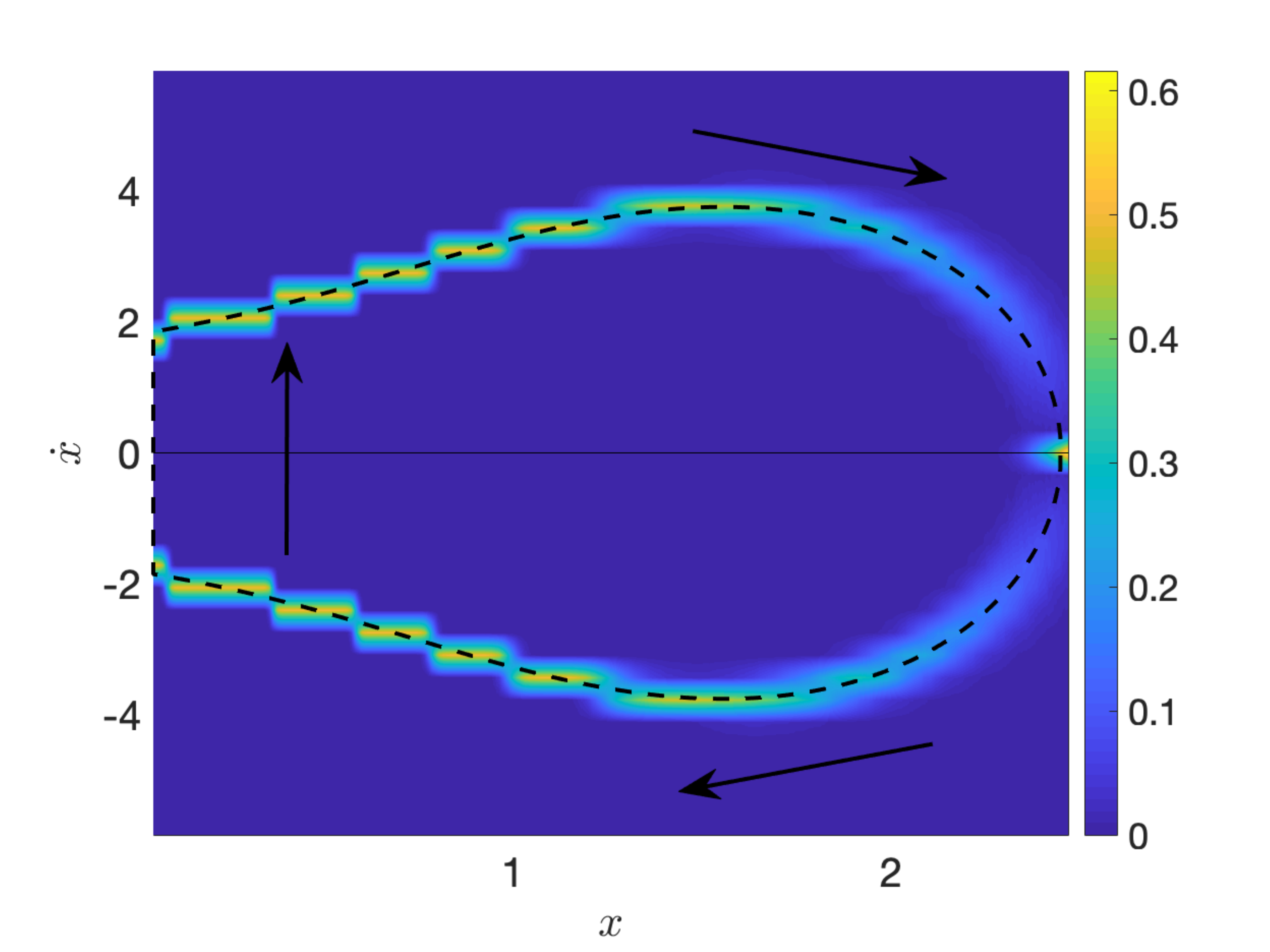}
		\caption{The function $H(x)$, defined by \eqref{eq:sphere:H:inhom}, intersecting the (dashed red) curve $\frac{2p_{0}}{\rho B^2}\left[\left(x^3+1\right)^{1/3}-1\right]$, with $p_{0}=15$ (left), and the associated velocity, given by \eqref{eq:sphere:dotx:forced} (right), for a dynamic sphere of stochastic radially inhomogeneous neo-Hookean-like material under dead-load traction, when $\rho=1$, $B=1$, and, for any fixed $R$, $\mu(R)$, given by \eqref{eq:mur}, follows a Gamma distribution with  $\rho_1=405/R^6$ and $\rho_2=0.01\cdot R^6$. The dashed black lines correspond to the expected values based only on mean parameter values. Each distribution was calculated from the average of $1000$ stochastic simulations.}\label{fig:stoch-inhomsphere}
	\end{center}
\end{figure}
%%%%%%%%%%%%

Considering the parameters $C_{1}$ and $C_{2}$, we have the following two cases:
\begin{itemize}
\item[(i)] When $C_{1}>2C_{2}$, the right-hand side in \eqref{eq:sphere:p0:inhom:nonzero:app} is an increasing function of $x$, and therefore,
\begin{equation}\label{eq:sphere:p0:inhom:bound}
p_{0}>\frac{5C_{1}}{2}.
\end{equation}

\item[(ii)] When $C_{1}<2C_{2}$, the right-hand side in \eqref{eq:sphere:p0:inhom:nonzero:app} decreases as $x$ increases, hence,
\begin{equation}\label{eq:sphere:p0:inhom:bounds}
0<p_{0}<\frac{5C_{1}}{2}.
\end{equation}
However, since
\begin{equation}\label{eq:sphere:inhom:osc-no}
\frac{2p_{0}}{\rho B^2}\left[\left(x^3+1\right)^{1/3}-1\right]-H(x)\left\{
\begin{array}{ll}
<0 & \mbox{if}\ x_{1}< x< x_{2},\\
>0 & \mbox{if}\ x>x_{2},
\end{array}
\right.
\end{equation}
the sphere cannot oscillate.
\end{itemize}

In Figure~\ref{fig:stoch-inhomsphere}, we illustrate the stochastic function $H(x)$, defined by \eqref{eq:sphere:H:inhom}, intersecting the (dashed red) curve $\frac{2p_{0}}{\rho B^2}\left[\left(x^3+1\right)^{1/3}-1\right]$, with $p_{0}=15$, to find the two distinct solutions to \eqref{eq:sphere:HC}, and the associated velocity, given by \eqref{eq:sphere:dotx:forced}, for a unit sphere (with $B=1$) of stochastic radially inhomogeneous neo-Hookean-like material when the random field shear modulus $\mu(R)$ is given by \eqref{eq:mur}, with $\rho_1=405/R^6$ and $\rho_2=0.01\cdot R^6$ (see Figure~\ref{fig:muR-gpdf}). In particular, $\rho_{1}=405$ and $\rho_{2}=0.01$ at $R=B$, hence, by \eqref{eq:mur}, $C_{1}=\underline{\mu}=\rho_{1}\rho_{2}=4.05$ and $C_{2}=\mu(B)-C_{1}$. For these functions, although the mean values, represented by black dashed lines, are the same as for those depicted in Figure~\ref{fig:stoch-NHsphere}, their stochastic behaviours are quite different compared to the case of a stochastic neo-Hookean sphere. Specifically, as we observe from

%%%%%%%%%%%%%%%%%%%%%%%%%%%%%%%%%%%%%%%%%%%%%%%%%%%%%%%%%%%%
\subsection{Static deformation of a stochastic radially inhomogeneous sphere under dead-load traction}\label{sec:sphere:inhom:static}

For the static inhomogeneous sphere, when the cavity surface is traction-free, at the outer surface, by analogy to \eqref{eq:sphere:T:NH}, and using \eqref{eq:mu:inhom}, we have
\begin{equation}\label{eq:sphere:Tint:inhom}
\begin{split}
T_{rr}(b)&=\frac{2}{3}\int_{x^3+1}^{\infty}\left(\beta_{1}-\beta_{-1}\frac{r^2}{R^2}\right)\frac{1+u}{u^{7/3}}du\\
&=\frac{2C_{1}}{3}\int_{x^3+1}^{\infty}\frac{1+u}{u^{7/3}}du
+C_{2}x^3\int_{x^3+1}^{\infty}\frac{1+u}{u^{7/3}\left(u-1\right)}\left[\frac{2}{3}-\frac{2u^3-3u^{7/3}+u}{\left(u-1\right)^2\left(u+1\right)}\right]du.
\end{split}
\end{equation}
Equivalently, after evaluating the integrals,
\begin{equation}\label{eq:sphere:T:inhom}
\begin{split}
T_{rr}(b)&=2C_{1}\left[\frac{1}{\left(x^3+1\right)^{1/3}}+\frac{1}{4\left(x^3+1\right)^{4/3}}\right]\\
&-\frac{C_{2}}{2}x^3\frac{2\left(x^3+1\right)^{4/3}+4\left(x^3+1\right)+3\left(x^3+1\right)^{2/3}+2\left(x^3+1\right)^{1/3}+1}{\left(x^3+1\right)^{8/3}+2\left(x^3+1\right)^{7/3}+3\left(x^3+1\right)^{2}+2\left(x^3+1\right)^{5/3}+\left(x^3+1\right)^{4/3}}.\\
\end{split}
\end{equation}
After multiplying by $\left(x^3+1\right)^{2/3}$ the above Cauchy stress, and denoting $\lambda_{b}=b/B=\left(x^3+1\right)^{1/3}$, the required tensile dead load at the outer surface, $R=B$, in the reference configuration, takes the form
\begin{equation}\label{eq:sphere:P:inhom}
P=2C_{1}\left(\lambda_{b}+\frac{1}{4\lambda_{b}^{2}}\right)
-\frac{C_{2}}{2}\left(\lambda_{b}^3-1\right)\frac{2\lambda_{b}^4+4\lambda_{b}^3+3\lambda_{b}^2+2\lambda_{b}+1}{\lambda_{b}^6+2\lambda_{b}^5+3\lambda_{b}^4+2\lambda_{b}^3+\lambda_{b}^2}.
\end{equation}
The critical dead-load traction for the initiation of cavitation is
\begin{equation}\label{eq:sphere:P0:inhom}
P_{0}=\lim_{\lambda_{b}\to1}P=\frac{5C_{1}}{2}=\frac{5\mu(0)}{2},
\end{equation}
and is the same as for the homogeneous sphere made entirely from the material found at its centre \cite{Sivaloganathan:1991}, as found also for the dynamic sphere.

\begin{figure}[htbp]
	\begin{center}
		\includegraphics[width=0.49\textwidth]{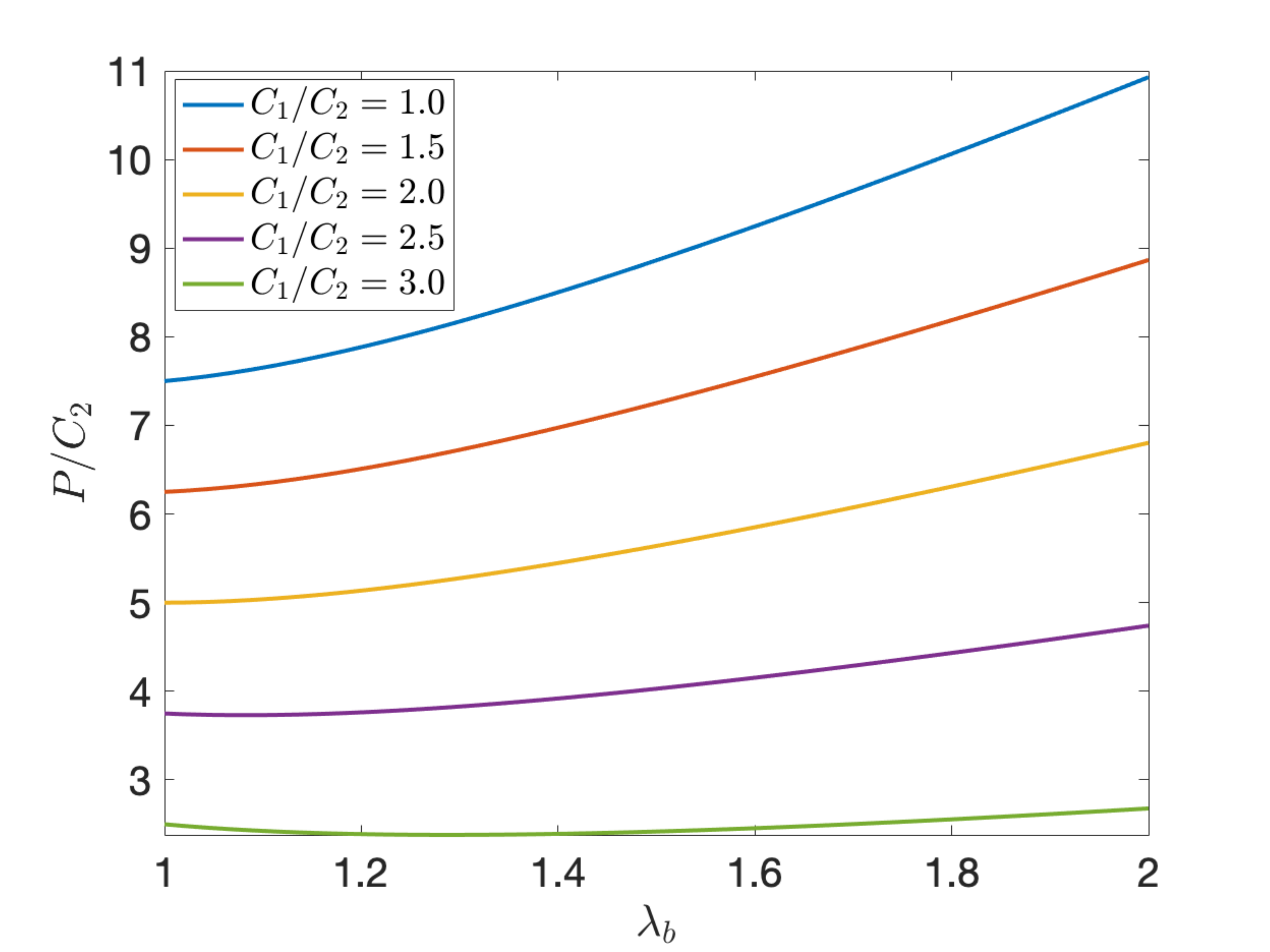}
		\includegraphics[width=0.49\textwidth]{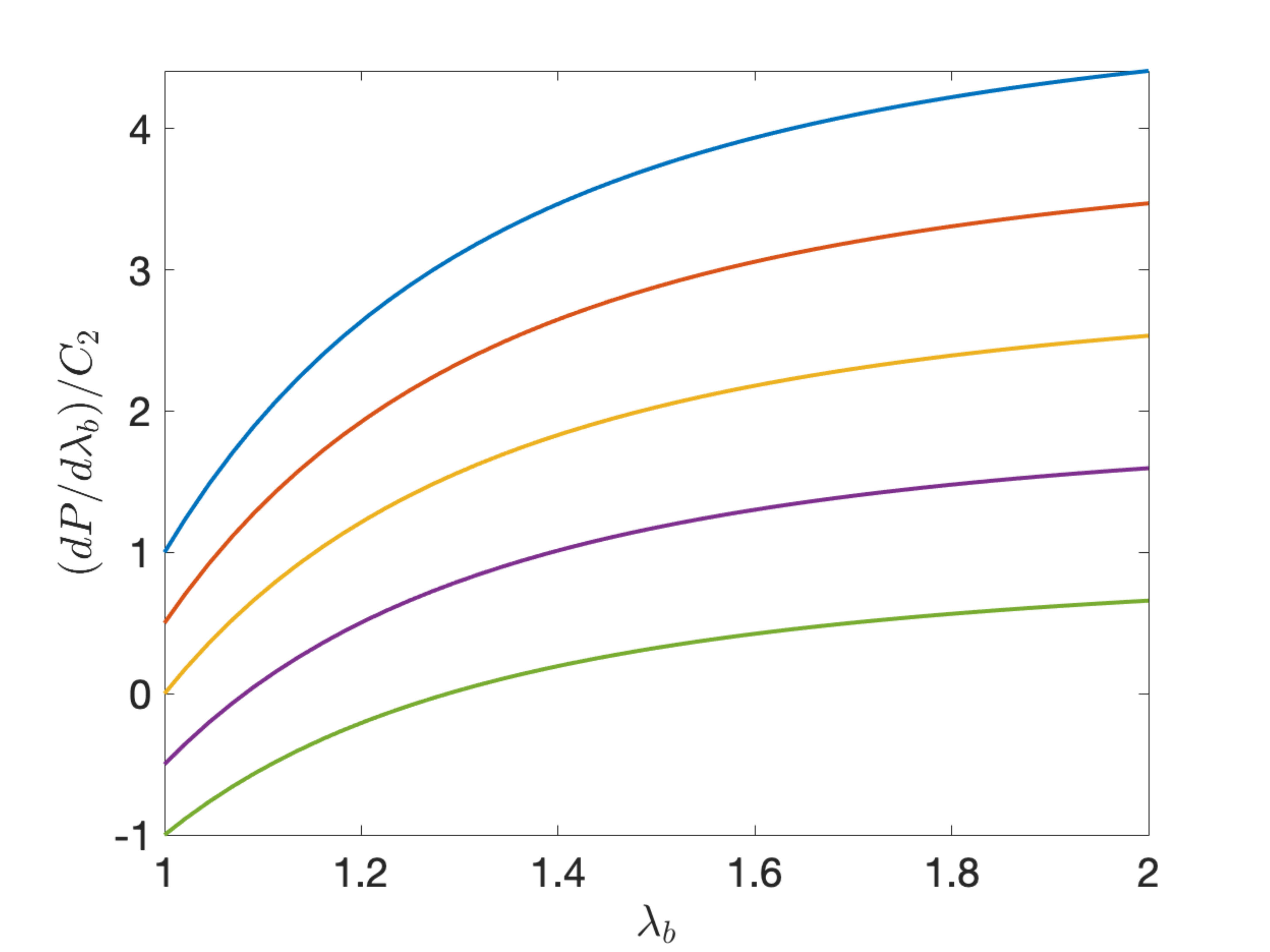}
		\caption{Examples of the scaled dead load, $P/C_{2}$ (left), with $P$ given by \eqref{eq:sphere:P:inhom}, and its derivative with respect to $\lambda_{b}$ (right), for a radially inhomogeneous sphere with different values of the ratio $C_{1}/C_{2}$.}\label{fig:PdPinhomdet}
	\end{center}
\end{figure}

To examine the post-cavitation behaviour, we take the first derivative of $P$, given by \eqref{eq:sphere:P:inhom}, with respect to $\lambda_{b}$,
\begin{equation}\label{eq:sphere:dP:inhom}
\begin{split}
\frac{\text{d}P}{\text{d}\lambda_{b}}&=2C_{1}\left(1-\frac{1}{2\lambda_{b}^{3}}\right)\\
&-\frac{3C_{2}}{2}\frac{2\lambda_{b}^4+4\lambda_{b}^3+3\lambda_{b}^2+2\lambda_{b}+1}{\lambda_{b}^4+2\lambda_{b}^3+3\lambda_{b}^2+2\lambda_{b}+1}\\
&-C_{2}\left(\lambda_{b}^3-1\right)\frac{4\lambda_{b}^3+6\lambda_{b}^2+3\lambda_{b}+1}{\lambda_{b}^6+2\lambda_{b}^5+3\lambda_{b}^4+2\lambda_{b}^3+\lambda_{b}^2}\\
&-C_{2}\left(\lambda_{b}^3-1\right)\frac{\left(2\lambda_{b}^4+4\lambda_{b}^3+3\lambda_{b}^2+2\lambda_{b}+1\right)\left(3\lambda_{b}^5+5\lambda_{b}^4+6\lambda_{b}^3+3\lambda_{b}^2+\lambda_{b}\right)}{\left(\lambda_{b}^6+2\lambda_{b}^5+3\lambda_{b}^4+2\lambda_{b}^3+\lambda_{b}^2\right)^2}.
\end{split}
\end{equation}
Letting $\lambda_{b}\to1$ in \eqref{eq:sphere:dP:inhom}, we obtain
\begin{equation}\label{eq:sphere:dP0:inhom}
\lim_{\lambda_{b}\to1}\frac{\text{d}P}{\text{d}\lambda_{b}}=C_{1}-2C_{2}.
\end{equation}
In Figure~\ref{fig:PdPinhomdet}, we represent the scaled dead load, $P/C_{2}$, with $P$ given by \eqref{eq:sphere:P:inhom} as a function of $\lambda_{b}$, and its derivative with respect to $\lambda_{b}$, for different values of the ratio $C_{1}/C_{2}$.

%%%%%%%%%%%%%%
%%%%%%%%%%%%%

%%%%%%%%%%%%%%
\begin{figure}[htbp]
	\begin{center}
		\includegraphics[width=\textwidth]{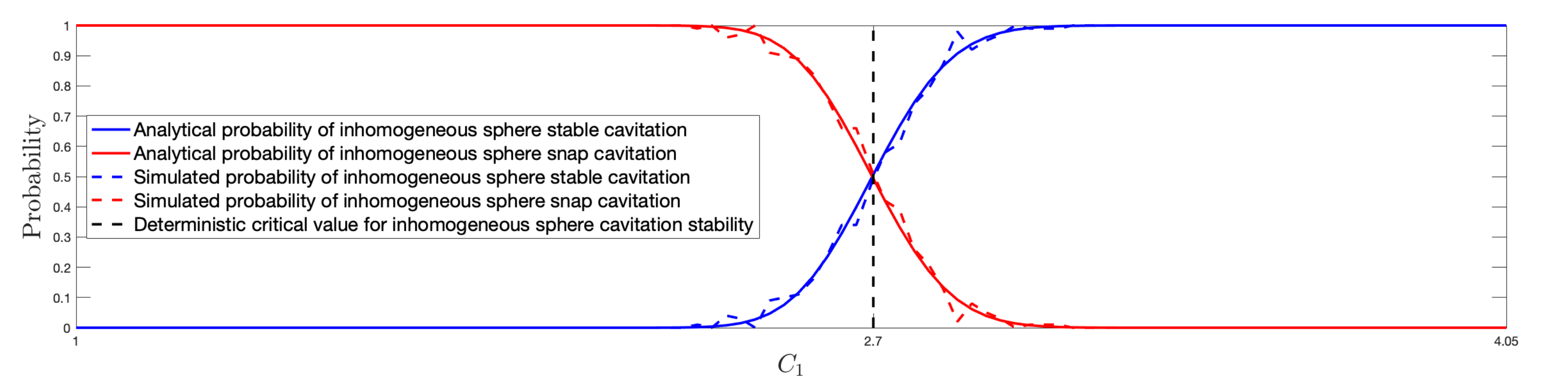}
		\caption{Probability distributions of stable or unstable cavitation for a static stochastic sphere of radially inhomogeneous neo-Hookean-like material, under dead-load traction, when $\mu(B)=C_{1}+C_{2}$, given by \eqref{eq:mur}, follows a Gamma distribution with $\rho_{1}^{(1)}=405$ and $\rho_{2}^{(1)}=0.01$. Continuous coloured lines represent analytically derived solutions, given by equations \eqref{eq:sphere:inhom:P1}-\eqref{eq:sphere:inhom:P2}, and the dashed versions represent stochastically generated data. The vertical line at the critical value, $2\underline{\mu}/3=2.7$, separates the expected regions based only on mean value, $\underline{\mu}=\rho_{1}\rho_{2}=4.05$. The probabilities were calculated from the average of 100 stochastic simulations.}\label{fig:inhom-sphere-pdfs}
	\end{center}
\end{figure}
%%%%%%%%%%%%%%%

%%%%%%%%%%%%%%
\begin{figure}[htbp]
	\begin{center}
		\includegraphics[width=0.5\textwidth]{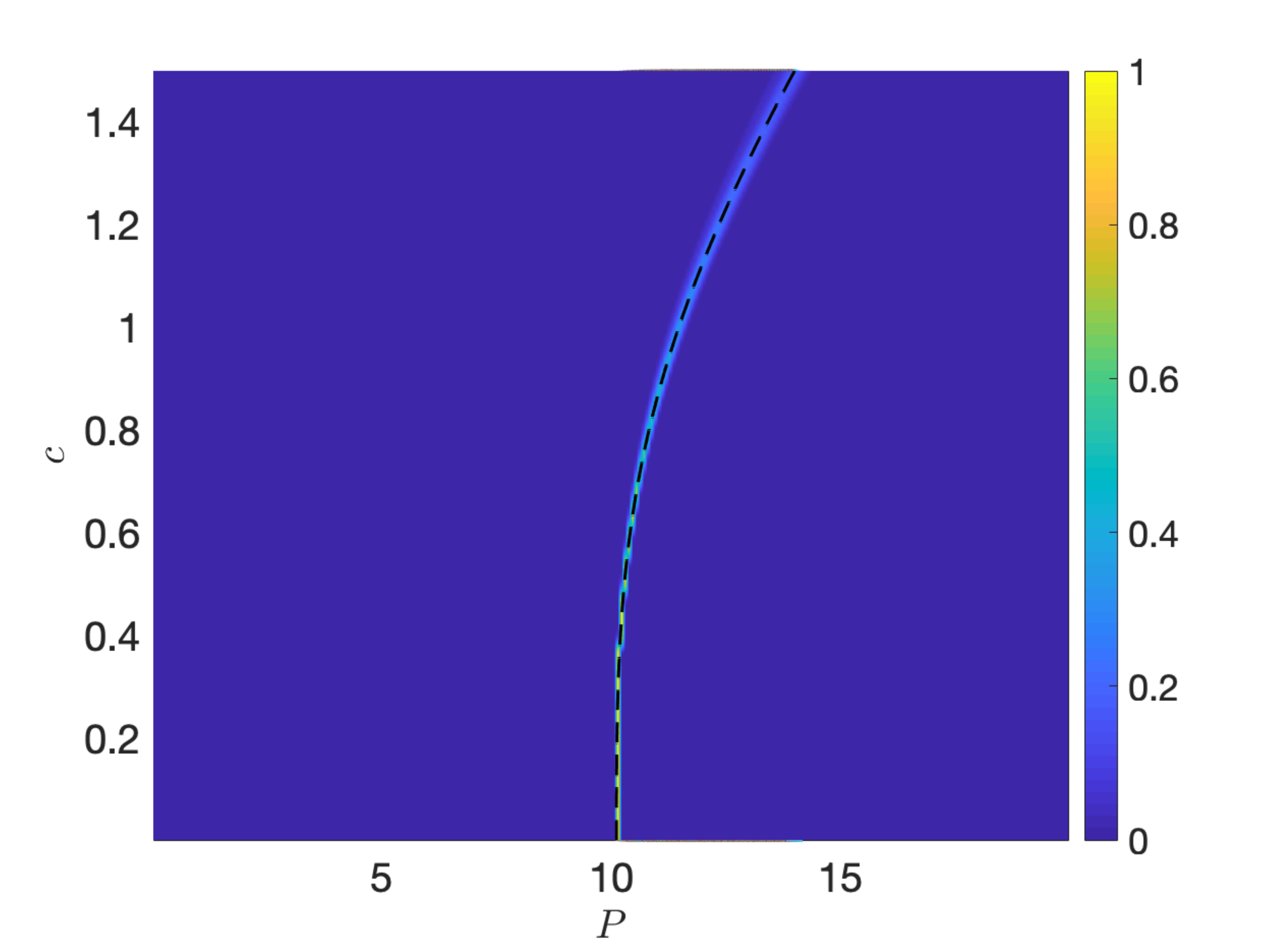}
		\caption{Probability distribution of the applied dead-load traction, $P$, causing cavitation of radius $c$ in a radially inhomogeneous sphere, when $\rho=1$, $B=1$, and, for any fixed $R$, $\mu(R)$, given by \eqref{eq:mur}, follows a Gamma distribution with  $\rho_1=405/R^6$ and $\rho_2=0.01\cdot R^6$. The dashed black line corresponds to the expected bifurcation based only on mean value, $\underline{\mu}=C_{1}=4.05$.}\label{fig:Pinhomstoch}
	\end{center}
\end{figure}
%%%%%%%%%%%%%

The following two types of cavitated solution are now possible:
\begin{itemize}
\item[(i)] If $C_{1}>2C_{2}$, or equivalently, if $\mu(B)=C_{1}+C_{2}<3C_{1}/2$, then cavitation is stable;
\item[(ii)] If $C_{1}<2C_{2}$, or equivalently, if $\mu(B)=C_{1}+C_{2}>3C_{1}/2$, then snap cavitation occurs.
\end{itemize}
Thus, the probability of stable cavitation (and also of oscillatory motion for the dynamic sphere) is equal to (see the example in Figure~\ref{fig:inhom-sphere-pdfs})
\begin{equation}\label{eq:sphere:inhom:P1}
P_{1}(C_{1})=\int_{0}^{3C_{1}/2}g(u;\rho_{1},\rho_{2})du,
\end{equation}
and that of snap cavitation (and also of non-oscillatory motion for the dynamic sphere) is
\begin{equation}\label{eq:sphere:inhom:P2}
P_{2}(C_{1})=1-P_{1}(C_{1})=1-\int_{0}^{3C_{1}/2}g(u;\rho_{1},\rho_{2})du.
\end{equation}

An example of post-cavitation stochastic behaviour of the static inhomogeneous unit sphere (with $B=1)$ is shown in Figure~\ref{fig:Pinhomstoch}, for the case when the shear modulus $\mu(R)$ is given by \eqref{eq:mur}, with $\rho_1=405/R^6$ and $\rho_2=0.01\cdot R^6$ (see Figure~\ref{fig:muR-gpdf}). Thus, $\rho_{1}=405$ and $\rho_{2}=0.01$ at $R=B$, and by \eqref{eq:mur}, $C_{1}=\underline{\mu}=\rho_{1}\rho_{2}=4.05$ and $C_{2}=\mu(B)-C_{1}$.

%%%%%%%%%%%%%%%%%%%%%%%%%%%%%%%%%%%%%%%%%%%%%%%%%%%%%%%%%%%%
\subsection{Alternative modelling formulations of radially inhomogeneous spheres}\label{sec:sphere:inhom:others}
In the foregoing analysis, by following the approach of \cite{Soize:2006,Staber:2018:SG}, we have assumed that, for the inhomogeneous shear modulus $\mu(R)$, the mean value $\underline{\mu}(R)=\underline{\mu}$ is independent of $R$. Then, the shear modulus can be described by \eqref{eq:mur}, where $C_{1}$ is a deterministic constant and $C_{2}$ is a (shifted) Gamma-distributed random variable with zero mean value. In this case, at the centre of the sphere, the shear modulus is a deterministic constant, while throughout the rest of the sphere, this modulus has constant mean value and non-constant variance \cite{Riggs:2017:RL}. 

An alternative model, where the mean value of the shear modulus varies throughout the sphere (see also \cite{Soize:2006}, can be constructed by taking $C_{2}$ as a Gamma-distributed random variable (with non-zero mean value, i.e., $\underline{C}_{2}\neq 0$), while $C_{1}$ remains a deterministic constant. Then,  by \eqref{eq:mur}, for any fixed $R$, $\mu(R)$ is a shifted Gamma-distributed random variable, with the mean value and variance, respectively, equal to
\begin{equation}\label{eq:mur:C12:R}
\underline{\mu}(R)=C_{1}+\underline{C}_{2}\frac{R^3}{B^3},\qquad \text{Var}[\mu(R)]=\text{Var}[C_{2}]\frac{R^6}{B^6}.
\end{equation}

A more general modelling approach, where the shear modulus is characterised by a probability distribution at the centre of the sphere as well, is to assume that $C_{1}$ and $C_{2}$ are described by independent Gamma distributions, with hyperparameters $\rho^{(1)}_{1}$, $\rho^{(1)}_{2}$, and $\rho^{(2)}_{1}$, $\rho^{(2)}_{2}$, respectively. Then, for any fixed $R$, the distribution of $\mu(R)$ is a linear combination of two Gamma distributions (see Theorem~1 of \cite{Moschopoulos:1985}), while at the centre of the sphere, the shear modulus $\mu(0)=C_{1}$ is Gamma-distributed. 

In all of these cases, the analysis follows the exposition of Sections~\ref{sec:sphere:inhom:osc} and \ref{sec:sphere:inhom:static}, but the numerical results will differ in each case. 

%%%%%%%%%%%%%%
\begin{figure}[htbp]
	\begin{center}
		\includegraphics[width=\textwidth]{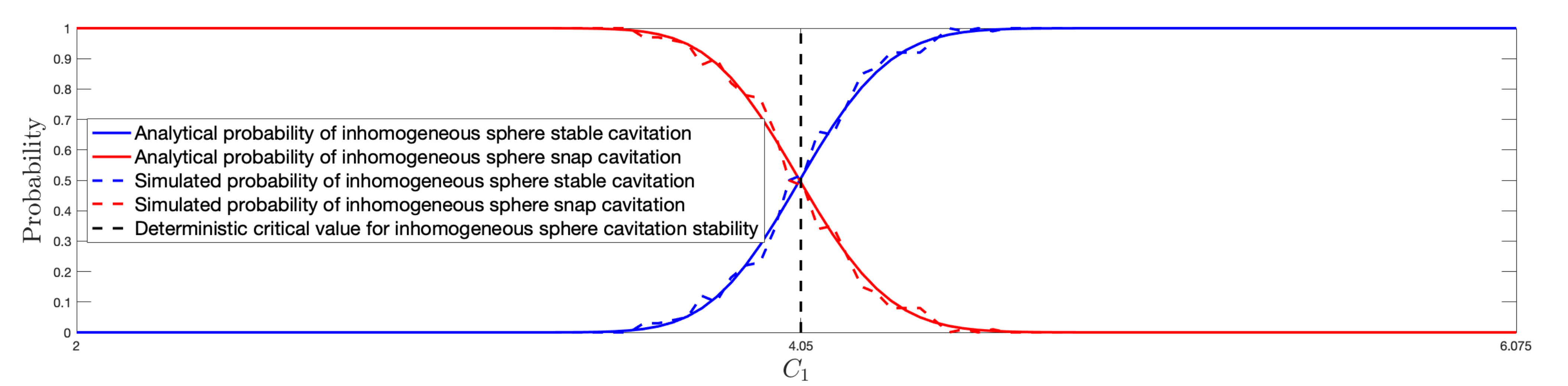}
		\caption{Probability distributions of stable or unstable cavitation for a static stochastic sphere of radially inhomogeneous neo-Hookean-like material under dead-load traction, with $\mu(R)=C_{1}+C_{2}R^3/B^3$ given by \eqref{eq:mur}, where $C_{2}$ follows a Gamma distribution with $\rho^{(2)}_{1}=405$ and $\rho^{(2)}_{2}=0.05$, while $C_{1}$ may be either deterministic or Gamma-distributed. Continuous coloured lines represent analytically derived solutions, given by equations \eqref{eq:sphere:inhomR:P1}-\eqref{eq:sphere:inhomR:P2}, and the dashed versions represent stochastically generated data. The vertical line at the critical value, $2\underline{C}_2=4.05$, separates the expected regions based only on mean value, $\underline{C}_{2}=\rho_{1}\rho_{2}=2.025$. The probabilities were calculated from the average of 100 stochastic simulations.}\label{fig:inhomR-sphere-pdfs}
	\end{center}
\end{figure}
%%%%%%%%%%%%%%%

%%%%%%%%%%%%%%
\begin{figure}[htbp]
	\begin{center}
		\includegraphics[width=0.47\textwidth]{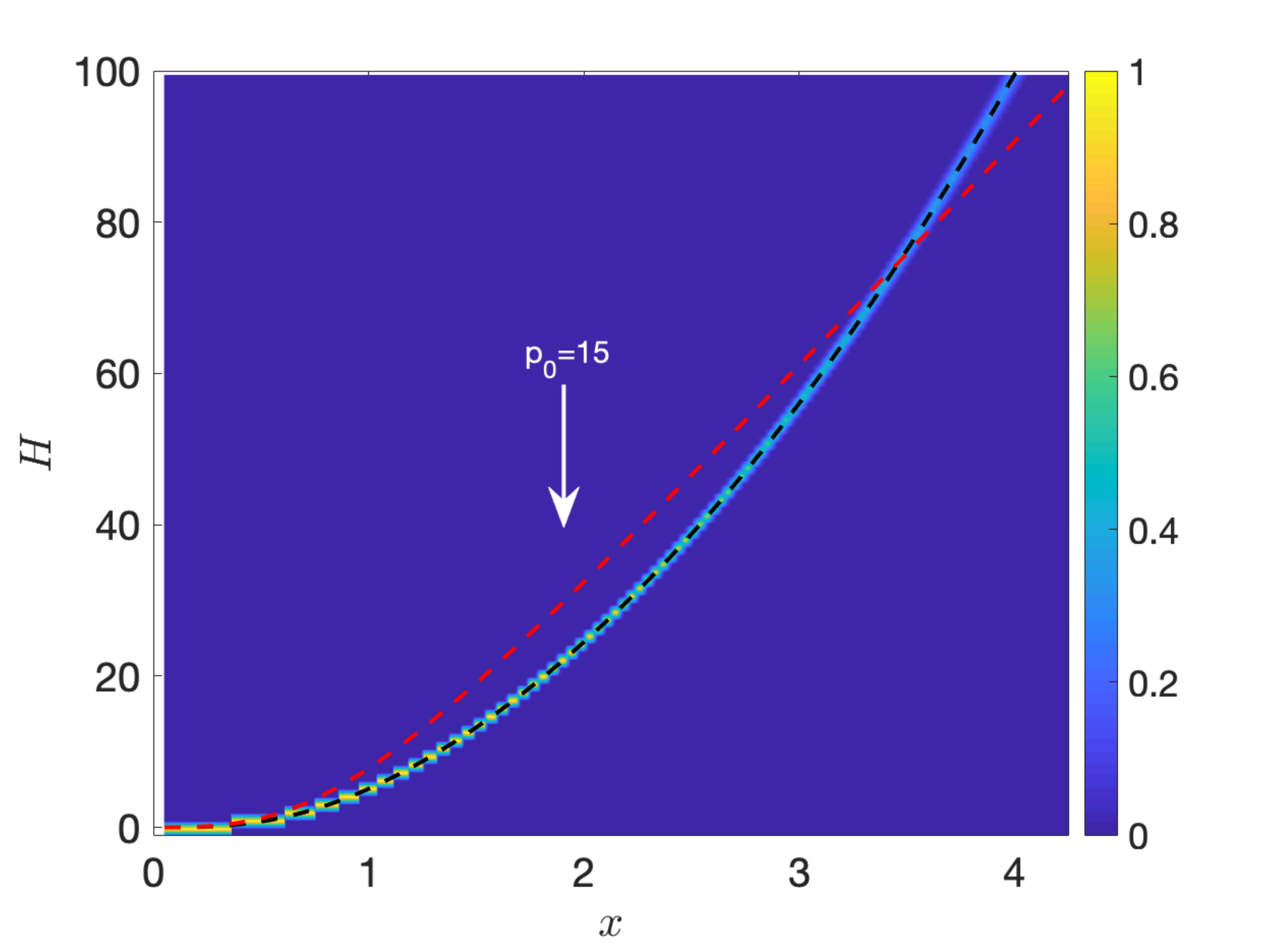}\qquad
		\includegraphics[width=0.47\textwidth]{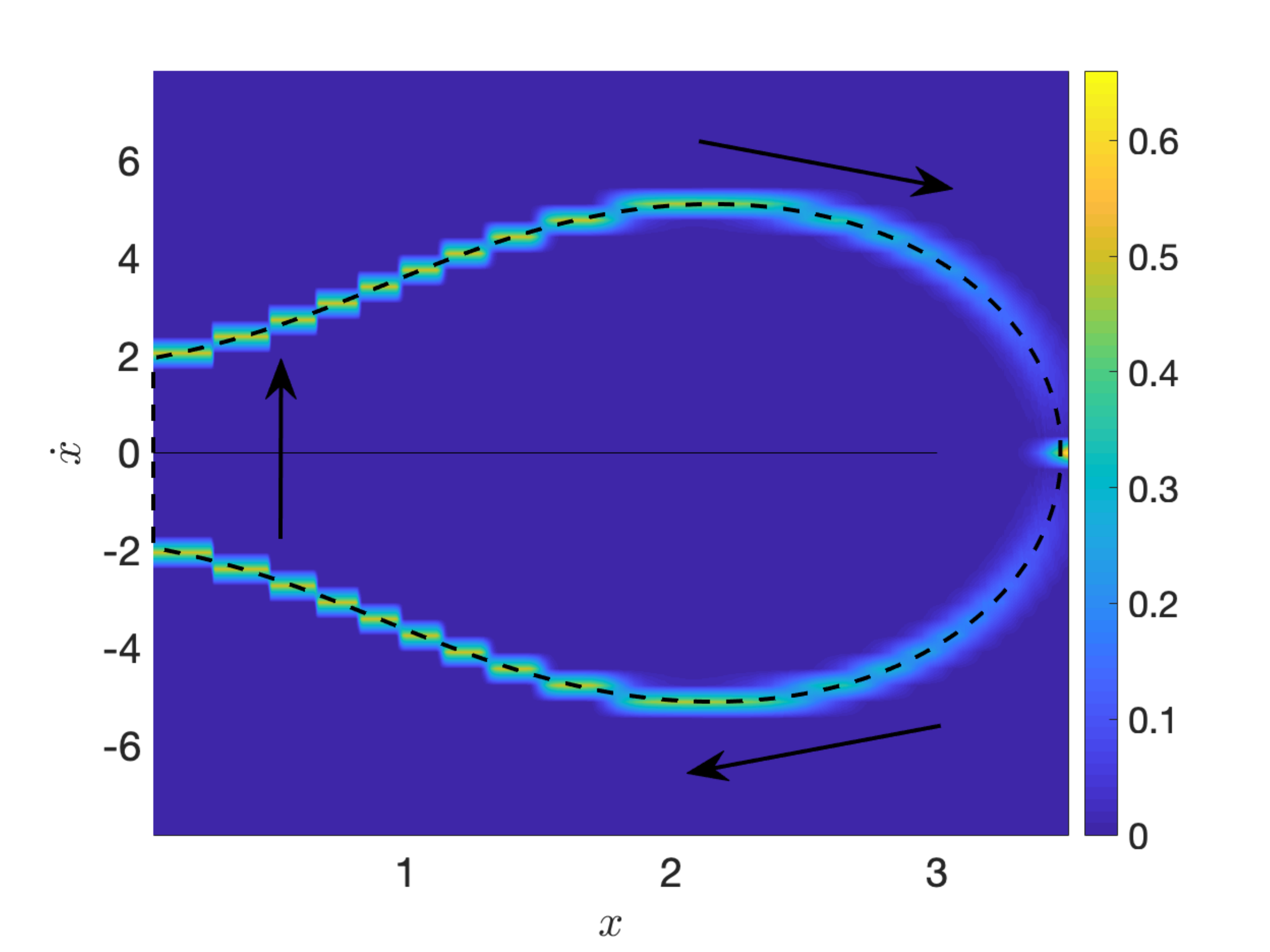}
		\caption{The function $H(x)$, defined by \eqref{eq:sphere:H:inhom}, intersecting the (dashed red) curve $\frac{2p_{0}}{\rho B^2}\left[\left(x^3+1\right)^{1/3}-1\right]$, with $p_{0}=15$ (left), and the associated velocity, given by \eqref{eq:sphere:dotx:forced} (right), for a dynamic sphere of stochastic radially inhomogeneous neo-Hookean-like material under dead-load traction, when $\rho=1$, $B=1$, and $\mu(R)$ is given by \eqref{eq:mur} , where $C_{1}=4.05$ and $C_{2}$ follows a Gamma distribution with $\rho^{(2)}_{1}=405$ and $\rho^{(2)}_{2}=0.005$. The dashed black lines correspond to the expected values based only on mean parameter values. Each distribution was calculated from the average of $1000$ stochastic simulations.}\label{fig:stoch-inhomsphereR1}
	\end{center}
\end{figure}
%%%%%%%%%%%%

%%%%%%%%%%%%%%
\begin{figure}[htbp]
	\begin{center}
		\includegraphics[width=0.47\textwidth]{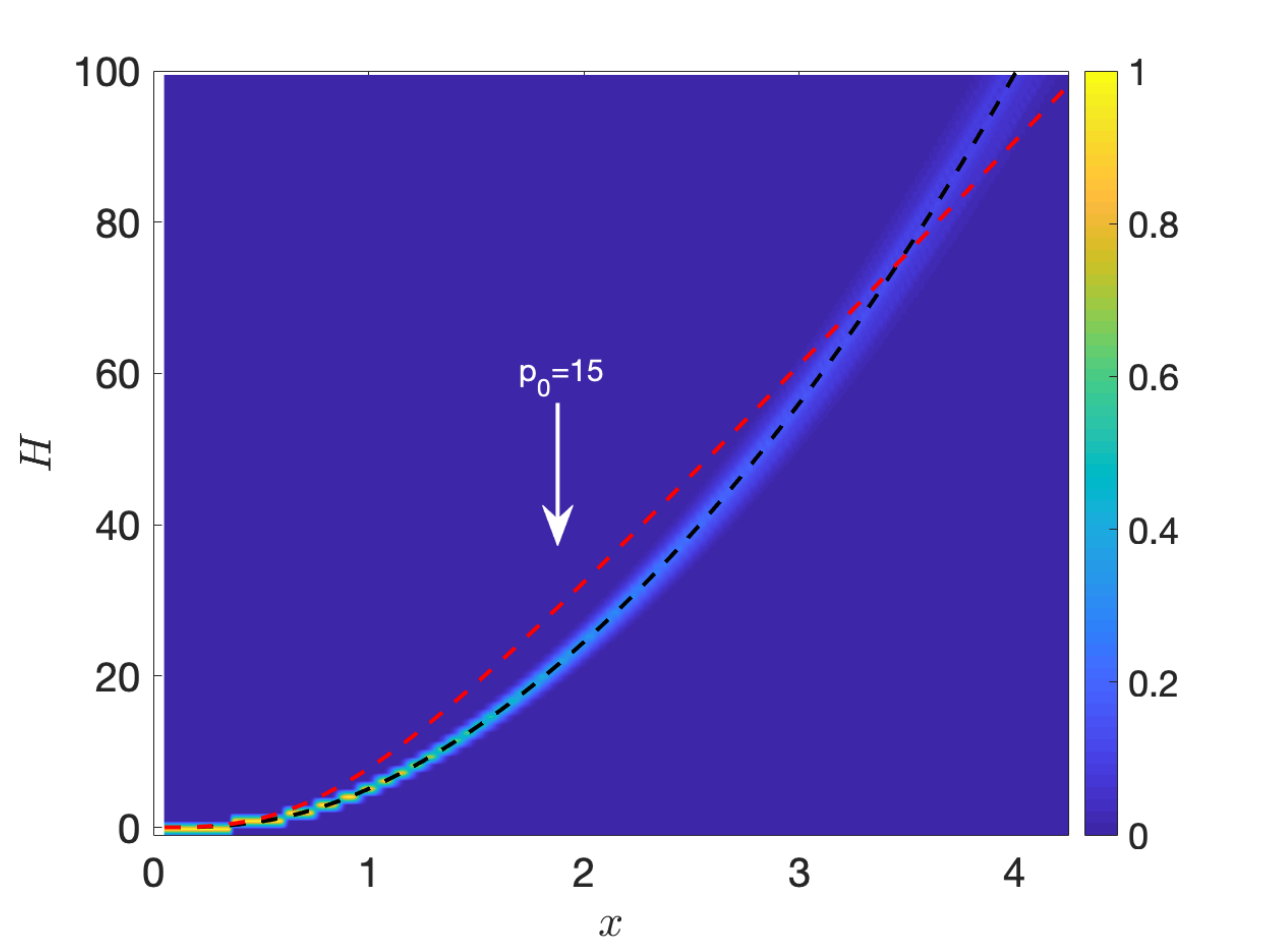}\qquad
		\includegraphics[width=0.47\textwidth]{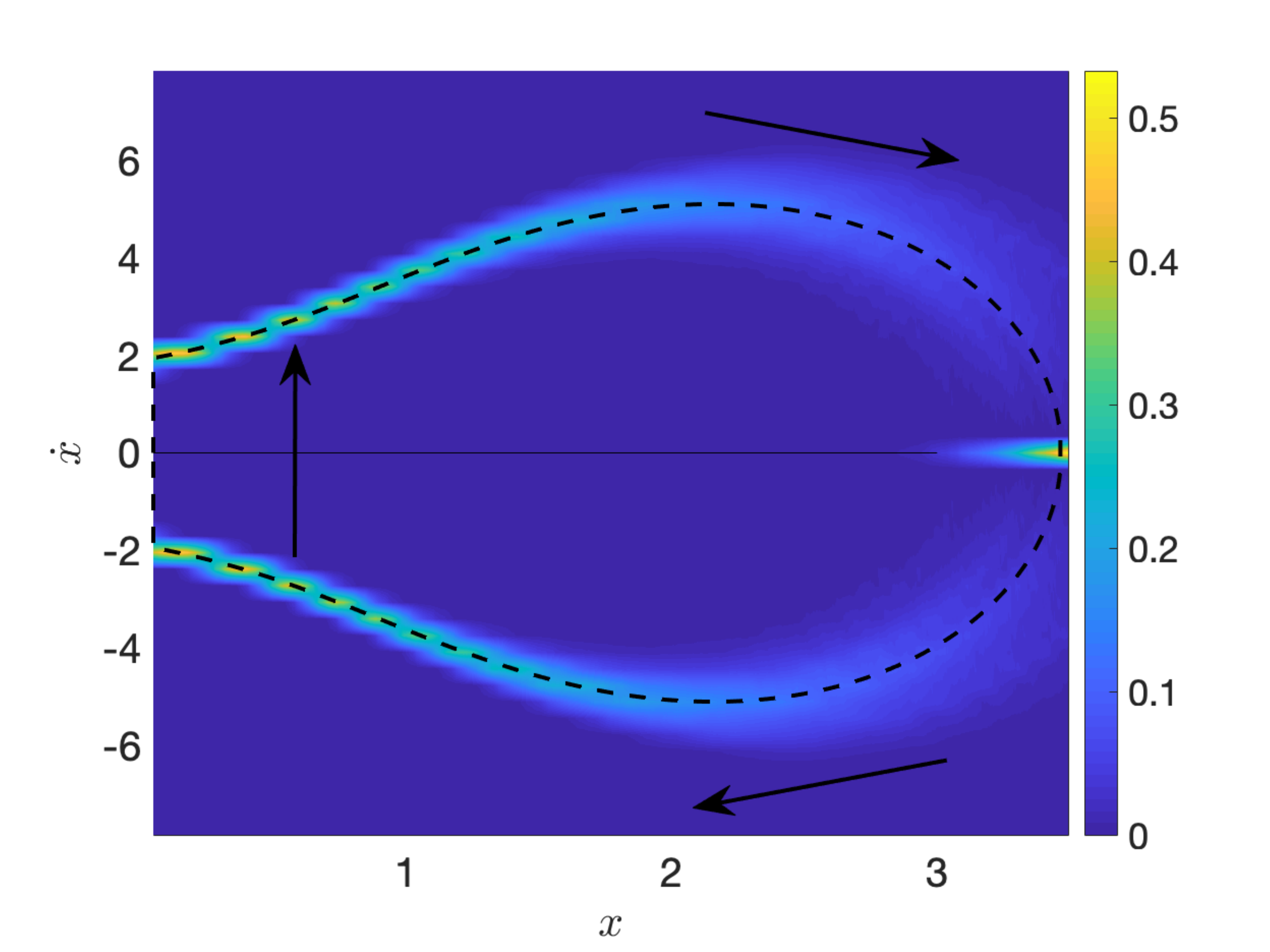}
		\caption{The function $H(x)$, defined by \eqref{eq:sphere:H:inhom}, intersecting the (dashed red) curve $\frac{2p_{0}}{\rho B^2}\left[\left(x^3+1\right)^{1/3}-1\right]$, with $p_{0}=15$ (left), and the associated velocity, given by \eqref{eq:sphere:dotx:forced} (right), for a dynamic sphere of stochastic radially inhomogeneous neo-Hookean-like material under dead-load traction, when $\rho=1$, $B=1$, and $\mu(R)$ is given by \eqref{eq:mur} , where $C_{1}$ follows a Gamma distribution with $\rho^{(1)}_{1}=405$ and $\rho^{(1)}_{2}=0.01$, while $C_{2}$ follows a Gamma distributions with $\rho^{(2)}_{1}=405$ and $\rho^{(2)}_{2}=0.005$. The dashed black lines correspond to the expected values based only on mean parameter values. Each distribution was calculated from the average of $1000$ stochastic simulations.}\label{fig:stoch-inhomsphereR2}
	\end{center}
\end{figure}
%%%%%%%%%%%%

%%%%%%%%%%%%%%
\begin{figure}[htbp]
	\begin{center}
		\includegraphics[width=0.47\textwidth]{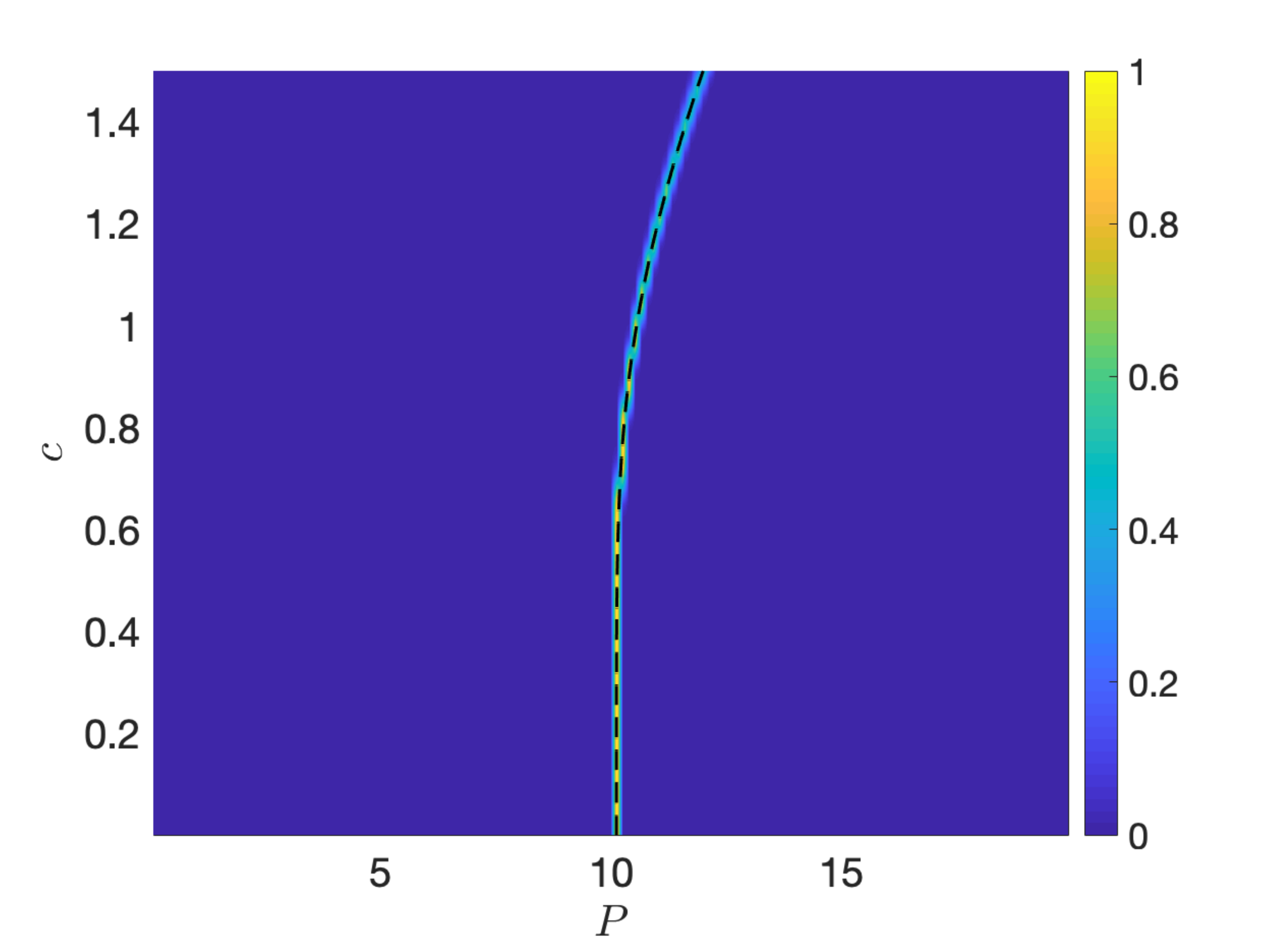}\qquad
		\includegraphics[width=0.47\textwidth]{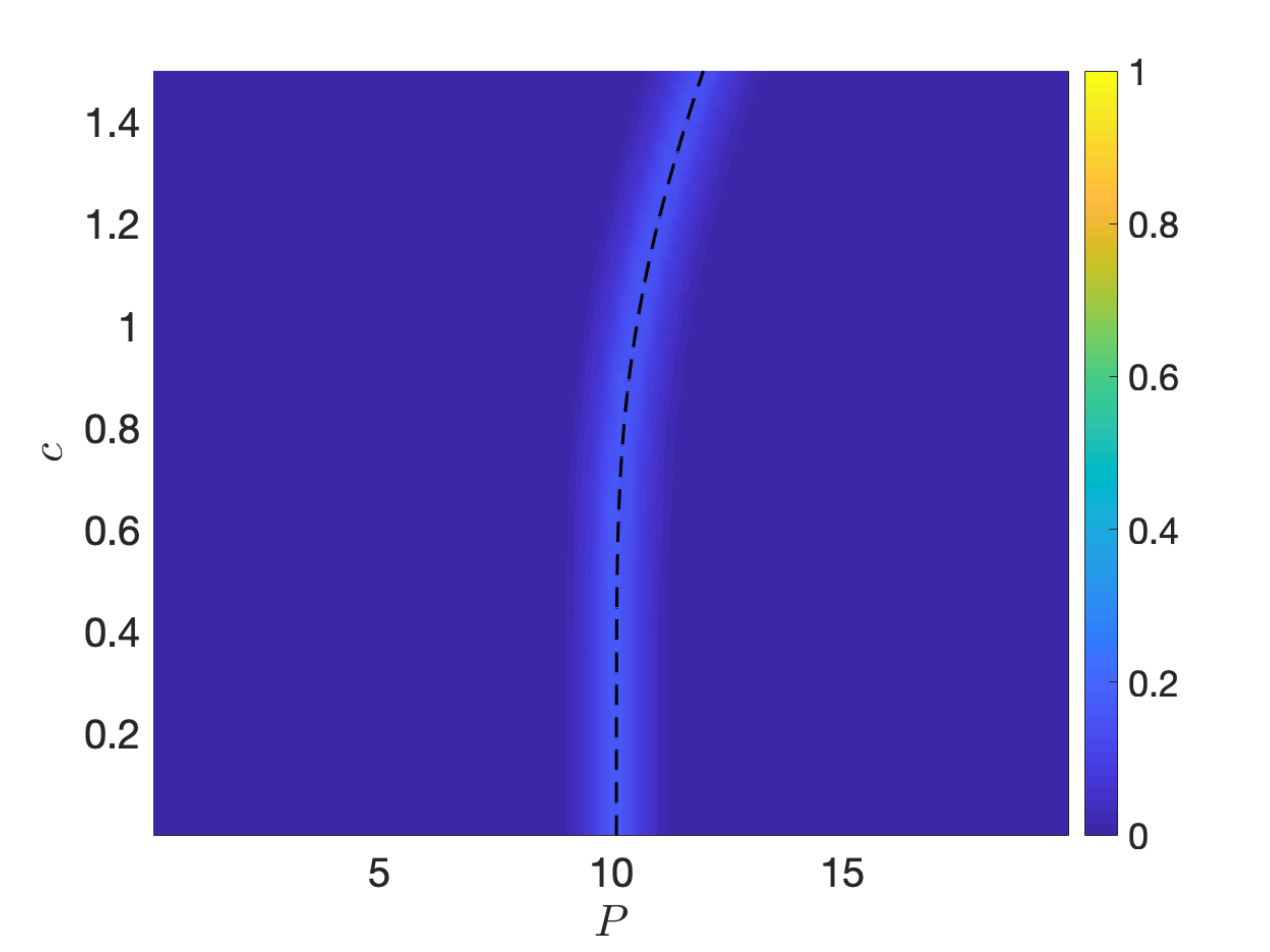}
		\caption{Probability distribution of the applied dead-load traction, $P$, causing cavitation of radius $c$ in a radially inhomogeneous sphere, when $\rho=1$, $B=1$, and $\mu(R)$ is given by \eqref{eq:mur}, where $C_{2}$ follows a Gamma distribution with $\rho^{(2)}_{1}=405$ and $\rho^{(2)}_{2}=0.05$, while $C_{1}=4.05$ (left), or $C_{1}$ follows a Gamma distribution with $\rho^{(1)}_{1}=405$ and $\rho^{(1)}_{2}=0.01$ (right). The dashed black line corresponds to the expected bifurcation based only on mean value parameters.}\label{fig:PinhomstochR}
	\end{center}
\end{figure}
%%%%%%%%%%%%%

For the two alternative formulations, the probability of stable cavitation is equal to (see Figure~\ref{fig:inhomR-sphere-pdfs})
\begin{equation}\label{eq:sphere:inhomR:P1}
P_{1}(C_{1})=\int_{0}^{C_{1}/2}g(u;\rho^{(2)}_{1},\rho^{(2)}_{2})du,
\end{equation}
and that of snap cavitation is
\begin{equation}\label{eq:sphere:inhomR:P2}
P_{2}(C_{1})=1-P_{1}(C_{1})=1-\int_{0}^{C_{1}/2}g(u;\rho^{(2)}_{1},\rho^{(2)}_{2})du.
\end{equation}

If $C_{1}>2C_{2}$, then the probability distribution of oscillatory motions occurring is given by \eqref{eq:sphere:NH:P1} and that of non-oscillatory motions by \eqref{eq:sphere:NH:P2}, with $C_{1}$ instead of $\mu$ and $g(u;\rho^{(1)}_{1},\rho^{(1)}_{2})$ instead of $g(u;\rho_{1},\rho_{2})$ (see Figure~\ref{fig:intpdfs-NHsphere}). 

For illustration, we choose a unit sphere with shear modulus $\mu(R)$ given by \eqref{eq:mur}, where $C_{2}$ follows a Gamma distribution with shape and scale parameters $\rho^{(2)}_{1}=405$ and $\rho^{(2)}_{2}=0.05$, respectively, while $C_{1}$ is either the deterministic constant $C_{1}=4.05$, or a Gamma-distributed random variable, with hyperparameters $\rho^{(1)}_{1}=405$ and $\rho^{(1)}_{2}=0.01$. For the dynamic sphere, the associated function $H(x)$, defined by \eqref{eq:sphere:H:inhom}, and the velocity, given by \eqref{eq:sphere:dotx:forced}, are represented in Figures~\ref{fig:stoch-inhomsphereR1} and \ref{fig:stoch-inhomsphereR2}, respectively. The corresponding post-cavitation behaviour of the static sphere is shown in Figure~\ref{fig:PinhomstochR}. 

%%%%%%%%%%%%%%%%%%%%%%%%%%%%%%%%%%%%%%%%%%%%%%%%%%%%%%%%%%%%
\subsection{Non-oscillatory motion of a stochastic radially inhomogeneous sphere under impulse traction}\label{sec:sphere:inhom:nosc}

We now consider the cavitation of radially inhomogeneous, incompressible spheres of stochastic hyperelastic material with the shear modulus of the form \eqref{eq:mur}, such that, for any fixed $R$, both the mean value, $\underline{\mu}$, and variance, $\text{Var}[\mu]$, of $\mu=\mu(R)$ may depend on $R$. 

Setting the initial conditions $x_{0}=x(0)=0$ and $\dot{x}_{0}=\dot{x}(0)=0$, we obtain equation \eqref{eq:sphere:ode:imp}, where 
\begin{equation}\label{eq:sphere:H:inhom:imp}
\begin{split}
H(x)&=\frac{C_{1}}{\rho B^2}\left[2\left(x^3+1\right)^{2/3}-\frac{1}{\left(x^3+1\right)^{1/3}}-1\right]\\
&-\frac{C_{2}}{\rho B^2}\int_{0}^{x}\xi^5\frac{2\left(\xi^3+1\right)^{4/3}+4\left(\xi^3+1\right)+3\left(\xi^3+1\right)^{2/3}+2\left(\xi^3+1\right)^{1/3}+1}{\left(\xi^3+1\right)^{8/3}+2\left(\xi^3+1\right)^{7/3}+3\left(\xi^3+1\right)^{2}+2\left(\xi^3+1\right)^{5/3}+\left(\xi^3+1\right)^{4/3}}d\xi,\\
&=\frac{C_{1}}{\rho B^2}\left[2\left(x^3+1\right)^{2/3}-\frac{1}{\left(x^3+1\right)^{1/3}}-1\right]\\
&-\frac{C_{2}}{\rho B^2}\left[\sqrt{3}\arctan\frac{2\left(x^3+1\right)^{1/3}+1}{\sqrt{3}}-\frac{\pi}{\sqrt{3}}\right]\\
&+\frac{3C_{2}}{2\rho B^2}\left\{\ln\left[\left(x^3+1\right)^{2/3}+\left(x^3+1\right)^{1/3}+1\right]-\ln 3\right\}\\
&-\frac{C_{2}}{\rho B^2}\left[\left(x^3+1\right)^{2/3}+\frac{1}{\left(x^3+1\right)^{1/3}}-2\right].
\end{split}
\end{equation}
Then, by substitution of \eqref{eq:sphere:H:inhom:imp} in \eqref{eq:sphere:HC:imp} and \eqref{eq:sphere:impulse}, we have
\begin{equation}\label{eq:sphere:T:inhom:imp}
\begin{split}
\frac{2T}{3}x^3&=C_{1}\left[2\left(x^3+1\right)^{2/3}-\frac{1}{\left(x^3+1\right)^{1/3}}-1\right]\\
&-C_{2}\left[\sqrt{3}\arctan\frac{2\left(x^3+1\right)^{1/3}+1}{\sqrt{3}}-\frac{\pi}{\sqrt{3}}\right]\\
&+\frac{3C_{2}}{2}\left\{\ln\left[\left(x^3+1\right)^{2/3}+\left(x^3+1\right)^{1/3}+1\right]-\ln 3\right\}\\
&-C_{2}\left[\left(x^3+1\right)^{2/3}+\frac{1}{\left(x^3+1\right)^{1/3}}-2\right].
\end{split}
\end{equation}
Equation \eqref{eq:sphere:T:inhom:imp} has a solution at $x_{1}=0$, while the  second solution, $x_{2}>0$, is a root of
\begin{equation}\label{eq:sphere:T:inhom:nonzero}
\begin{split}
T&=\frac{3C_{1}}{2}\left[\frac{\left(x^3+1\right)^{1/3}+1}{\left(x^3+1\right)^{2/3}+\left(x^3+1\right)^{1/3}+1}+\frac{1}{\left(x^3+1\right)^{1/3}}\right]\\
&-\frac{C_{2}}{6}\left(x^3-3\right)-\frac{3C_{2}}{2}\left[2\frac{\left(x^3+1\right)^{1/3}+1}{\left(x^3+1\right)^{2/3}+\left(x^3+1\right)^{1/3}+1}-\frac{1}{\left(x^3+1\right)^{1/3}}\right].
\end{split}
\end{equation}
To obtain the right-hand side of the above equation we have assumed that, in \eqref{eq:sphere:T:inhom:imp}, $x$ is sufficiently small such that after expanding the respective functions to the second order in $x^3$,
\begin{equation}
\left[\sqrt{3}\arctan\frac{2\left(x^3+1\right)^{1/3}+1}{\sqrt{3}}-\frac{\pi}{\sqrt{3}}\right]\frac{1}{x^3}\approx\frac{2-x^3}{12}
\end{equation}
and
\begin{equation}
\frac{\ln\left[\left(x^3+1\right)^{2/3}+\left(x^3+1\right)^{1/3}+1\right]-\ln 3}{x^3}\approx\frac{18-7x^3}{54}.
\end{equation}
Next, expanding the right-hand side of the equation \eqref{eq:sphere:T:inhom:nonzero} to first order in $x^3$ gives
\begin{equation}\label{eq:sphere:T:inhom:nonzero:app}
T\approx\frac{5C_{1}}{2}-\frac{x^3}{3}\left(2C_{1}+C_{2}\right).
\end{equation}
The critical dead load for the onset of cavitation is then
\begin{equation}\label{eq:sphere:T:inhom:crit}
T_{0}=\lim_{x\to0_{+}}T=\frac{5C_{1}}{2},
\end{equation}
and is the same as for the homogeneous sphere made entirely from the material found at the centre of the inhomogeneous sphere.

Noting that $2C_{1}+C_{2}=\mu(0)+\mu(B)>0$, the right-hand side in \eqref{eq:sphere:T:inhom:nonzero:app} decreases as $x$ increases, hence, 
\begin{equation}\label{eq:sphere:T:inhom:bounds}
0<T<\frac{5C_{1}}{2},
\end{equation}
or equivalently, by \eqref{eq:sphere:impulse},
\begin{equation}\label{eq:sphere:p0:inhom:bounds:imp}
0<p_{0}<\frac{5C_{1}}{\rho B^2}.
\end{equation}
However,
\begin{equation}\label{eq:sphere:inhom:osc-no:imp}
\frac{2p_{0}}{\rho B^2}x^3-H(x)\left\{
\begin{array}{ll}
<0 & \mbox{if}\ x_{1}< x< x_{2},\\
>0 & \mbox{if}\ x>x_{2},
\end{array}
\right.
\end{equation}
and the sphere cannot oscillate.

%%%%%%%%%%%%%%%%%%%%%%%%%%%%%%%%%%%%%%%%%%%%%%%%%%%%%%%%%%%%
\subsection{Static deformation of a stochastic radially inhomogeneous sphere under impulse traction}\label{sec:sphere:inhom:static:imp}

For the static inhomogeneous sphere, when the cavity surface is traction-free, at the outer surface, the tensile traction takes the form
\begin{equation}\label{eq:sphere:T:inhom:static}
\begin{split}
T&=2C_{1}\left[\frac{1}{\left(x^3+1\right)^{1/3}}+\frac{1}{4\left(x^3+1\right)^{4/3}}\right]\\
&-\frac{C_{2}}{2}x^3\frac{2\left(x^3+1\right)^{4/3}+4\left(x^3+1\right)+3\left(x^3+1\right)^{2/3}+2\left(x^3+1\right)^{1/3}+1}{\left(x^3+1\right)^{8/3}+2\left(x^3+1\right)^{7/3}+3\left(x^3+1\right)^{2}+2\left(x^3+1\right)^{5/3}+\left(x^3+1\right)^{4/3}}.\\
\end{split}
\end{equation}
Denoting $\lambda_{b}=b/B=\left(x^3+1\right)^{1/3}$, the required tensile load at the outer surface, $R=B$, can be written equivalently as follows,
\begin{equation}\label{eq:sphere:T:inhom:lambda}
T=2C_{1}\left(\frac{1}{\lambda_{b}}+\frac{1}{4\lambda_{b}^{4}}\right)
-\frac{C_{2}}{2}\left(\lambda_{b}^3-1\right)\frac{2\lambda_{b}^4+4\lambda_{b}^3+3\lambda_{b}^2+2\lambda_{b}+1}{\lambda_{b}^8+2\lambda_{b}^7+3\lambda_{b}^6+2\lambda_{b}^5+\lambda_{b}^4}.
\end{equation}
Thus the critical load for the initiation of cavitation is
\begin{equation}\label{eq:sphere:P0:inhom:imp}
T_{0}=\lim_{\lambda_{b}\to1}T=\frac{5C_{1}}{2}=\frac{5\mu(0)}{2},
\end{equation}
and is the same as for the homogeneous sphere made entirely from the material found at its centre, as found also for the dynamic sphere.

%%%%%%%%%%%%%%
\begin{figure}[htbp]
	\begin{center}
		\includegraphics[width=0.5\textwidth]{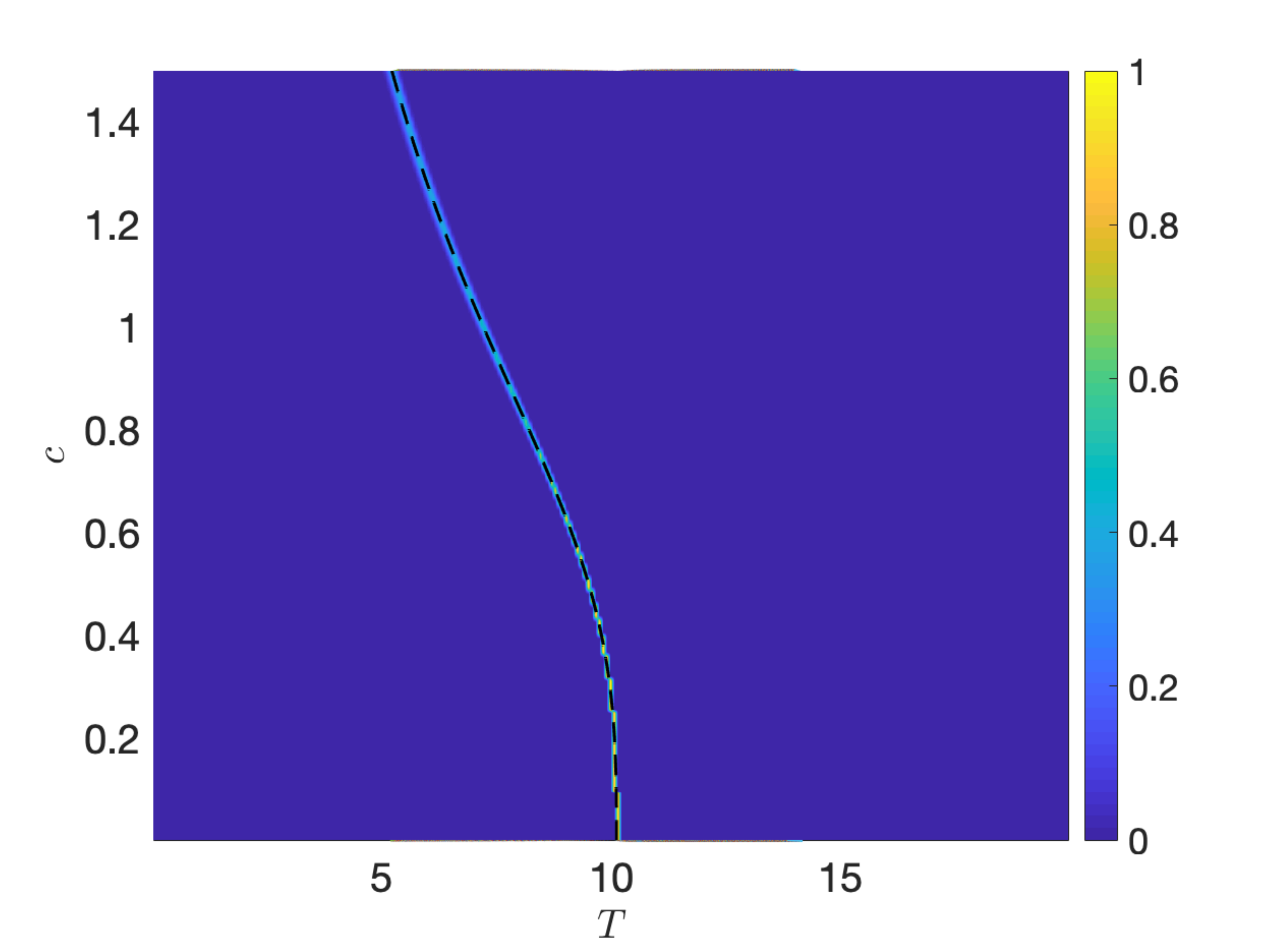}
		\caption{Probability distribution of the applied impulse traction, $T$, causing cavitation of radius $c$ in a radially inhomogeneous sphere, when $\rho=1$, $B=1$, and, for any fixed $R$, $\mu(R)$, given by \eqref{eq:mur}, follows a Gamma distribution with  $\rho_1=405/R^6$ and $\rho_2=0.01\cdot R^6$. The dashed black line corresponds to the expected bifurcation based only on mean value, $\underline{\mu}=C_{1}=4.05$.}\label{fig:Tinhomstoch}
	\end{center}
\end{figure}
%%%%%%%%%%%%%

%%%%%%%%%%%%%%
\begin{figure}[htbp]
	\begin{center}
		\includegraphics[width=0.45\textwidth]{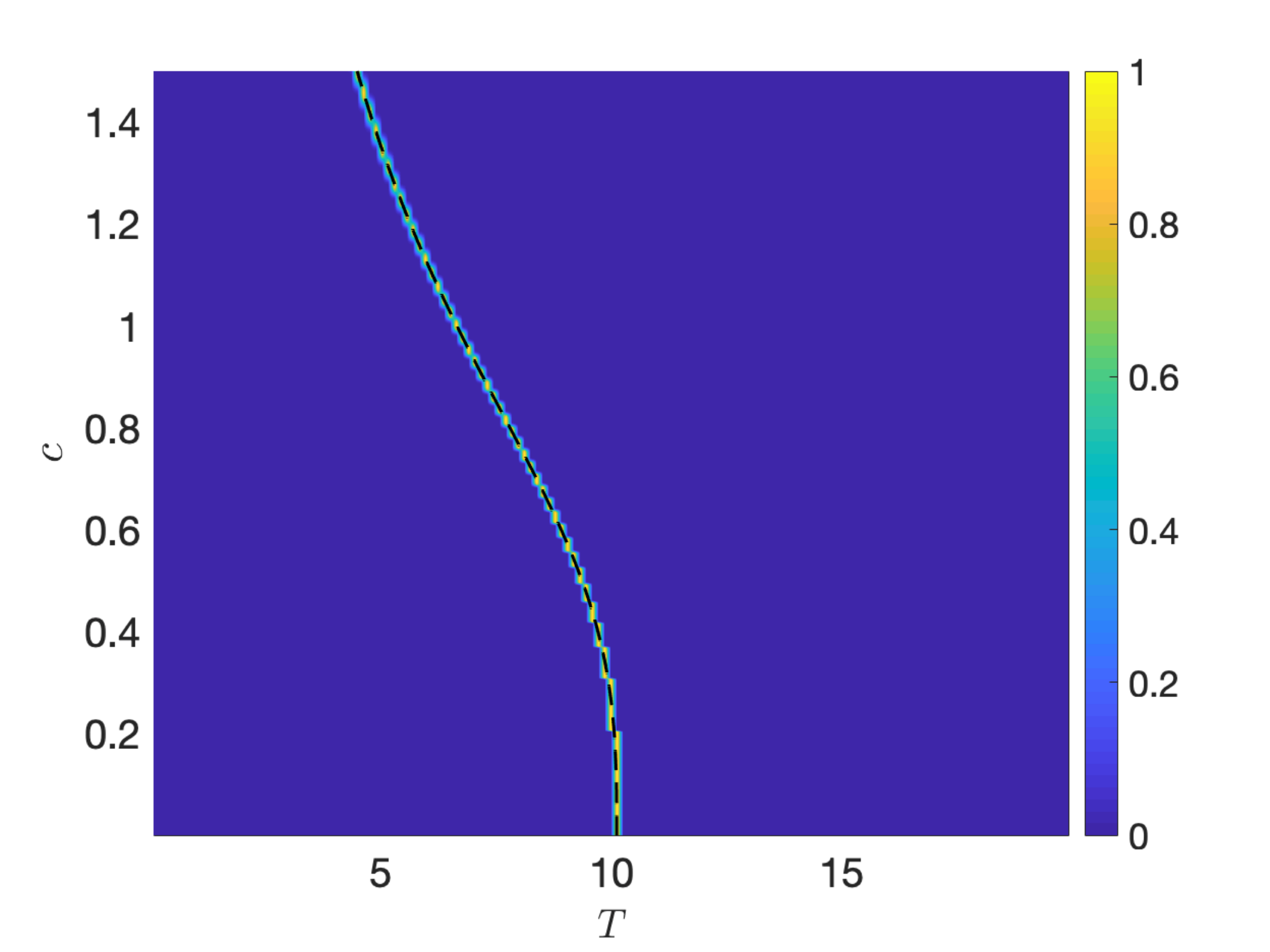}\qquad
		\includegraphics[width=0.45\textwidth]{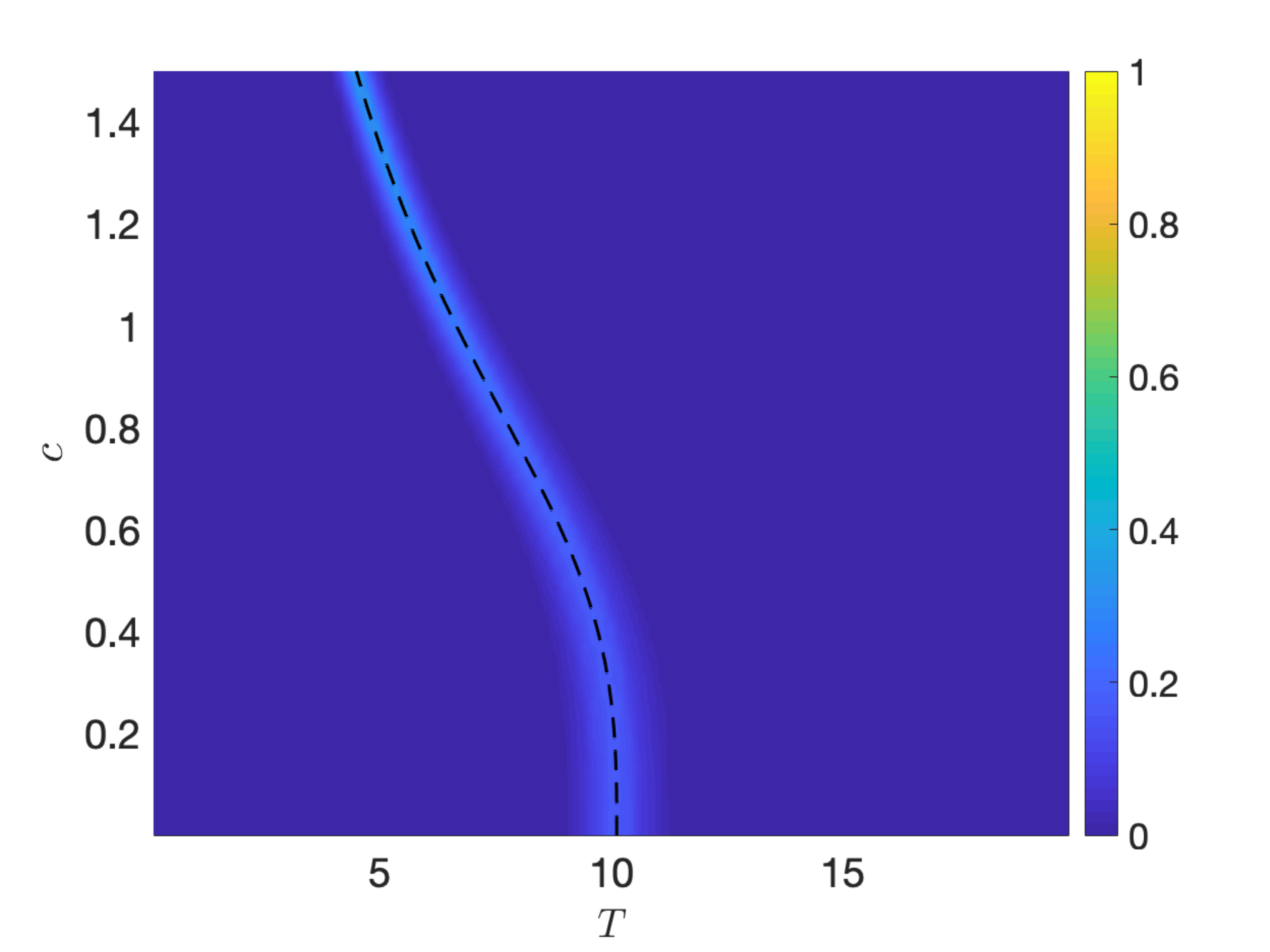}
		\caption{Probability distribution of the applied impulse traction, $T$, causing cavitation of radius $c$ in a radially inhomogeneous sphere, when $\rho=1$, $B=1$, and $\mu(R)$ is given by \eqref{eq:mur}, where $C_{2}$ follows a Gamma distribution with $\rho^{(2)}_{1}=405$ and $\rho^{(2)}_{2}=0.05$, while $C_{1}=4.05$ (left), or $C_{1}$ follows a Gamma distribution with $\rho^{(1)}_{1}=405$ and $\rho^{(1)}_{2}=0.01$ (right). The dashed black line corresponds to the expected bifurcation based only on mean value parameters.}\label{fig:TinhomstochR}
	\end{center}
\end{figure}
%%%%%%%%%%%%%

To examine the post-cavitation behaviour, we take the first derivative of $T$, given by \eqref{eq:sphere:T:inhom:lambda}, with respect to $\lambda_{b}$,
\begin{equation}\label{eq:sphere:dP:inhom:imp}
\begin{split}
\frac{\text{d}T}{\text{d}\lambda_{b}}&=-2C_{1}\left(\frac{1}{\lambda_{b}^2}+\frac{1}{\lambda_{b}^{5}}\right)\\
&-\frac{3C_{2}}{2}\frac{2\lambda_{b}^4+4\lambda_{b}^3+3\lambda_{b}^2+2\lambda_{b}+1}{\lambda_{b}^6+2\lambda_{b}^5+3\lambda_{b}^4+2\lambda_{b}^3+\lambda_{b}^2}\\
&-C_{2}\left(\lambda_{b}^3-1\right)\frac{4\lambda_{b}^3+6\lambda_{b}^2+3\lambda_{b}+1}{\lambda_{b}^8+2\lambda_{b}^7+3\lambda_{b}^6+2\lambda_{b}^5+\lambda_{b}^4}\\
&-C_{2}\left(\lambda_{b}^3-1\right)\frac{\left(2\lambda_{b}^4+4\lambda_{b}^3+3\lambda_{b}^2+2\lambda_{b}+1\right)\left(4\lambda_{b}^7+7\lambda_{b}^6+9\lambda_{b}^5+5\lambda_{b}^4+2\lambda_{b}^3\right)}{\left(\lambda_{b}^8+2\lambda_{b}^7+3\lambda_{b}^6+2\lambda_{b}^5+\lambda_{b}^4\right)^2}.
\end{split}
\end{equation}
Letting $\lambda_{b}\to1$ in \eqref{eq:sphere:dP:inhom:imp}, we obtain
\begin{equation}\label{eq:sphere:dT0:inhom}
\lim_{\lambda_{b}\to1}\frac{\text{d}T}{\text{d}\lambda_{b}}=-4C_{1}-2C_{2}=-2C_{1}-2\left(C_{1}+C_{2}\right)=-2\mu(0)-2\mu(B)<0.
\end{equation}
Hence, only snap cavitation can occur.

In Figures~\ref{fig:Tinhomstoch} and \ref{fig:TinhomstochR}, we show examples of post-cavitation stochastic behaviour of the static inhomogeneous unit sphere (with $B=1)$. 

%%%%%%%%%%%%%%%%%%%%%%%%%%%%%%%%%%%%%%%%%%%%%%%%%%%%%%%%%%%%
%%%%%%%%%%%%%%%%%%%%   NEW SECTION  %%%%%%%%%%%%%%%%%%%%%%%%
%%%%%%%%%%%%%%%%%%%%%%%%%%%%%%%%%%%%%%%%%%%%%%%%%%%%%%%%%%%%
\section{Conclusion}\label{sec:conclusion}

We have presented here a theoretical study of static and dynamic cavitation, and possible oscillatory motions, of homogeneous spheres of stochastic neo-Hookean material, of composite spheres with two concentric stochastic neo-Hookean phases, and of inhomogeneous spheres of neo-Hookean-like material with the material parameter varying continuously in the radial direction. For these materials, the elastic parameters are either spatially-independent random variables, or spatially-dependent random fields, described by Gamma probability density functions at a continuum level. 

In our approach, the radially-symmetric dynamic deformation of each sphere was treated as a quasi-equilibrated motion. Then, the static deformation was examined by reducing the quasi-equilibrated motion to a static equilibrium state. Universally, we find that the critical load at which a nontrivial cavitated solution appears is the same as for the homogeneous sphere composed entirely of the material found at its centre. However, there are important differences in the post-cavitation behaviours under the different constitutive formulations and loading conditions. These different conditions give us a variety of different behaviours, specifically:
\begin{itemize}
	\item For a neo-Hookean sphere, cavitation is stable and the motion is oscillatory under dead-load traction, while under the impulse traction, cavitation is unstable and the radial motion is non-oscillatory; 
	\item In composite spheres with two concentric neo-Hookean phases subject to either dead-load or impulse traction, both stable and unstable cavitation can be obtained, depending on the material parameters of the two phases. However, oscillatory motion only occurs when cavitation is stable under dead-load traction; 
	\item For radially inhomogeneous spheres under dead-load traction, both stable and unstable cavitation occur, and oscillatory motion is found when cavitation is stable under dead-load traction. Under impulse traction,  cavitation is unstable and the motion is non-oscillatory.
\end{itemize} 

Usually, in many application areas, the Gaussian (normal) distribution is assumed to describe the random variation that occurs in the available data. However, many material measurements show a skewed distribution where mean values are low, variances are large, and values cannot be negative (see, e.g., \cite{Fitt:2019:FWWM}). These are important features of the data that need to be incorporated in models. For small variability, the Gaussian and Gamma distributions both fit well to experimental data collected from material tests \cite{Fitt:2019:FWWM,Mihai:2019a:MDWG}. This guides us to choose the distribution that is adequate also for cases exhibiting large variability, i.e., the Gamma distribution. If the data cannot be smaller than a certain bound different from zero, then a shift parameter can be incorporated to define a three-parameter distribution. To compare directly our stochastic results with the deterministic ones, we sampled from distributions where the parameters were set to have mean values corresponding to the deterministic system. Thus, the mean value of the distribution is guaranteed to converge to the expected value.

We found that, in contrast to the deterministic problem where a single critical value strictly separates the stable and unstable cases, for the stochastic problem, there is a probabilistic interval where the stable and unstable states compete, in the sense that both have a quantifiable chance to be obtained with a given probability. This means that solutions exist in the stochastic case where deterministically they should not. Generally, for the stochastic solution, the variance changes non-uniformly around the mean value, suggesting that, as the variance increases, the average value becomes less significant from the physical point of view. A natural question that is yet to be consider is: what is the effect of individual parameters on the variance in the stochastic solution?

For the finite dynamic deformations, attention was given to the periodic (oscillatory) motion and the time-dependent stresses, while taking into account the stochastic model parameters. In this case, the amplitude and period of the oscillation of the stochastic bodies are characterised by probability distributions, and there is a parameter interval where both the oscillatory and non-oscillatory motions can occur with a given probability. For the oscillating spheres, a cavity would form and expand until its radius reaches a maximum value, then contract again to zero periodically. An immediate extension of this could be to consider similar spheres where there is pressure at the cavity surface as well. 

One could also apply the stochastic approach to the dynamic cavitation of spheres where the `base' neo-Hookean material is replaced, for example, by a material with the following strain-energy function \cite{Mihai:2019c:MDWG},
\begin{equation}\label{eq:W:stoch:2term}
\mathcal{W}(\lambda_{1},\lambda_{2},\lambda_{3})=\frac{\mu_{1}}{2m^2}\left(\lambda_{1}^{2m}+\lambda_{2}^{2m}+\lambda_{3}^{2m}-3\right)
+\frac{\mu_{2}}{2n^2}\left(\lambda_{1}^{2n}+\lambda_{2}^{2n}+\lambda_{3}^{2n}-3\right),
\end{equation}
where the exponents $m$ and $n$ are fixed, and the coefficients $\mu_{1}$ and $\mu_{2}$ are given, such that the shear modulus in infinitesimal deformation satisfies $\mu=\mu_{1}+\mu_{2}>0$. As is well know, in this case, assuming that the BE inequalities \eqref{eq:W:BE} hold, cavitation occurs if and only if \cite{Ball:1982,ChouWang:1989a:CWH,Horgan:1995:HP,Mihai:2019c:MDWG}
\begin{equation}\label{eq:sphere:mnbounds}
-\frac{3}{4}< m, n<\frac{3}{2}.
\end{equation}
Therefore, cavitation is expected in a neo-Hookean sphere (with $m=1$ and $n=0$), but not in a Mooney-Rivlin sphere (with $m=1$ and $n=-1$). The cases where $m\in\{-1/2, 1\}$ and $n=0$ are discussed also as particular examples in \cite{Ball:1982}, and the spheres where $m\in\{1/2, 3/4, 1, 5/4\}$ and $n=0$ are  analysed explicitly in \cite{ChouWang:1989a:CWH}. For these well-behaved spheres, under uniform dead-load traction, the cavitated solution is always stable, and it makes little difference whether the model coefficients are stochastic or not, although, in the stochastic case, the solution appears somewhat 'hazy' (albeit with the variance changing non-uniformly) around the mean value \cite{Mihai:2019c:MDWG}. 

Similarly, for a sphere made of a material described by \eqref{eq:W:stoch:2term} subject to uniform radial impulse traction, if the BE inequalities and \eqref{eq:sphere:mnbounds} are satisfied, then cavitation always occurs and making the model parameters stochastic only generates some dispersion around the usual deterministic solution. However, in this sphere, cavitation can also happen through a subcritical bifurcation where unstable (snap) cavitation takes place for values less than the critical dead load that causes cavitation. Hence, a sudden jump to a cavitated solution with a finite internal radius can be observed. For example, when the strain-energy density takes the form \cite{Mihai:2019c:MDWG} 
\begin{equation}\label{eq:W:stoch:2term:snap}
\mathcal{W}(\lambda_{1},\lambda_{2},\lambda_{3})=\frac{\mu_{1}}{2}\left(\lambda_{1}^{2}+\lambda_{2}^{2}+\lambda_{3}^{2}-3\right)
+2\mu_{2}\left(\lambda_{1}^{-1}+\lambda_{2}^{-1}+\lambda_{3}^{-1}-3\right),
\end{equation}
both stable and unstable cavitation can occur, depending on the values of the material coefficients $\mu_{1}$ and $\mu_{2}$ (see Figure~3 of \cite{Mihai:2019c:MDWG}). Then, in a stochastic sphere, the stable and unstable cases compete \cite{Mihai:2019c:MDWG}. 

We further note that the theorem on quasi-equilibrated motion cannot be applied when the material is compressible \cite[pp.~208-209]{TruesdellNoll:2004}. As our analysis relies on the notion of quasi-equilibrated motion, the same approach cannot be employed for compressible spheres. However, while stochastic versions of compressible hyperelastic materials can also be constructed \cite{Staber:2016:SG}, there are not many theoretical results available on oscillatory motions of finitely deformed compressible hyperelastic spheres \cite{Akyuz:1998:AE}. Nevertheless, elastodynamic problems involving more complex constrained, or unconstrained materials, can still be formulated, where the motion is not quasi-equilibrated, and the corresponding solution can then be tackled computationally. More generally, radial asymmetry, various behaviour laws, with or without damping, and thermal effects may also be of great theoretical and physical interest.

%%%%%%%%%%%%%%%%%%%%%%%%%%%%%%%%%%%
\paragraph{Acknowledgement.} The support by the Engineering and Physical Sciences Research Council of Great Britain under research grants EP/R020205/1 to Alain Goriely and EP/S028870/1 to L. Angela Mihai is gratefully acknowledged.

%%%%%%%%%%%%%%%%%%%%%%%%%%%%%

%%%%%%%%%%%%%%%%%%%

\begin{thebibliography}{9}

\bibitem{Abramowitz:1964} Abramowitz M, Stegun IA. 1964. Handbook of Mathematical Functions with Formulas, Graphs, and Mathematical Tables, National Bureau of Standards, Applied Mathematics Series, vol. 55, Washington.

\bibitem{Akyuz:1998:AE} Aky\"{u}z U, Ertepinar A. 1998. Stability and asymmetric vibrations of pressurized compressible hyperelastic cylindrical shells, International Journal of Non-Linear Mechanics 34, 391-404.

\bibitem{Aranda:2015:etal} Aranda-Iglesias D, Vadillo G, Rodr\'{i}guez-Martínez JA. 2015. Constitutive sensitivity of the oscillatory behaviour of hyperelastic cylindrical shells, Journal of Sound and Vibration 358, 199-216.

\bibitem{Aranda:2018:etal} Aranda-Iglesias D, Rodr\'{i}guez-Martínez JA, Rubin MB. 2018. Nonlinear axisymmetric vibrations of a hyperelastic orthotropic cylinders, International Journal of Non-Linear Mechanics 99, 131-143.

\bibitem{BakerEricksen:1954} Baker M, Ericksen JL. 1954. Inequalities restricting the form of stress-deformation relations for isotropic elastic solids and Reiner-Rivlin fluids, Journal of the Washington Academy of Sciences 44, 24-27.

\bibitem{Balakrishnan:1978:BS} Balakrishnan R, Shahinpoor M. 1978. Finite amplitude oscillations of a hyperelastic spherical cavity, International Journal of Non-Linear Mechanics 13, 171-176.

\bibitem{Ball:1982} Ball JM. 1982. Discontinuous equilibrium solutions and cavitation in nonlinear elasticity, Philosophical Transactions of the Royal Society A 306, 557-611.

\bibitem{Bayes:1763} Bayes T. 1763. An essay toward solving a problem in the doctrine of chances, Philosophical Transactions of the Royal Society 53, 370-418.

\bibitem{Beatty:2007} Beatty MF. 2007. On the radial oscillations of incompressible, isotropic, elastic and limited elastic thick-walled tubes, International Journal of Non-Linear Mechanics 42, 283-297.

\bibitem{Beatty:2011} Beatty MF. 2011. Small amplitude radial oscillations of an incompressible, isotropic elastic spherical shell, Mathematics and Mechanics of Solids 16, 492-512.

\bibitem{Calderer:1983} Calderer C. 1983. The dynamical behaviour of nonlinear elastic spherical shells, Journal of Elasticity 13, 17-47.

\bibitem{ChouWang:1989a:CWH} Chou-Wang M-S, Horgan CO. 1989. Void nucleation and growth for a class of incompressible nonfinearly elastic materials, International Journal of Solids and Structures 25, 1239-1254.

\bibitem{ChouWang:1989b:CWH} Chou-Wang M-S, Horgan CO.1989. Cavitation in nonlinear elastodynamics for neo-Hookean materials, International Journal of Engineering Science 27, 967-973.

\bibitem{Cidonio:2019:CGDO} Cidonio G, Glinka M, Dawson JI, Oreffo ROC. 2019. The cell in the ink: Improving biofabrication by printing stem cells for skeletal regenerative medicine, Biomaterials, doi: 10.1016/j.biomaterials.2019.04.009.

\bibitem{DePascalis:etal:2018} De Pascalis R, Parnell WJ, Abrahams ID, Shearer T, Daly DM, Grundy D. 2018. The inflation of viscoelastic balloons and hollow viscera, Proceedings of the Royal Society A 474, 20180102.

\bibitem{DeSimone:2009:dST} DeSimone A, Teresi L. 2009. Elastic energies for nematic elastomers, The European Physical Journal E 29, 191-204.

\bibitem{Elishakoff:2012:ES} Elishakoff I, Soize C (eds). 2012. Nondeterministic Mechanics, Springer, New York.
	
\bibitem{Ertepinar:1976:EA} Ertepinar A, Akay HU. 1976. Radial oscillations of nonhomogeneous, thick-walled cylindrical and spherical shells subjected to finite deformations, International Journal of Solids and Structures 12, 517-524.

\bibitem{Fitt:2019:FWWM} Fitt D, Wyatt H, Woolley TE, Mihai LA. 2019. Uncertainty quantification of elastic material responses: testing, stochastic calibration and Bayesian model selection, Mechanics of Soft Materials (doi: 10.1007/s42558-019-0013-1).

\bibitem{Fond:2001} Fond, C. 2001. Cavitation criterion for rubber materials: a review of void-growth models, Journal of Polymer Science: Part B 39, 2081-2096.

\bibitem{Freiling:1986} Freiling C. 1986. Axioms of symmetry: Throwing darts at the real number line, The Journal of Symbolic Logic 51(1), 190-200.

\bibitem{Fried:2004:FS} Fried E, Sellers S. 2004. Free-energy density functions for nematic elastomers, Journal of Mechanics and Physics of Solids 52, 1671-1689.

\bibitem{Gent:1991} Gent AN. 1991. Cavitation in rubber: a cautionary tale, Rubber Chemistry and Technology 63, G49-G53.

\bibitem{Gent:1959:GL} Gent AN, Lindley PB. 1959. Internal rupture of bonded rubber cylinders in tension, Proceedings of the Royal Society of London A 249, 195-205.

\bibitem{Ghanem:2017:GHOR} Ghanem R, Higdon D, Owhadi H (Eds.). 2017. Handbook of Uncertainty Quantification, Springer, New-York, 2017.

\bibitem{goriely17} Goriely A. 2017. The Mathematics and Mechanics of Biological Growth, Springer-Verlag, New York.

\bibitem{Green:1970:GA} Green AE, Adkins JE. 1970. Large Elastic Deformations (and Non-linear Continuum Mechanics), 2nd ed, Oxford University Press, Oxford.

\bibitem{Grimmett:2001:GS} Grimmett GR, Stirzaker DR. 2001. Probability and Random Processes, 3rd ed, Oxford University Press, Oxford.

\bibitem{Heng:1963:HS} Heng GZ, Solecki R. 1963. Free and forced finite amplitude oscillations of an elastic thick-walled hollow
sphere made of incompressible material, Archiwum Mechaniki Stosowanej 3, 427-433.

\bibitem{Holmes:2019} Holmes DP. 2019. Elasticity and stability of shape-shifting structures, Current Opinion in Colloid \& Interface Science 40, 118-137.

\bibitem{Horgan:1989:HP} Horgan CO, Pence TJ. 1989. Cavity formation at the center of a composite incompressible nonlinearly elastic sphere, Journal of Applied Mechanics 56, 302-308.

\bibitem{Horgan:1995:HP} Horgan CO, Polignone DA. 1995. Cavitation in nonlinearly elastic solids: a review, Applied Mechanics Reviews 48, 471-485.

\bibitem{Hughes:2010:HH} Hughes I, Hase TPA. 2010. Measurements and Their Uncertainties: A Practical Guide to Modern Error Analysis, Oxford University Press, Oxford.

\bibitem{Hutchens:2016:HFC} Hutchens SB, Fakhouri S, Crosby AJ. 2016. Elastic cavitation and fracture via injection, Soft Matter 12, 2557.

\bibitem{Jaynes:1957a} Jaynes ET. 1957. Information theory and statistical mechanics i, Physical Review 108, 171-190.

\bibitem{Jaynes:1957b} Jaynes ET. 1957. Information theory and statistical mechanics ii, Physical Review 106, 620-630.

\bibitem{Jaynes:2003} Jaynes ET. 2003. Probability Theory: The Logic of Science, Cambridge University Press, Cambridge, UK.

\bibitem{Johnson:1994:JKB} Johnson NL, Kotz S, Balakrishnan N. 1994. Continuous Univariate Distributions, Vol 1, 2nd edition, John Wiley \& Sons, New York.

\bibitem{Kaminski:2018:KL} Kaminski M, Lauke B. 2018. Probabilistic and stochastic aspects of rubber hyperelasticity, Meccanica 53, 2363-2378.

\bibitem{Kang:2018:KWT} Kang J, Wang C, Tan H. 2018. Cavitation in inhomogeneous soft solids, Soft Matter 14,7979-7986.

\bibitem{Knowles:1960} Knowles JK. 1960. Large amplitude oscillations of a tube of incompressible elastic material, Quarterly of Applied Mathematics 18, 71-77.

\bibitem{Knowles:1962} Knowles JK. 1962. On a class of oscillations in the finite-deformation theory of elasticity, Journal of Applied Mechanics 29, 283-286.

\bibitem{Knowles:1965:KJ} Knowles JK, Jakub MT. 1965. Finite dynamic deformations of an incompressible elastic medium containing a spherical cavity, Archive of Rational Mechanics and Analysis 18, 376-387.

\bibitem{Marzano:1983} Marzano M. 1983. An interpretation of Baker-Ericksen inequalities in uniaxial deformation and stress, Meccanica 18, 233-235.

\bibitem{McGrayne:2012} McGrayne SB. 2012. The Theory That Would Not Die: How Bayes' Rule Cracked the Enigma Code, Hunted Down Russian Submarines, an Emerged Triumphant from Two Centuries of Controversy, Paperback ed., Yale University Press, New Haven.

\bibitem{Mihai:2019a:MDWG} Mihai LA, Fitt D, Woolley TE, Goriely A. 2019. Likely equilibria of stochastic hyperelastic spherical shells and tubes, Mathematics and Mechanics of Solids, 24(7), 2066-2082 (doi: 10.1177/1081286518811881).

\bibitem{Mihai:2019b:MDWG} Mihai LA, Fitt D, Woolley TE, Goriely A. 2019. Likely oscillatory motions of stochastic hyperelastic solids, Transactions of Mathematics and Its Applications 3(1), tnz003 (doi: 10.1093/imatrm/tnz003).

\bibitem{Mihai:2019c:MDWG} Mihai LA, Fitt D, Woolley TE, Goriely A. 2019. Likely cavitation in stochastic elasticity, Journal of Elasticity 137(1), 27-42 (doi: 10.1007/s10659-018-9706-1).

\bibitem{Mihai:2017:MG} Mihai LA, Goriely A. 2017. How to characterize a nonlinear elastic material? A review on nonlinear constitutive parameters in isotropic finite elasticity, Proceedings of the Royal Society A 473, 20170607 (doi: 10.1098/rspa.2017.0607).

\bibitem{Mihai:2018:MWG} Mihai LA, Woolley TE, Goriely A. 2018. Stochastic isotropic hyperelastic materials: constitutive calibration and model selection, Proceedings of the Royal Society A 474, 20170858.

\bibitem{Mihai:2019a:MWG} Mihai LA, Woolley TE, Goriely A. 2019. Likely equilibria of the stochastic Rivlin cube, Philosophical Transactions of the Royal Society A 377, 20180068 (doi: 10.1098/rsta.2018.0068).

\bibitem{Mihai:2019b:MWG} Mihai LA, Woolley TE, Goriely A. 2019. Likely chirality of stochastic anisotropic hyperelastic tubes, International Journal of Non-Linear Mechanics 114, 9-20 (doi: 10.1016/j.ijnonlinmec.2019.04.004).

\bibitem{Moschopoulos:1985} Moschopoulos PG. 1985. The distribution of the sum of independent Gamma random variables 37, 541-544.

\bibitem{Mumford:2000} Mumford D. 2000. The dawning of the age of stochasticity, Mathematics: Frontiers and Perspectives,  
V. Arnold, M. Atiyah, P. Lax, B. Mazur (eds.), American Mathematical Society, 197-218. 

\bibitem{Oden:2018} Oden JT. 2018. Adaptive multiscale predictive modelling, Acta Numerica 27, 353-450.

\bibitem{Ogden:1997} Ogden RW. 1997. Non-Linear Elastic Deformations, 2nd ed, Dover, New York.

\bibitem{Ostoja:2007} Ostoja-Starzewski M. 2007. Microstructural Randomness and Scaling in Mechanics of Materials, Chapman and Hall, CRC Press.

\bibitem{Poulain:2017:PLLPRC} Poulain X, Lef\`{e}vre V, Lopez-Pamies O, Ravi-Chandar K. 2017. Damage in elastomers: nucleation and growth of cavities, micro-cracks, and macro-cracks, International Journal of Fracture 205, 1-21.

\bibitem{Poulain:2018:PLPRC} Poulain X, Lopez-Pamies O, Ravi-Chandar K. 2018. Damage in elastomers: Healing of internally nucleated cavities and micro-cracks, Soft Matter 14, 4633-4640.

\bibitem{Raayai:2019:etal} Raayai-Ardakani S,  Earla DR,  Cohen T. 2019. The intimate relationship between cavitation and fracture, Soft Matter 15, 4999.

\bibitem{Ren:2008} Ren J.-s.. 2008. Dynamical response of hyper-elastic cylindrical shells under periodic load, Applied Mathematics and Mechanics 29, 1319-1327.

\bibitem{Ren:2009} Ren J.-s.. 2009.  Dynamics and destruction of internally pressurized incompressible hyper-elastic spherical shells, International Journal of Engineering Science 47, 745-753.

\bibitem{Riggs:2017:RL} Riggs JD, Lalonde TL. 2017. Handbook for Applied Modeling: Non-Gaussian and Correlated Data, Cambridge University Press, Cambridge, UK.

\bibitem{Robert:2007} Robert CP. 2007. The Bayesian Choice: From Decision-Theoretic Foundations to Computational Implementation, 2nd ed, Springer, New York.

\bibitem{Martinez:2015:etal} Rodri\'{i}guez-Marti\'{i}nez JA, Fern\'{a}ndez-S\'{a}aez J, aera R. 2015. The role of constitutive relation in the stability of hyper-elastic spherical membranes subjected to dynamic inflation, International Journal of Engineering Science 93, 31-45.

\bibitem{Shannon:1948} Shannon CE. 1948. A mathematical theory of communication, Bell System Technical Journal 27, 379-423, 623-659.

\bibitem{Sivaloganathan:1991} Sivaloganathan, I. 1991. Cavitation, the incompressible limit, and material inhomogeneity, Quarterly of Applied Mathematics 49, 521-541.

\bibitem{Soares:etal:2019} Soares RM, Amaral PFT, Silva FMA, Gon\c{c}alves PB. 2019. Nonlinear breathing motions and instabilities of a pressure-loaded spherical hyperelastic membrane, Nonlinear Dynamics 1-22 (doi: 10.1007/s11071-019-04855-4). 

\bibitem{Soize:2000} Soize C. 2000. A nonparametric model of random uncertainties for reduced matrix models in structural dynamics, Probabilistic Engineering Mechanics 15, 277-294.

\bibitem{Soize:2001} Soize C. 2001. Maximum entropy approach for modeling random uncertainties in transient elastodynamics, Journal of the Acoustical Society of America 109, 1979-1996.

\bibitem{Soize:2006} Soize C. 2006. Non-Gaussian positive-definite matrix-valued random fields for elliptic stochastic partial differential operators, Computer Methods in Applied Mechanics and Engineering 195, 26-64.

\bibitem{Soize:2013} Soize C. 2013. Stochastic modeling of uncertainties in computational structural dynamics - Recent theoretical advances, Journal of Sound and Vibration 332, 2379-2395.

\bibitem{Soize:2017} Soize C. 2017. Uncertainty Quantification: An Accelerated Course with Advanced Applications in Computational Engineering, Interdisciplinary Applied Mathematics Book 47, Springer, New York.

\bibitem{Soni:2017:SG} Soni J, Goodman R. 2017. A Mind at Play: How Claude Shannon Invented the Information Age, Simon \& Schuster, New York.

\bibitem{Staber:2015:SG} Staber B, Guilleminot J. 2015. Stochastic modeling of a class of stored energy functions for incompressible hyperelastic materials with uncertainties, Comptes Rendus M\'{e}canique 343, 503-514.

\bibitem{Staber:2016:SG} Staber B, Guilleminot J. 2016. Stochastic modeling of the Ogden class of stored energy functions for hyperelastic materials: the compressible case, Journal of Applied Mathematics and Mechanics/Zeitschrift f\"{u}r Angewandte Mathematik und Mechanik 97, 273-295.

\bibitem{Staber:2017:SG} Staber B, Guilleminot J. 2017. Stochastic hyperelastic constitutive laws and identification procedure for soft biological tissues with intrinsic variability, Journal of the Mechanical Behavior of Biomedical Materials 65, 743-752.

\bibitem{Staber:2018:SG} Staber B, Guilleminot J. 2018. A random field model for anisotropic strain energy functions and its application for uncertainty quantification in vascular mechanics, Computer Methods in Applied Mechanics and Engineering 333, 94-113.

\bibitem{Staber:2019:SGSMI} Staber B, Guilleminot J, Soize C, Michopoulos J, Iliopoulos A. 2019. Stochastic modeling and identification of an hyperelastic constitutive model for laminated composites, Computer Methods in Applied Mechanics and Engineering  347, 425-444.

\bibitem{Sullivan:2015} Sullivan TJ. 2015. Introduction to Uncertainty Quantification, Springer-Verlag, New York.

\bibitem{Truesdell:1962} Truesdell C. 1962. Solutio generalis et accurata problematum quamplurimorum de motu corporum elasticorum incomprimibilium in deformationibus valde magnis, Archive of Rational Mechanics and Analysis 11, 106-113.

\bibitem{TruesdellNoll:2004} Truesdell C, Noll W. 2004. The Non-Linear Field Theories of Mechanics, 3rd ed, Springer-Verlag, New York.

\bibitem{Verron:1999:VKDR} Verron E, Khayat RE, Derdouri A, Peseux B. 1999. Dynamic inflation of hyperelastic spherical membranes, Journal of Rheology 43, 1083-1097.

\bibitem{Wang:1965} Wang CC. 1965. On the radial oscillations of a spherical thin shell in the finite elasticity theory, Quarterly of Applied Mathematics 23, 270-274.

\bibitem{Wang:1972:WE} Wang CC, Ertepinar A. 1972. Stability and vibrations of elastic thick-walled cylindrical and spherical shells subjected to pressure, International Journal of Non-Linear Mechanics 7, 539-555.

\bibitem{Warner:2007:WT} Warner M, Terentjev EM. 2007. Liquid Crystal Elastomers, paper back, Oxford University Press, Oxford, UK.

\end{thebibliography}
\end{document}